\documentclass[preprint,aps,11pt,nofootinbib]{revtex4}
\usepackage{mathtools}
\usepackage{graphics}
\usepackage{graphicx}
\usepackage{dcolumn}
\usepackage{bm}
\usepackage{dsfont} 
\usepackage{amsmath,amssymb}
\usepackage{hyperref}
\usepackage{tabularx}
\usepackage[normalem]{ulem}
\hypersetup{
	colorlinks=true,
	linkcolor=blue,
	filecolor=magenta,      
	urlcolor=blue,
	citecolor=blue
}
\usepackage{float}
\usepackage{comment}
\newcommand\be{\begin{equation}}
\newcommand\ba{\begin{eqnarray}}
\newcommand\ee{\end{equation}}
\newcommand\ea{\end{eqnarray}}

\begin{document}

\title{Tree-level Graviton Scattering in the Worldline Formalism}

\author{Yuchen Du}
\affiliation{
 Department of Physics, University of Virginia, Charlottesville, Virginia 22904, USA
}

\author{Diana Vaman}
\affiliation{
 Department of Physics, University of Virginia, Charlottesville, Virginia 22904, USA
}

\date{\today}

\begin{abstract} 
	We use the worldline formalism to study tree-level scattering processes involving gravitons.  
A massless spin 2 particle is described by an $N=4$ supersymmetric worldline action which is also $O(4)$ symmetric. More generally,  $N=2S$ supersymmetric worldline actions exhibiting $O(N)$ symmetry describe free spin $S$ particles.  
Recently a BRST approach was used to construct the on-shell background graviton emission vertex from a graviton worldline.  Nonetheless, an action describing the coupling  of higher spin ($S\geq 2$) particles with generic background gravity is unknown.
In this paper we found that in order to reproduce Einstein's general relativity 3-point graviton vertex,
interpreted as  the emission of an off-shell graviton from the worldline,  the coupling to background gravity must break the $O(4)$ symmetry to $O(2) \times O(2)$.  
In addition to this symmetry-breaking feature, we also found that the coefficient $\beta$ of the worldline action counterterm $\beta R$ differs from previous results in the literature.

By comparing the linearized graviton and photon emission vertex operators from different worldlines, we noticed that they obey a squaring relation.  For MHV (Maximal Helicity Violating) amplitudes, these squaring relations among the linearized vertex operators directly result in double-copy-like relations between the scattering amplitudes. 
\end{abstract}

\maketitle

{
    \hypersetup{linkcolor=blue}
    \tableofcontents
}

\section{Introduction}
\noindent In the 1990s, Bern and Kosower \cite{Bern:1991aq} successfully studied the infinite tension limit of string amplitudes (which is the limit that restricts the infinite tower of string states to a handful of massless states). They gave a master formula and a set of rules, which allowed the calculation of one-loop N-point amplitudes efficiently.
Soon after, Strassler \cite{Strassler:1992zr} derived their formula and rules, without using string theory or Feynman diagrams but through a rewriting of the field theory loop amplitudes in terms of a % first-quantized 
path integral over point-particle coordinates, i.e. the integral along the worldline of a particle. This was the beginning of the so-called ”worldline formalism”.

String theory identifies relations between the scattering amplitude among particles with different spins. This arises because the closed string is (almost) the ”square” of an open string.  Most notable is the Kawai-Lewellen-Tye (KLT) relation \cite{Kawai:1985xq}, which states that gravity amplitudes can be decomposed as sums of products of the amplitudes of gauge bosons. This, combined with the Bern-Carrasco-Johanson (BCJ) relation \cite{Bern:2008qj}, yields the double copy relation \cite{Bern:2010ue}.  The latter states that the numerator of each pole structure in a gravity amplitude is exactly the square of those in the corresponding gauge boson amplitude. However, these relations are nowhere to be seen in the field theory since the Lagrangians of general relativity and Yang-Mills theory are completely different. When compared with the string amplitude calculation, which is a first quantized calculation, one can see that the worldline approach shares a similar structure.  This is not surprising since a worldline can be considered the skeleton of a string worldsheet. The similarity between string theory calculation and worldline formalism  motivates us to explore whether worldline formalism allows us to relate scattering amplitudes among different particles.

Our work has several purposes.  First, we construct a worldline Lagrangian which describes the emission of an off-shell  graviton from the worldline for spins $S=0,1,2$.  This is relevant for the classical limit of scattering amplitudes where,  for example, the gravitational interaction between two particles is computed by integrating out the soft background off-shell gravitons that mediate the interaction between the particles' dressed worldlines \cite{Mogull:2020sak}.  

In worldline formalism, the vertex emission operators can be extracted from the worldline Lagrangian  of a particle coupled to background fields and Taylor-expanding in the background fields.  For example, the linearized vertex operator,  which is obtained as the leading order term in the difference between the interacting and non-interacting Lagrangians, describes the emission of an off-shell background quantum from the worldline.  In this paper, we construct an action that yields the cubic field-theoretical interaction vertex between on-shell particles with spin $S=0,1,2$ and off-shell background gravitons. From a first-quantized perspective,  particles with spin are described by supersymmetric worldlines. The spin degrees of freedom are realized on the worldline by Grassmann-odd coordinates.  A free, massless spin 2 particle is described by an $N=4$ supersymmetric worldline action which is also $O(4)$ symmetric \cite{Howe:1988ft}. Spin $S$ free particles are described through $N=2S$ supersymmetric worldline actions exhibiting $O(N)$ symmetry \cite{Bastianelli:2011cc}.  Coupling with background gravity imposes restrictions for the background geometry if worldline supersymmetry is to be preserved (in \cite{Howe:1988ft} the conclusion was that $N=4$ supersymmetry constrains the background curvature to vanish; in \cite{Kuzenko:1995mg} it was shown that  $N\geq 4$ supersymmetry could be preserved in an anti de-Sitter background; in \cite{Bastianelli:2008nm} it was demonstrated that the $N$-supersymmetric particle can be consistently coupled with a conformally flat background). 
Recently, in \cite{Bonezzi:2018box} a BRST approach was used to construct the on-shell background graviton emission vertex from a graviton worldline.  Nonetheless, an action describing the coupling  of higher spin ($S\geq 2$) particles with generic background gravity is unknown.

Our approach was to constrain the worldline action by requiring that the term linear in the off-shell background field  yields a 3-point vertex that matches the 3-point QFT vertex. 
In particular, we found that to reproduce general relativity's cubic graviton vertex,
interpreted as  the emission of an off-shell graviton from the worldline,  the coupling to background gravity must break the $O(4)$ symmetry to $O(2) \times O(2)$.  

Another difference with past results in the literature pertains to the so-called conterterms.  When gravity is involved, applying the worldline formalism requires a regularization scheme (to properly define the products of distributions arising from Wick contractions on the worldline) followed by a renormalization (matching) condition.  This results in adding terms of the form $\beta R$ on the worldline, called counterterms. We calculated the 3-point vertex (with the background particle taken off-shell) and 4-point amplitudes for several different worldlines coupled with background gravity. We matched the 3-point vertex with the corresponding field theory result to determine the coefficients of the counterterms; we also found the coefficients of the background gravitational interaction with the graviton worldline, which are responsible for the breaking of the $O(4)$ symmetry to $O(2)\times O(2)$. We verified them with a calculation of 4-point amplitudes.

We noticed that the coefficient $\beta$ of the counterterms needed for the calculation of %tree-level 
scattering amplitudes in a background gravitational field differs from previous results in the literature. {This difference is the result of imposing a different matching condition. }

There have been extensive applications of the worldline formalism in studying loop diagrams in QFT, e.g. %calculating one-loop amplitudes, effective actions, and beta functions
\cite{Bastianelli:2002fv}, \cite{Bastianelli:2019xhi}, \cite{Ahmadiniaz:2020htz}, \cite{Edwards:2021vhg}.  But very few works are related to tree-level diagrams when gravity is involved: in \cite{Bonezzi:2018box}, the 3-point graviton scattering amplitude was calculated, while in \cite{Ahmadiniaz:2019ppj}, the calculation involves only one external graviton. The counterterms were irrelevant in the previous works \cite{Bonezzi:2018box},\cite{Ahmadiniaz:2019ppj} since the background gravitons were put on-shell (which implies that $R=0$ to linear order in the background metric fluctuation).

Recently, there has been a lot of interest in applying the worldline formalism to general relativity. With some twists, the formalism becomes the so-called Worldline Quantum Field Theory (WQFT). In WQFT, a spinning, massive compact object (e.g. black holes, neutron stars) is described by a spinning particle. A correspondence between the eikonal phase and the free energy of  WQFT was established in \cite{Mogull:2020sak}. This correspondence then allows the study of different phenomena, such as black hole scattering \cite{Mogull:2020sak}, gravitational bremsstrahlung \cite{Jakobsen:2021smu}, light bending \cite{Bastianelli:2021nbs} and radiation-reaction \cite{Jakobsen:2022psy} with relative ease. The counterterms were ignored in \cite{Mogull:2020sak} due to the classical limit taken in that paper.  {However, for spinning particles, the spin effects are captured by couplings of the type $R_{\mu\nu\rho\sigma}S^{\mu\nu}S^{\rho\sigma}$ where $R_{\mu\nu\rho\sigma}$ is the background curvature and $S^{\mu\nu}$ is the spin operator. These couplings are of the same order as the counterterms $\beta R$, which implies that these counterterms could be important when considering spin effects within the framework of WQFT.}

%One is to determine the counterterms needed in the worldline action to ensure the correctness of the tree amplitudes, which is important for future application of the formalism to tree-level related calculations involving gravity. 

Another purpose of our paper is to inquire whether the worldline formalism leads naturally to relationships among the scattering amplitudes involving different interactions. These relations are far from evident from a field theory perspective.  We noticed that the  photon and graviton linearized vertex operators, which are derived from the worldline action, 
obey a squaring relation. We computed tree-level 4-point amplitudes and higher n-point MHV (Maximal Helicity Violating) amplitudes in the worldline formalism and found that the squaring relations, together with a choice of reference twistors in writing the positive and negative helicity external-line factors,  yield the double-copy relation between gravity and Yang-Mills scattering amplitudes.

The paper is organized as follows. In section \ref{wlformalism}, we describe the general formalism and analyze scalar QED as an example. In section \ref{wlwithgravity}, we discuss how particles with different spins interact with background gravity and with background gauge fields. We also calculate the  3-point vertex and 4-point amplitudes to determine the counterterms needed for each worldline and the coupling of the Grassmann-odd worldline coordinates with background gravity.  
In section \ref{mhv}, we consider the MHV amplitudes and show that the worldline formalism easily leads to the double copy relation between the amplitudes. We conclude in section \ref{conclusions3}. We relegate some of the more technical details to appendices.

%%%%%%%%%%%%%%%%%%%%%%%%%%%%%%%%%%%%%%%%%%%%%%%
\section{General formalism}\label{wlformalism}
\subsection{Path integral formulation of worldline formalism }
\noindent We begin by reviewing a few elements of the worldline formalism. Consider the worldline action of a spin 0 (scalar) particle:
\be
S=\int d\tau \mathcal{L}(x(\tau), \dot{x}(\tau))\,
\ee
with the worldline Lagrangian decomposed in a free part ${\cal L}_0$ and an interacting part $V$ describing the interactions between the scalar particle and a background field.
The position-space propagator can be expressed as a path integral
\ba
G(X', X)&=&\langle X'|\frac{1}{\Box-m^2+i\epsilon}|X\rangle=\int_0^\infty dT \langle X'|\exp(i T(\Box-m^2+i\epsilon))|X\rangle\nonumber\\
&=&\int_0^\infty dT  \int_{-\infty}^\infty \frac{d^4p}{(2\pi)^4} e^{i p\cdot (X'-X)}e^{-iT(p^2+m^2-i\epsilon)}\nonumber\\
&=&\int _0^\infty dT \int_{\stackrel{ x^\mu(T)= X'{}^\mu}{ x^\mu(0)= X^\mu}} {\cal D}x(\tau)\int {\cal D}p(\tau)e^{-i \int_0^T d\tau( p^2(\tau)+m^2 - p\cdot \dot x-i\epsilon)}\nonumber\\
&=&\int _0^\infty dT\int_{\stackrel{\tilde x^\mu(T)=X'{}^\mu}{ x^\mu(0)= X^\mu}} {\cal D}x(\tau)
\exp(i \int_0^T d\tau {\cal L}_0[x,\dot x])\,.
\ea
Accounting for the interactions with the background field, the propagator takes the form
\be
\Gamma[X',X] = \int^{\infty}_0 dT \int_{\stackrel{ x^\mu(T)=X'{}^\mu}{ x^\mu(0)= X^\mu}} \mathcal{D}x(\tau)\ e^{-\int^T_0 d\tau \mathcal{L}(x(\tau), \dot{x}(\tau))}\,,
\ee
where we have also performed a Wick-rotation $\tau\to i \tau$.
The relationship between the dressed propagator and scattering amplitudes can be seen after performing a plane-wave expansion for the background field
\be
	V= \sum^{N-1}_{i=2} V_i(x(\tau),\dot{x}(\tau))
\ee
where $N$ is the number of all the particles involved in the interaction and $(N-2)$ is the number of background particles. From now on, we abbreviate $V_i(x(\tau), \dot{x}(\tau))$ as $V_i(\tau)$. Expanding the dressed propagator and keeping the term corresponding to an interaction with $N-2$ background particles, we define
\be
\Gamma_N[X',X] = \int^{\infty}_0 dT \int^{x(T)=x}_{x(0)=x'} \mathcal{D}x(\tau)\ e^{-\int^T_0 d\tau \mathcal{L}_0} \prod^{N-1}_{i=2}\bigg( \int^T_0 d\tau_i\ V_i(\tau_i)\bigg)\,,
\ee
and in momentum space
\be
\Gamma_N[p,p'] = \int d^4 x \int d^4 x' e^{i p\cdot x} 
e^{i p' \cdot x'} \Gamma[x',x] \,.
\ee
Anticipating, the final expression will be of the form
\be
\Gamma_N[p,p'] = \frac{1}{p^2 + m^2} \mathcal{A}_N \frac{1}{p'^2 + m^2}
\ee
where $\mathcal{A}_N$ is the scattering amplitude that we need.

Although it is possible to directly compute the path integral, the calculation can be quite tedious. So we choose to turn the path integral into another equivalent form which involves using operators. On the one hand, with operators, we could use tricks such as Wick contraction just as in quantum field theory to simplify the calculation. On the other hand, the expression in terms of operators takes a similar form to that in string theory and has some nice features which we will exploit.

%%%%%%%%%%%%%%%%%%%%%%%%%%%%%%%%%%%%%%%%%%%%%%%%%%%%%
\subsection{From the path integral to the operator approach}
\noindent 
Defining the transition amplitude for the particle at $x$ at $\tau=0$ and at $x'$ at $\tau=T$
\be
K(x',T; x,0)=\int_{\stackrel{\tilde x^\mu(T)=X'{}^\mu}{ x^\mu(0)= X^\mu}} \mathcal{D}x(\tau)\ e^{-\int^T_0 d\tau \mathcal{L}(x(\tau), \dot{x}(\tau))}=\langle x',T|U_I(T;0)|x,0\rangle\,,\label{transition}
\ee
where $U_I(T,0)$ is the interaction-picture unitary time-evolution operator between the worldline times $\tau=0$ and $\tau=T$,
the dressed propagator is the integrated  transition amplitude
\be
\Gamma[x',x]=\int_0^\infty d T \,K(x',T; x,0)\label{dprop}
\ee
Expanding the time-evolution operator and keeping the  term corresponding to $N-2$  interactions, the dressed propagator $\Gamma_N[x,x']$ can be written as 
\ba
\Gamma_N[x,x'] = \int^{\infty}_{0} dT \bigg(\prod_{i=2}^{N-1}\int^{T}_{0} d\tau_{i}\bigg) \langle x, \tau_N=T|\mathcal{T} \{V_{N-1}(\tau_{N-1}) V_{N-2}(\tau_{N-2})...V_{2}(\tau_{2})\}|x', \tau_1=0\rangle \, , \nonumber\\
\ea
where $\mathcal T$ denotes time-ordering of the interaction picture $V_i(\tau_i)$ operators, which have the interpretation of background-particle emission vertex operators.
After the Fourier transform, we have
\ba
	\Gamma_N[p,p'] &=& \int d^4x \int d^4{x'}\ e^{i p \cdot x} e^{i p' \cdot x'} \Gamma_N[x, x']\nonumber\\
	&=& \int d^4x \int d^4{x'} \langle p|x\rangle   \Gamma_N[x,x'] \langle x'|p'\rangle \nonumber\\ 
	&=&  \int^{+\infty}_{0}dT\ (\prod_{i=2}^{N-1}\int^{T}_{0} d\tau_{i}) \langle p, \tau_N=T|\mathcal{T}\{V_{N-1}(\tau_{N-1}) V_{N-2}(\tau_{N-2})...V_{2}(\tau_{2})\}|p', \tau_1=0\rangle  \nonumber\\
\ea
For simplicity and without losing generality, we take $N=4$ case 
as an example
\ba
	\Gamma_4[p,p'] & = &\int^{+\infty}_{0}dT \int^{T}_{0} d\tau_{3} \int^{\tau_3}_{0} d\tau_{2} \langle p, \tau_4=T|\mathcal{T}\{V_{3}(\tau_3)V_{2}(\tau_2)\}|p', \tau_1=0\rangle  \nonumber\\
	&=& \int^{+\infty}_{0}dT \bigg[ \int_0^T d \tau_{32} \int_0^T d\tau_{21} \Theta(T-\tau_{32}-\tau_{21})\langle p|e^{-H_0 \tau_{43}}V_3(0) e^{-H_0 \tau_{32}}V_2(0)e^{-H_0\tau_{21}}|p'\rangle \nonumber \\
 &&+\int_0^T d \tau_{23} \int_0^T d\tau_{31} \Theta(T-\tau_{23}-\tau_{31})\langle p|e^{-H_0 \tau_{42}}V_2(0) e^{-H_0 \tau_{23}}V_3(0)e^{-H_0\tau_{31}}|p'\rangle \bigg]\nonumber \\
	&=& \int^{+\infty}_{0}d\tau_{43} \int^{+\infty}_{0} d\tau_{32} \int^{+\infty}_{0} d\tau_{21} \langle p|e^{-H_{0}\tau_{43}}\ V_{3}(0)\ e^{-H_{0}\tau_{32}}\ V_{2}(0)\ e^{-H_{0}\tau_{21}}|p'\rangle  \nonumber
 \\
 &&+\int^{+\infty}_{0}d\tau_{42} \int^{+\infty}_{0} d\tau_{23} \int^{+\infty}_{0} d\tau_{31} \langle p|e^{-H_{0}\tau_{42}}\ V_{2}(0)\ e^{-H_{0}\tau_{23}}\ V_{3}(0)\ e^{-H_{0}\tau_{31}}|p'\rangle
\nonumber\\
 &=& \frac{1}{p^2+m^2} \bigg(
\int^{+\infty}_{0} d\tau_{32} \langle p|V_{3} e^{-H_{0}\tau_{32}} V_{2}|p'\rangle  +\int^{+\infty}_{0} d\tau_{23} \langle p|V_2 e^{-H_{0}\tau_{23}} V_{3}|p'\rangle \bigg) 
 \frac{1}{p'^2+m^2}\label{g4}\,,
\ea
where in the intermediate steps we used the notation $\tau_4=T, \tau_1=0, \tau_{ij}=\tau_i-\tau_j$. We also wrote $V_i(0)$ as $V_i$ to simplify notation. $H_0$ is the free Hamiltonian that governs the evolution on the worldline, and we performed the Wick rotation $\tau \rightarrow i\tau$.\par 
Clearly, the left and right factors are just propagators. Amputating the propagators, and setting the momenta $p$ and $p'$ on-shell, yields the scattering amplitude
\ba
	\mathcal{A}_4 &=& \int^{+\infty}_{0} d\tau_{32} \langle p|V_{3}\ e^{-H_{0}\tau_{32}}\ V_{2}|p'\rangle+\int^{+\infty}_{0} d\tau_{23} \langle p|V_{2}\ e^{-H_{0}\tau_{23}}\ V_{3}|p'\rangle  \nonumber\\
	&=& \langle p| V_{3} \bigg(\int^{+\infty}_{0} d\tau_{32}\ e^{-H_{0}\tau_{32}}\bigg)\ V_{2}|p'\rangle + \langle p|V_{2} \bigg(\int^{+\infty}_{0} d\tau_{23}\ e^{-H_{0}\tau_{23}}\bigg)\ V_{3}|p'\rangle\,.
\ea
This expression shares a lot of similarities with the string scattering amplitude $\mathcal{A}=\langle \phi_4|V_3 \Delta V_2|\phi_1\rangle  $, where $\Delta = \int d\tau\ e^{-(L_0 -1)\tau}$ with $(L_0-1)$ being the Hamiltonian for the string. Since $|p\rangle  $ and $|p'\rangle  $ are on-shell states, we can transform the expression a bit more by introducing vertex operators which create these on-shell states in the infinite past and infinite future 
\ba
	\mathcal{A}_4 &=& \langle p|e^{-H_{0}\tau_{43}}\ V_{3} (\int^{+\infty}_{0} d\tau_{32}\ e^{-H_{0}\tau_{32}})\ V_{2}\ e^{-H_{0}\tau_{21}}|p'\rangle \nonumber\\&&+ \langle p|e^{-H_{0}\tau_{42}}\ V_{2} (\int^{+\infty}_{0} d\tau_{23}\ e^{-H_{0}\tau_{23}})\ V_{3}\ e^{-H_{0}\tau_{31}}|p'\rangle\nonumber \\
	%&=& \langle 0|V_{4}\ e^{-H_{0}\tau_{43}}\ V_{3} (\int^{+\infty}_{0} d\tau_{32}\ e^{-H_{0}\tau_{32}})\ V_{2}\ e^{-H_{0}\tau_{21}}\ V_{1}|0\rangle  \\
	%&= &\int^{+\infty}_{0} d\tau_{32}\langle 0|V_{4}\ e^{-H_{0}\tau_{43}}\ V_{3}\  e^{-H_{0}\tau_{32}}\ V_{2}\ e^{-H_{0}\tau_{21}}\ V_{1}|0\rangle  \\
	&= &\int^{+\infty}_{-\infty} d\tau_{32}\langle {\cal T}\{V_{4}(\tau_4)V_{3}(\tau_3) V_{2}(\tau_2) V_{1}(\tau_1)\}\rangle  
\ea
where $\tau_1<\tau_{2,3}, \tau_4>\tau_{2,3}$ and where $V_4$ and $V_1$ are two vertex operators that create $|p\rangle  $ and $|p'\rangle  $ when acting on the vacuum. Just as in string theory, we can easily compute this expression once we know the vacuum correlation function $\langle x^{\mu}(\tau)x^{\nu}(\tau')\rangle  $ which is the Green's function on the (infinite) line %\footnote{\textcolor{red}{This is in contrast with the finite interval Green's function used in Plefka and Steinhoff...}.}
  $
 \partial^{2}_{\tau} \langle x^{\mu}(\tau) x^{\nu}(\tau')\rangle  =-\eta^{\mu\nu}\delta(\tau-\tau')
$ with the solution
\begin{equation}
 \langle x^{\mu}(\tau) x^{\nu}(\tau')\rangle=-\frac{1}{2}\eta^{\mu\nu}|\tau-\tau'|\,.
\end{equation}
\par
\noindent This expression only depends on the difference between $\tau_3$ and $\tau_2$.  This is a manifestation of reparametrization invariance and an equivalent symmetry to conformal symmetry exhibited by the worldline. Thus, just as in string theory \cite{Green:1987sp}, we can fix three of the worldline times to  arbitrary values  \cite{Dai:2008bh}. In practice,  by choosing 1 and 4 to be the asymptotic states,  we  send $\tau_4$ and $\tau_1$ to $+\infty$ and $-\infty$ respectively. We could also set either $\tau_3$ or $\tau_2$ to be at any value, but it is convenient to set one of them to be 0.\par
So, in general, we have,
\begin{align}\label{An}
	\mathcal{A}_{N} = \lim_{\substack{\tau_N \rightarrow +\infty\\ \tau_1 \rightarrow -\infty}}(\prod_{i=3}^{N-1}\int^{\infty}_{-\infty} d\tau_{i}) \langle \mathcal{T}\{V_{N}(\tau_N)V_{N-1}(\tau_{N-1}) V_{N-2}(\tau_{N-2})...V_{2}(0)V_{1}(\tau_1)\}\rangle 
\end{align}
Similar to Feynman diagrams in QFT, we can represent the expression with a specific ordering of $\{\tau_i\}$ diagrammatically as
\begin{figure}[H]
	\centering
	\includegraphics{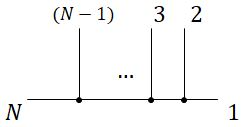}
	\caption{A part of $\mathcal{A}_N$, with ordering $\tau_{N-1} > \tau_{N-2}... > \tau_3 > 0$ }
	\label{fig:An_general}
\end{figure}
\noindent The whole expression can also be thought of as the same diagram with leg $3,4...(N-1)$ freely sliding on the worldline. 
For particles with spins, we will add fermions to the worldline action. The fermion correlation functions on the infinite line are 
\be\langle \bar{\psi}^{a}(\tau) \psi^{b}(\tau')\rangle  = \eta^{ab}\Theta(\tau-\tau')\,.\ee %besides $\langle x^{\mu}(\tau)x^{\nu}(\tau')\rangle =-\frac{1}{2}\eta^{\mu\nu}|\tau-\tau'|$.
From now on, we will represent the worldlines with different spins as
\begin{figure}[H]
	\centering
	\includegraphics{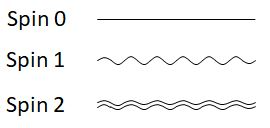}
	\caption{Worldlines of particles with different spins}
	\label{fig:worldlines}
\end{figure}
\noindent We will also represent the linear vertex operators of particles with different spins as
\begin{figure}[H]
	\centering
	\includegraphics{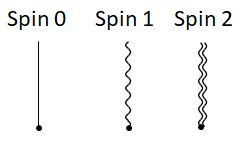}
	\caption{Linear vertex operators of particles with different spins}
	\label{fig:linear operator}
\end{figure}

\subsection{Example: scalar QED}
\noindent We will take scalar QED as a warm-up example and compute the 3-point and 4-point amplitudes.
Start with the worldline action for a scalar interacting with a background photon field,
\be
S=\int d\tau\ \bigg(\frac{1}{2}\dot{x}^2(\tau) - \frac{i}{2} \dot{x}^{\mu}(\tau) A_{\mu}(x(\tau))\bigg)
\ee
Then separate the interaction part of the Lagrangian and perform a plane wave expansion for the background field $A_{\mu}$,
\ba
	&&\mathcal{L}_{int} = -\frac{i}{2} \dot{x}^{\mu}(\tau) A_{\mu}(x(\tau))\\
	&&A_{\mu}(x(\tau)) = \sum^{N-1}_{i=2} \epsilon_{i\mu} e^{i k_i \cdot x(\tau)}\,.
\ea
We can extract the photon emission vertex operator $V_i(\tau)$,
\be
V_j(\tau) = -\frac{i}{2} (\epsilon_j \cdot \dot{x}(\tau)) e^{i k_j \cdot x(\tau)}, \ j=2,3...(N-1)\,.
\ee
For the scalar particle, the asymptotic states are created by $V_1(\tau) = e^{i k_1 \cdot x(\tau)}$ and $V_N(\tau) = e^{i k_N \cdot x(\tau)}$.
\subsubsection{3-point amplitude}
\noindent Since $\langle x^{\mu}(\tau) x^{\nu}(\tau') \rangle = -\frac{1}{2}\eta^{\mu\nu} |\tau - \tau'|$, we have $\langle \dot{x}^{\mu}(\tau) x^{\nu}(\tau') \rangle = -\frac{1}{2}\eta^{\mu\nu} sign(\tau - \tau')$.
Using the master equation \eqref{An}
\ba
	\mathcal{A}_3 &= &\langle V_3(+\infty) V_2(0) V_1(-\infty)\rangle\nonumber\\
	&= &\langle e^{i k_3 \cdot x(+\infty)}\ (-\frac{i}{2})\epsilon_2 \cdot \dot{x}(0) e^{i k_2 \cdot x(0)}\ e^{i k_1 \cdot x(-\infty)}\rangle\nonumber \\
	&=& -\frac{1}{2}\epsilon_2 \cdot (-\frac{1}{2} k_3 + \frac{1}{2} k_1) e^{\sum_{i>j} \frac{1}{2}k_i \cdot k_j (\tau_i - \tau_j)}\nonumber\\
	&=& -\frac{1}{4}\epsilon_2 \cdot (k_1-k_3)\,,
\ea
where in the last step we used  momentum conservation $k_1+k_2+k_3=0$, and the on-shell conditions for the scalar particle ($k_1^2=0, k_3^2=0$). 
If we strip off $\epsilon_2$, then we recover the 3-point vertex of scalar QED.

Diagrammatically, the amplitude is represented as
\begin{figure}[H]
	\centering
	\includegraphics{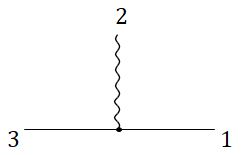}
	\caption{3-point amplitude of scalar QED}
	\label{fig:scalar QED 3pt}
\end{figure}
\noindent For details of the calculation, please see  Appendix \ref{appendix: scalar QED}.
\subsubsection{4-point amplitudes}
\noindent Since $\langle \dot{x}^{\mu}(\tau) x^{\nu}(\tau') \rangle = -\frac{1}{2}\eta^{\mu\nu} sign(\tau - \tau')$, we further have $\langle \dot{x}^{\mu}(\tau) \dot{x}^{\nu}(\tau') \rangle = \eta^{\mu\nu} \delta(\tau - \tau')$. Using \eqref{An} we have:
\ba
	\mathcal{A}_4 &=& \int^{\infty}_{-\infty} d\tau\langle {\cal T}\{V_4(+\infty) V_3(\tau) V_2(0) V_1(-\infty)\}\rangle\nonumber\\
	&=& \int^{\infty}_{-\infty} d\tau \langle {\cal T}\{e^{i k_4 \cdot x(+\infty)}\ (-\frac{i}{2})\epsilon_3 \cdot \dot{x}(\tau) e^{i k_3 \cdot x(\tau)}\ (-\frac{i}{2})\epsilon_2 \cdot \dot{x}(0) e^{i k_2 \cdot x(0)}\ e^{i k_1 \cdot x(-\infty)}\}\rangle \nonumber\\
	&=& \frac{1}{4}(\epsilon_3 \cdot k_4)(\epsilon_2 \cdot k_1) \int^{+\infty}_0 d\tau \langle   e^{i k_4 \cdot x(+\infty)}\ e^{i k_3 \cdot x(\tau)}\ e^{i k_2 \cdot x(0)}\ e^{i k_1 \cdot x(-\infty)}\rangle\nonumber\\
	&& +\frac{1}{4}(\epsilon_2 \cdot k_4)(\epsilon_3 \cdot k_1) \int^0_{-\infty} d\tau \langle  e^{i k_4 \cdot x(+\infty)}\ e^{i k_2 \cdot x(0)}\ e^{i k_3 \cdot x(\tau)}\ e^{i k_1 \cdot x(-\infty)}\rangle\nonumber\\
	&& -\frac{1}{4}(\epsilon_3 \cdot \epsilon_2) \int^{+\infty}_{-\infty} d\tau\ \delta(\tau) \langle  {\cal T}\{e^{i k_4 \cdot x(+\infty)}\ e^{i k_3 \cdot x(\tau)}\ e^{i k_2 \cdot x(0)}\ e^{i k_1 \cdot x(-\infty)}\}\rangle\label{a4interm}\\
	&=& -\frac{1}{2s}(\epsilon_3 \cdot k_4)(\epsilon_2 \cdot k_1) -\frac{1}{2u}(\epsilon_2 \cdot k_4)(\epsilon_3 \cdot k_1)-\frac{1}{4}(\epsilon_3\cdot\epsilon_2)\,,\label{a4sgfinal}
\ea
where $s = -(k_1 + k_2)^2$, $u = -(k_1 + k_3)^2$ and $t = -(k_1 + k_4)^2$ are Mandelstam variables, and we used the on-shell conditions for all four particles. In equation \eqref{a4interm}, the two factors $\int^{+\infty}_{0} d\tau \langle e^{i k_4 \cdot x(+\infty)} e^{i k_3 \cdot x(\tau)} e^{i k_2 \cdot x(0)} e^{i k_1 \cdot x(-\infty)}\rangle$ and $\int^{0}_{-\infty} d\tau \langle e^{i k_4 \cdot x(+\infty)} e^{i k_2 \cdot x(0)} e^{i k_3 \cdot x(\tau)} e^{i k_1 \cdot x(-\infty)}\rangle $  yield the poles  $-\frac{2}{s}$ and $-\frac{2}{u}$ respectively.

Diagrammatically, the three terms in \eqref{a4interm} which give the amplitude can be represented as
\begin{figure}[H]
	\centering
	\includegraphics[width=15cm]{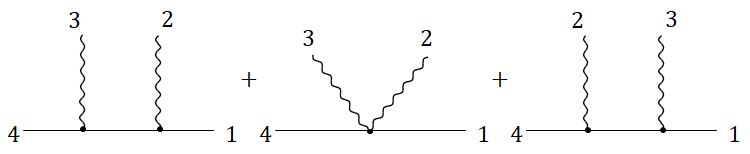}
	\caption{4-point amplitude of scalar QED}
	\label{fig:scalar QED 4pt}
\end{figure}
Notice that in QFT, we will have to calculate each diagram separately and the 4-point vertex used for the second diagram will also need to be derived separately from the 3-point vertex. However, in worldline formalism, one single expression naturally generates all three diagrams. We simply get the second diagram by contracting two linear vertex operators. This is a feature of the worldline formalism that is similar to string theory: the higher point QFT vertex can emerge through the contraction of the worldline linear vertex operators.
	
%%%%%%%%%%%%%%%%%%%%%%%%%%%%%%%%%%%%%%%%%%%%%%%%%%%%%
%%%%%%%%%%%%%%%%%%%%%%%%%%%%%%%%%%%%%%%%%%%%%%%%%%%%
	\section{Worldline coupled with background gravity}\label{wlwithgravity}

\noindent The Euclidean classical action of the $O(N)$ spinning particle in a curved background is given by \cite{Howe:1988ft, Bastianelli:2011cc}
\begin{equation}
S=\int d\tau\ \bigg[\frac{1}{2}g_{\mu\nu}(\dot{x}^{\mu}\dot{x}^{\nu}+b^{\mu}c^{\nu}+a^{\mu}a^{\nu})+\frac{1}{2}\psi_{ia}D_{\tau}\psi_i^a+\alpha R_{abcd}\psi_i^a\psi_i^b\psi_j^c\psi_j^d\bigg]\label{spinningspart}
\end{equation}
where $D_{\tau} \psi_i^a=\partial_{\tau}\psi_i^a+\dot{x}^{\mu}\omega_{\mu}^{ab}\psi_{ib}$ and where the Grassmann-even $a^\mu$ and the Grassmann-odd $b^\mu, c^\mu $ are ghosts introduced to make up for the $\sqrt{-g}$ factor in the general-covariant path integral measure $\int {\cal D} x^\mu \sqrt{g(x)}$ (see for example \cite{Bastianelli:2006rx}).  We use Greek letters $\mu,\nu$ for curved indices, Latin letters $a,b...$ for flat indices, and $i,j=1\dots N$ for fermion species.  For example,    if $N=0$ this action describes a spin 0 massless particle (though a mass term can be added straightforwardly),  if $N=1$ the action describes the worldline of a spin 1/2 particle,   and if $N=2$ it describes a massless spin 1 particle.
The curvature tensor indices have been flattened by contraction with the vielbein $e_\mu^a$ with $g_{\mu\nu}=e_\mu^a e_\nu^a$. The spin connection $\omega_\mu^{ab}$ is determined from the vielbein postulate $\partial_\mu e_\nu^a-\Gamma_{\mu\nu}^\rho e_\rho^a+\omega_{\mu}^{ab}e_{\nu b}=0$.

To allow for an easier comparison with previous literature, we add a few words about dimensions. We are working in units where the speed of light $c=1$. The action should have dimensions of Planck's constant $[\hbar]=\text{ Mass}\times \text{Length}$. This means that the worldline parameter $\tau$ has units $\text{Length/Mass} $.  The worldline fermions have dimension $[\psi]=\sqrt{\text{Length}\times \text{Mass}}$ and the coefficient $\alpha$ is dimensionless.

It is also known \cite{Bastianelli:2006rx} that upon quantization we need to add counterterms to the action in order to get matching results with the quantum field theory corresponding to the first-quantized worldline action. 
In computing in the worldline formalism one encounters products of distributions $\delta(\tau-\tau') \Theta(\tau-\tau')$ which result from contractions of the type  $\langle \dot X \dot X\rangle \langle X\dot X\rangle$ etc.  These expressions are defined through a regularization scheme. The counterterms depend on the regularization scheme.
% The form of counterterms depends on the regularization scheme we use to handle the divergence and ambiguity. 
If we use dimensional regularization, the counterterms will be in a general-covariant form $V_{DR}=\beta R$ \cite{Bastianelli:2000nm}\cite{Bastianelli:2000pt}.  Then, the worldline action in dimensional regularization is:
\begin{equation}
S_{DR}=\int d\tau\ \bigg[\frac{1}{2}g_{\mu\nu}(\dot{x}^{\mu}\dot{x}^{\nu}+b^{\mu}c^{\nu}+a^{\mu}a^{\nu})+\frac{1}{2}\psi_{ia}D_{\tau}\psi_i^a+\alpha R_{abcd}\psi_i^a\psi_i^b\psi_j^c\psi_j^d+\beta R\bigg]\label{actionN}
\end{equation}
where the coefficient $\beta$ has units of $[\hbar^2]=\text{Mass}^2\times\text{Length}^2$.  From now on, though, for simplicity we will work in units where $\hbar=1$.

Previous results in the literature have been derived starting from the $O(N)$ supersymmetric spinning particle action \eqref{spinningspart} (see for example \cite{Bastianelli:2011cc}).
These concern the so-called transition amplitude, which is the probability amplitude for the particle in some initial state, specified by $x^\mu(\tau{=}0)=X^\mu, \psi^i (\tau{=}0)=\Psi^i$,  to evolve at some later time $\tau=T$ into a final state specified by $x^\mu(\tau{=}T)= X'{}^\mu, \bar \psi'{}^i (\tau{=}T)=
\bar\Psi'{}^i$:  
\begin{equation}
K(X', \bar\Psi'; X,\Psi)=\langle  X',\bar\Psi'\,|e^{i T \hat H}|\,X,\Psi\rangle\,,
\end{equation}
where the fermionic Grassmann coordinates were doubled by the doubling trick \cite{Bastianelli:2011cc}.
On the other hand what we are computing is, essentially, the field theory dressed propagator, which when further amputated and setting the external lines on-shell yields scattering amplitudes. The dressed propagator and the transition amplitudes are related through \eqref{dprop} and \eqref{transition}.

Previously, the coefficients $\alpha$ and $\beta$ were determined for $N=0,1,2$  as follows.
The coefficient $\alpha$ was determined by requiring that worldline supersymmetry is preserved in the presence of the background fields \cite{Howe:1988ft}.  
The coefficient $\beta$ was found by requiring that the worldline calculations are properly regularized (i.e. products of distributions are rendered well defined either by time slicing, or mode regularization or dimensional regularization; however non-general-covariant additional terms in the form of products of Christoffel symbols arise if the regularization scheme breaks general covariance) and renormalized (this is done by imposing a matching condition). 

In \cite{Bastianelli:2011cc} the following values for $\alpha$ and $\beta$ are derived for $N=0,1,2$ supersymmetry:
%\begin{center}
\begin{table}[H]
	\centering
	\begin{tabular}{|c|c|c|} \hline
		N & $\alpha$ & $\beta$ \\ \hline
		0 & NA   & $-\frac{1}{8}$ \\ \hline
		1 &  $-\frac{1}{4}$  &  0 \\ \hline
		2 & $-\frac{1}{8}$  & 0 \\ \hline
	\end{tabular}
	\caption{$\alpha$ and $\beta$ coefficients in past literature}\label{them}
\end{table}
%\end{center}
\noindent 

Here, we take a straightforward strategy. In section \ref{3ptVertex}, we will first consider non-dynamical background gravity. We fix the coupling with background gravity and the counterterms by computing the 3-point vertex with the background graviton off-shell. Then, in section \ref{4ptAmplitude}, we verify these coefficients by computing 4-point scattering amplitudes. %As we shall see in the following sections, the counterterms for tree amplitudes are not the same as those in the loop case. 
By matching with the corresponding field theory tree diagrams (e.g.   minimally coupled scalar  or Yang-Mills theory  in a curved background)    we obtained the following results:
%\begin{center}
\begin{table}[H]
	\centering
	\begin{tabular}{|c|c|c|} \hline
		N & $\alpha$ & $\beta$ \\ \hline
		0 & NA   & $-\frac{1}{8}$ \\ \hline
		2 & $-\frac{1}{8}$  & $-\frac{1}{8}$ \\ \hline
	\end{tabular}
	\caption{$\alpha $ and $\beta$ coefficients for $N=0,2$ determined by matching with the 3-point QFT vertex}\label{us}
\end{table} 
%\end{center}
{We would like to comment on the difference in the values for the counterterm $\beta$ for the  $N=2$ supersymmetric action coupled to background gravity in  Tables \ref{them} and \ref{us}. Our $\beta$ is determined by matching the 3-point photon-photon-graviton vertex with an off-shell graviton with the corresponding QFT expression \eqref{ppgvertex}. Note in particular the term $\frac 18 \eta^{\mu\nu}(k_3+k_1)^2(\epsilon_1 \cdot \epsilon_3)$ in \eqref{ppgvertex} 
which matches the term $\frac 18 \eta^{\mu\nu}(k_3+k_1)^2$ in \eqref{ssgvertex}. Both these terms originate in the $\beta R$ counterterm in the worldline action, and they are required by  matching with the corresponding QFT vertex. The same matching that fixes $\beta=-1/8$ for the $N=0$ action (see also \cite{Mogull:2020sak}) can be used to fix $\beta=-1/8$ for the $N=2$ action. The matching used by \cite{Bastianelli:2011cc} which resulted in $\beta=0$ for $N=2$ was done as follows: the transition amplitude was computed in path integral by expanding in the transition time\footnote{To avoid confusion, we'd like to point out that the transition time in \cite{Bastianelli:2011cc} was denoted by $\beta$ and not $T$ as in our paper, and that the worldline time in \cite{Bastianelli:2011cc} was rescaled by $\beta$ such that the range of $\tau$ is $(0,1)$. } $T$, and matched with the evaluation of the transition amplitude in the operator approach, where the time-evolution is generated by the spinning particle Hamiltonian 
$H$ given in equations (1.1) and (1.2) in \cite{Bastianelli:2011cc}. More precisely, on the one hand, the authors computed the small-$T$ expansions of the matrix element of $\exp(-T H)$  on  $|x\rangle \otimes |\xi\rangle$ states,  where $|\xi\rangle $ are coherent fermionic states. On the other hand, starting from  the classical action given in equation (1.3) and allowing for possible counterterms of the type $\beta R$ (other non-general covariant counterterms can arise in path integral if a regularization scheme other than dimensional regularization is used), they computed the path integral transition amplitude (with the bosonic  coordinates $x$ and the coherent Grassmann eigenvalues $\xi$ specifying the bosonic and fermionic path integral boundary conditions). Then the coefficient $\beta$ was determined by matching these two calculations. In dimensional regularization this is given by $\beta_{DR}=-(1/8+ \alpha N/2)$, and it is zero for $\alpha=-1/8$ and $N=2$. Consequently, the path integral action of \cite{Bastianelli:2011cc} cannot yield  the desired term $\frac 18 \eta^{\mu\nu}(k_3+k_1)^2(\epsilon_1 \cdot \epsilon_3)$ in the QFT 3-point vertex \eqref{ppgvertex}. 
From our perspective, matching with the QFT 3-point interaction is more important than matching with the spinning particle Hamiltonian of \cite{Bastianelli:2011cc} and therefore we have a different $\beta$ counterterm\footnote{It is amusing to note that for $N=0$ both matching with the worldline Hamiltonian and the QFT matching yield the same counterterm.}. }
%In particular, for $N=2$, the Hamiltonian in equation (3.37) in \cite{Bastianelli:2011cc} when evaluated on spin states cannot yield a contraction of the type $\epsilon_1\cdot\epsilon_3$, since the only source of such terms is the Ricci-scalar term $R$ (which is multiplied by the identity matrix when acting on spin states). For $N=2$, the coefficient of this term,  $-(1/8+ \alpha N/2)$, is zero for $\alpha=-1/8$. This ultimately means that one cannot recover the desired term $\frac 18 \eta^{\mu\nu}(k_3+k_1)^2(\epsilon_1 \cdot \epsilon_3)$ in the QFT 3-point vertex \eqref{ppgvertex}.    }

We start with the free $N=4$ supersymmetric worldline action to compute graviton scattering amplitudes in worldline formalism, which describes a massless spin $S=2$ particle. Then we couple it to background gravity. As discussed in the introduction, there are constraints on the background placed by preserving $N=4$ supersymmetry. 
In previous works in the literature, the $O(4)$ symmetry acting on the fermionic coordinates was preserved by the background couplings. For example, in \cite{Bastianelli:2011cc}, Bastianelli et al. used the $O(N)$ symmetric worldline action and kept the curved background generic even though  supersymmetry was broken for $N\geq 4$. %This was used to evaluate transition amplitudes over a finite propagation time $T$ %\footnote{To avoid confusion, we'd like to point out that the transition time in \cite{Bastianelli:2011cc} was denoted by $\beta$ and not $T$ as in our paper, and that the worldline time in \cite{Bastianelli:2011cc} was rescaled by $\beta$ such that the range of $\tau$ is $(0,1)$. }
 %(recall footnote 1). 
 As discussed, the dimensional regularization counterterm $V_{DR}= -(\frac{1}{8}+\frac{\alpha N}{2})R$ was identified by a matching condition with the spinning particle Hamiltonian. %, e.g. the transition amplitude was required to satisfy a Schrodinger equation with the spinning particle Hamiltonian associated with the $O(N)$ action. 
It is tempting to use this formula and apply it for $N = 4$, but for this we need $\alpha$.
Since for generic backgrounds we cannot use supersymmetry to determine $\alpha$ (as it was done  for $N\leq 2$), we rely on a different input:
\cite{Bonezzi:2018box}.
Unlike \cite{Bastianelli:2011cc}, Bonezzi et al. \cite{Bonezzi:2018box} side-stepped the issues of the curved-background worldline action and constructed a nilpotent $N=4$ $O(4)$-symmetric BRST operator in a curved background. Then the authors of \cite{Bonezzi:2018box} used the BRST operator to write down the on-shell graviton vertex operator and verified it against the 3-point graviton tree-level amplitude.  Extrapolating a result from \cite{Bonezzi:2018box} (by comparing their linearized graviton vertex operator with the coupling that can be inferred from the $N=4$ worldline action)
fixes the coefficient $\alpha$ in the $O(N{=}4)$ action \eqref{actionN} to be $\alpha = -\frac{1}{8}$. We can now combine the two results and derive the coefficient $\beta_{DR}=
 \frac{1}{8}$. Nonetheless this is not what we find.
\par
 
 Instead of postulating the $O(4)$ symmetry for the worldline action of a graviton in a curved background, we follow the same strategy described earlier. We allow for the possibility of more general couplings,  extract the linearized off-shell graviton vertex,  compute the three-point vertex  and match it against general relativity's cubic graviton vertex (see for example Appendix A in \cite{Carone:2017mdw}).   We noticed that the $O(4)$ spinning particle action does not yield the correct result when one graviton is off-shell, though it does yield the correct 3-point amplitude. Instead, our starting point is an $O(2) \times O(2)$ spinning particle whose euclidean action is
\ba
	S_{DR}&=&\int d\tau\ \bigg[\frac{1}{2}g_{\mu\nu}(\dot{x}^{\mu}\dot{x}^{\nu}+b^{\mu}c^{\nu}+a^{\mu}a^{\nu}) +\frac{1}{2}\bar{\Psi}_a\partial_{\tau}\Psi^{a} +\frac{1}{2}\Psi_a\partial_{\tau}\bar{\Psi}^{a} +\frac{1}{2}\bar{\tilde{\Psi}}_a\partial_{\tau}\tilde{\Psi}^{a} +\frac{1}{2}\tilde{\Psi}_a\partial_{\tau}\bar{\tilde{\Psi}}^{a}\nonumber\\
	&&+\frac{1}{2}\dot{x}^{\mu}\omega_{\mu a b} (S^{ab}+\tilde{S^{ab}}) +2\alpha_1 R_{abcd}S^{ab}\tilde{S}^{cd} +\alpha_2 R_{abcd}(S^{ab}S^{cd}+\tilde{S}^{ab}\tilde{S}^{cd})+\beta R\bigg]\,,
\ea
where $S^{ab}=\bar{\Psi}^{a}\Psi^{b}-\bar{\Psi}^{b}\Psi^{a}$, $\tilde{S}^{ab}=\bar{\tilde{\Psi}}^{a}\tilde{\Psi}^{b}-\bar{\tilde{\Psi}}^{b}\tilde{\Psi}^{a}$.
If $\alpha_1\neq \alpha_2$, the $O(4)$ symmetry is broken to $O(2) \times O(2)$. Our calculation suggests that the following coefficients are needed:
\begin{table}[H]
	\centering
	\begin{tabular}{|c|c|c|c|} \hline
		N & $\alpha_1$ & $\alpha_2$ & $\beta$ \\ \hline
		4 & $-\frac{1}{8}$ & $+\frac{1}{8}$ & $+\frac{3}{8}$\\ \hline
	\end{tabular}
	\caption{Spin-coupling $\alpha_{1,2}$ and counterm $\beta$ coefficients for $N=4$ from matching with QFT }\label{usthree}
\end{table}

Although there is a difference in the coefficients of the counterterms and spin coupling given in Table \ref{usthree} and Table \ref{them}, the difference disappears when the background gravitons are dynamical and self-interacting, in which case only $\alpha_1$ matters. 
In the case of a graviton worldline, it is a bit subtle since it doesn't seem to make much sense to distinguish the interaction between the worldline graviton and background gravitons from the interaction among background gravitons themselves. After all, they are all gravitons! Thus, we will consider the background to be dynamical, which affects how we compute the 4-point amplitude. This is similar with how one computes scattering amplitudes in worldline formalism for gauge bosons: one needs to attach lower trees to the worldline, corresponding to solving the background Yang-Mills non-linear equations of motion with external particle sources \cite{Dai:2008bh}. On the other hand we can imagine processes where the kinematics (e.g. eikonal limit or classical limit) singles out one or two graviton worldlines in which case we can revert to treating differently the graviton worldline and the background field. The latter for example could be integrated out and yield an effective interaction potential between the two worldlines.

In section \ref{dynamicalbg}, we prove that once the self-interaction from the source-free background gravitons is included (for clarity we take the external background gravitational field to be a vacuum solution and expand in plane waves, with $k^2=0$ and with the polarization tensors transverse and traceless, which is equivalent to coupling the lower background graviton trees to the worldline), the coefficients $\alpha_2$ and $\beta$ can be arbitrarily chosen without affecting the final amplitudes. 

\subsection{Three point vertex}\label{3ptVertex}
\noindent Here, we will consider the case that the worldline particle is on-shell while the background field particle could be off-shell.
\subsubsection{Scalar worldline}
\noindent We have the following vertex operators
\ba
	&&V_1(\tau_1)=e^{i k_1 \cdot x(\tau_1)}\nonumber\\ 
	&&V_3(\tau_3)=e^{i k_3 \cdot x(\tau_3)}\nonumber\\
	&&V_2^{\mu\nu}(\tau_2)=-\frac{1}{2}\dot{x}^{\mu}(\tau_2)\dot{x}^{\nu}(\tau_2)e^{i k_2 \cdot x(\tau_2)}-\beta(R^{(1)})^{\mu\nu}\,,
\ea
where $(R^{(1)})^{\mu\nu}$ is defined by the linearized expansion of the background Ricci scalar: $R[\eta_{\mu\nu}+h_{\mu\nu}]=h_{\mu\nu} (R^{(1)})^{\mu\nu}+{\cal O}[h^2]$, and so $(R^{(1)})^{\mu\nu}=\eta_{\mu\nu} k^2-k_\mu k_\nu$ with $k^\mu$ the momentum of the off-shell graviton.

The 3-point vertex is
\begin{figure}[H]
	\centering
	\includegraphics{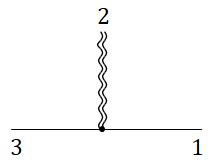}
	\caption{3-point vertex for the scalar-graviton interaction}
	\label{fig:scalar gravity 3pt}
\end{figure}
\noindent When $\beta = -\frac{1}{8}$,
\ba
	\langle V_3(\tau_3) V^{\mu\nu}(\tau_2) V_1(\tau_1)\rangle  &=& -\frac{1}{2}(-1)\frac{1}{4}(k_3-k_1)^{\mu}(k_3-k_1)^{\nu}+\frac{1}{8}(-1)[k_2^{\mu}k_2^{\nu}-\eta^{\mu\nu}k_2^2]\nonumber\\
	&=&-\frac{1}{8}[(k_3+k_1)^{\mu}(k_3+k_1)^{\nu}-\eta^{\mu\nu}(k_1+k_3)^2-(k_3-k_1)^{\mu}(k_3-k_1)^{\nu}]\nonumber\\
	&=&-\frac{1}{4}(k_1^{\mu}k_3^{\nu}+k_3^{\mu}k_1^{\nu}-\frac{1}{2}\eta^{\mu\nu}(k_1+k_3)^2)\,,\label{ssgvertex}
\ea
which matches the 3-point vertex of a minimally coupled scalar field theory.   This matching of the three-point vertex was also done in \cite{Mogull:2020sak}. A different value for $\beta$ can be used to describe the non-minimal coupling between scalar field and gravity. This matching with QFT serves as a renormalization condition for the counterterm. It can be used to replace the renormalization condition used earlier in \cite{Bastianelli:2000nm} (for a complete reference list please see \cite{Bastianelli:2006rx}), namely  that the transition amplitude has to satisfy the Schrodinger equation with the appropriate Hamiltonian (that of a scalar particle in a curved background). 
\subsubsection{Photon worldline}
\noindent We have the following vertex operators: $V_1$ and $V_3$ create the asymptotic photon states \cite{Dai:2008bh} and $V_2$ can be derived from the linear expansion in background gravitational fields of the $O(2)$ action $g_{\mu\nu}(x)=\eta_{\mu\nu}+h_{\mu\nu}(x)$ where we further decompose $h_{\mu\nu}$ into plane waves. Concretely,
\ba
	&&V_1(\tau_1)=\epsilon_{1\mu} \bar{\psi}^{\mu}(\tau_1) e^{i k_1 \cdot x(\tau_1)}\nonumber\\ 
	&&V_3(\tau_3)=\epsilon_{3\mu} \psi^{\mu}(\tau_3) e^{i k_3 \cdot x(\tau_3)}\nonumber\\
	&&V_2^{\mu\nu}(\tau_2)=-\frac{1}{2}\dot{x}^{\mu}(\tau_2)(\dot{x}^{\nu}(\tau_2)+i k_{2\sigma}S^{\nu\sigma}(\tau_2))e^{i k_2 \cdot x(\tau_2)}\nonumber\\
	&&\ \ \ \ \ \ \ \ \ \ \ \ -\alpha [R^{(1)}_{abcd}S^{ab}S^{cd}]^{\mu\nu}-\beta(R^{(1)})^{\mu\nu}\label{step1}\\
	&&\ \ \ \ \ \ \ \ \ \ =-\frac{1}{2}\dot{x}^{\mu}(\tau_2)(\dot{x}^{\nu}(\tau_2)+i k_{2\sigma}S^{\nu\sigma}(\tau_2))\nonumber\\
	&&\ \ \ \ \ \ \ \ \ \ \ \ +4\alpha [R^{(1)}_{ad}\bar{\Psi}^a \Psi^d]^{\mu\nu}
 %\nonumber\\
	%&&\ \ \ \ \ \ \ \ \ \ \ \ 
 -\beta(R^{(1)})^{\mu\nu}\,,\label{step2}
\ea
where $[R^{(1)}_{ad} \bar{\Psi}^{a}\Psi^{d}]^{\mu\nu}$ is the coefficient of $h_{\mu\nu}$ of the linearized $R_{ad}[\eta_{\mu\nu}+h_{\mu\nu}] \bar{\Psi}^{a} \Psi^{d}$ i.e. $\frac{1}{2}(k^2 \bar{\Psi}^{(\mu} \Psi^{\nu)} + k_a k_d \bar{\Psi}^a \Psi^d \eta^{\mu\nu} - k_a \bar{\Psi}^a k^{(\mu} \Psi^{\nu)} - k^{(\mu} \bar{\Psi}^{\nu)} k_a \Psi^a)$. In going from \eqref{step1} to \eqref{step2} the 4-Fermi term $R_{abcd}S^{ab}S^{cd}$ became $-4R_{ad}\bar\Psi^a\Psi^d$ since two of the fermions had to be contracted with one another in the computation of the 3-point function (note that the asymptotic states can soak at most one $\Psi$ and one $\bar\Psi$).
The 3-point vertex is
\begin{figure}[H]
	\centering
	\includegraphics{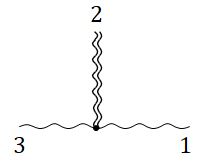}
	\caption{3-point vertex for photon-graviton interaction}
	\label{fig:photon gravity 3pt}
\end{figure}
\noindent When $\alpha = -\frac{1}{8}, \beta = -\frac{1}{8}$, the vertex operator becomes
\ba
	\langle V_3(\tau_3) V^{\mu\nu}(\tau_2) V_1(\tau_1)\rangle  &=&\frac{1}{4}(k_3-k_1)^{\mu}\bigg( \frac{1}{2}(k_3-k_1)^{\nu}(\epsilon_3\cdot \epsilon_1) -\epsilon_3^{\nu}\epsilon_1\cdot k_3 +\epsilon_1^{\nu}\epsilon_3\cdot k_1 \bigg)\nonumber\\
	&&\ \ +\frac{1}{4}\bigg(\epsilon_3\cdot k_1 \epsilon_1^{\mu}(k_1+k_3)^\nu+\epsilon_1\cdot k_3\epsilon_3^{\mu}(k_1+k_3)^{\nu}\nonumber\\
	&&\ \ \ \ \ \ \ \ -\epsilon_3\cdot k_1 \epsilon_1\cdot k_3\eta^{\mu\nu}-(k_1+k_3)^2\epsilon_3^{\mu}\epsilon_1^{\nu}\bigg)\nonumber\\
	&&\ \ -\frac{1}{8}\bigg((k_3+k_1)^{\mu}(k_3+k_1)^{\nu}-\eta^{\mu\nu}(k_3+k_1)^2\bigg)(\epsilon_1 \cdot \epsilon_3)\,,\label{ppgvertex}
\ea
which matches the field theory result (i.e. Yang-Mills theory in a curved background).\footnote{The symmetry in $\mu$ and $\nu$ indices here is implicit. But we could symmetrize the graviton vertex operator at the beginning or the vertex expression at the end to make it explicit.}

\subsubsection{Graviton worldline}
\noindent Here we have (since the ghosts don't contribute to the 3-point vertex, we simply ignore them):
\ba
	&&V_1(\tau_1)=\epsilon_{1\mu\nu} \bar{\Psi}^{\mu}(\tau_1) \tilde{\bar{\Psi}}^{\nu}(\tau_1) e^{i k_1 \cdot x(\tau_1)}\nonumber\\ 
	&&V_3(\tau_3)=\epsilon_{3\mu\nu} \Psi^{\mu}(\tau_3) \tilde{\Psi}^{\nu}(\tau_3) e^{i k_3 \cdot x(\tau_3)}\nonumber\\
	&&V_2^{\mu\nu}(\tau_2)=-\frac{1}{2}\dot{x}^{\mu}(\tau_2)\dot{x}^{\nu}(\tau_2)e^{i k_2 \cdot x(\tau_2)}-\frac{1}{2} \dot{x}^{\mu} (i k_{2\rho}) (S^{\nu\rho}+\tilde{S}^{\nu\rho})e^{i k_2 \cdot x(\tau_2)}\nonumber\\
	&&\ \ \ \ \ \ \ \ \ \ -2\alpha_1 [R^{(1)}_{abcd}(S^{ab}\tilde{S}^{cd})]^{\mu\nu}\nonumber\\
	&&\ \ \ \ \ \ \ \ \ \ +4\alpha_2 [R^{(1)}_{ad}(\bar{\Psi}^a \Psi^d+\bar{\tilde{\Psi}}^a \tilde{\Psi}^d)]^{\mu\nu}\nonumber\\
	&&\ \ \ \ \ \ \ \ \ \ -\beta(R^{(1)})^{\mu\nu}\,.\label{gcubed vertex}
\ea

Diagrammatically, the 3-point vertex is
represented as 
\begin{figure}[H]
	\centering
	\includegraphics{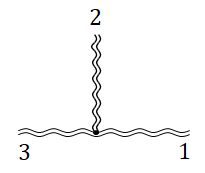}
	\caption{3-point vertex for graviton-graviton interaction}
	\label{fig:gravity 3pt}
\end{figure}
\noindent When $\alpha_1=-\frac{1}{8}, \alpha_2=+\frac{1}{8}, \beta=+\frac{3}{8}$
\ba
	V_2^{\mu\nu}(\tau_2)&=&-\frac{1}{2}(\dot{x}^{\mu}(\tau_2)+i k_{2\rho}S^{\mu\rho}(\tau_2))(\dot{x}^{\nu}(\tau_2)+i k_{2\sigma}\tilde{S}^{\nu\sigma}(\tau_2))e^{i k_2 \cdot x(\tau_2)}\nonumber\\
	& &+\frac{1}{2} [R^{(1)}_{ad}(\bar{\Psi}^a \Psi^d+\bar{\tilde{\Psi}}^a \tilde{\Psi}^d)]^{\mu\nu}\nonumber\\
	&&-\frac{3}{8}(R^{(1)})^{\mu\nu} 
\ea
We can now compute the 3-point vertex, with one graviton taken off-shell
\ba
	\langle V_3(\tau_3) V^{\mu\nu}(\tau_2) V_1(\tau_1)\rangle  &=& \frac{1}{2}\bigg(\frac{1}{2}(k_3-k_1)^{\mu}(\epsilon_3\cdot \epsilon_1)-\epsilon_3^{\mu}\epsilon_1\cdot k_3 +\epsilon_1^{\mu}\epsilon_3\cdot k_1\bigg)\nonumber\\
	&&\ \ \ \ \ \ \ \ \ \times\bigg(\frac{1}{2}(k_3-k_1)^{\nu}(\epsilon_3\cdot \epsilon_1) -\epsilon_3^{\nu}\epsilon_1\cdot k_3 +\epsilon_1^{\nu}\epsilon_3\cdot k_1\bigg)\nonumber\\
	&&-\frac{1}{2}(\epsilon_3\cdot\epsilon_1)\bigg(\epsilon_3\cdot k_1\epsilon_1^{\mu}(k_1+k_3)^\nu+\epsilon_1\cdot k_3 \epsilon_3^{\mu}(k_1+k_3)^{\nu}\nonumber\\
	&&\ \ \ \ \ \ \ \ \ \ \ \ -\epsilon_3\cdot k_1 \epsilon_1\cdot k_3\eta^{\mu\nu}-(k_1+k_3)^2\epsilon_3^{\mu}\epsilon_1^{\nu}\bigg)\nonumber\\
	&&+\frac{3}{8}(\epsilon_3 \cdot \epsilon_1)^2 \bigg((k_3+k_1)^{\mu}(k_3+k_1)^{\nu}-\eta^{\mu\nu}(k_3+k_1)^2\bigg)\,,
\ea
and recover the same result as in general relativity (see for example Appendix A of \cite{Carone:2017mdw}, and keep only one graviton off-shell).\par
%Note that if we consider an on-shell amplitude, which leads to $R_{\mu\nu}^{(1)}=0$ and $R^{(1)}=0$, and compare it with the calculation for gauge boson self-interaction, 

By imposing an on-shell condition on the linearized
graviton vertex operator $V_2^{\mu\nu}$
in \eqref{gcubed vertex} we arrive at the following expression
\be
V_2^{\mu\nu}(\tau_{2})=-\frac{1}{2}(\dot{x}^{\mu}(\tau_2)+i k_{2\rho}S^{\mu\rho}(\tau_2))(\dot{x}^{\nu}(\tau_2)+i k_{2\sigma}\tilde{S}^{\nu\sigma}(\tau_2))\label{onshellgcubed}\,,
\ee
where we used that the linearized Ricci tensor vanishes and the linearized term $R_{abcd}S^{ab}\tilde S^{cd}$ leads to the squaring we see in \eqref{onshellgcubed} if $\alpha_1=-\frac 18$. 
By comparing the on-shell linearized
graviton vertex operator \eqref{onshellgcubed} with the linearized gauge-boson vertex (describing the emission of a gluon from a gluon worldline) we have the double copy structure
%we immediately have the double copy structure (ignore the $e^{i k_i \cdot x(\tau_i)}$ part): 
\ba
	&&V_2^{\mu\nu}(\tau_{2})=-\frac{1}{2}(\dot{x}^{\mu}(\tau_2)+i k_{2\rho}S^{\mu\rho}(\tau_2))(\dot{x}^{\nu}(\tau_2)+i k_{2\sigma}\tilde{S}^{\nu\sigma}(\tau_2))=-2\bar{V}^{\mu}(\tau_2)\bar{V}^{\nu}(\tau_2)\,,
\ea
where $\bar{V}^{\mu}(\tau)=-\frac{1}{2}(\dot{x}^{\mu}(\tau)+i k_{2\rho}S^{\mu\rho}(\tau))$ is the vertex operator for gauge boson self-interaction. 
This squaring relation among the linearized  graviton and gauge boson vertex operators is reminiscent of the squaring relation between the graviton and gauge boson vertex operators in closed and open string theory.

This leads directly to 
the double copy relation between the 3-point amplitudes:
\ba &&\langle V_3(\tau_3)V_2(\tau_2)V_(\tau_1)\rangle _{gravity}=-2\langle \bar{V}_3(\tau_3)\bar{V}_2(\tau_2)\bar{V}_(\tau_1)\rangle _{gauge\ boson}^2\,.
\ea
Diagrammatically, this can be represented as
\begin{figure}[H]
	\centering
	\includegraphics{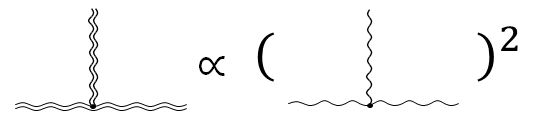}
	\caption{Double copy relation for the 3-point amplitude}
	\label{fig:gravity 3pt double copy}
\end{figure}
  
\subsection{Four point amplitudes}\label{4ptAmplitude}
\noindent In principle we could still treat the background field to be off-shell. But for simplicity, we will be considering all particles to be on-shell, which means we will be using $R_{\mu\nu}^{(1)}=0$ and $R^{(1)}=0$ to simplify expressions.
\subsubsection{Scalar worldline}
\noindent As usual, we have 
\[
\begin{aligned}
	&V_1(\tau_1)=e^{i k_1 \cdot x(\tau_1)}\\ 
	&V_4(\tau_4)=e^{i k_4 \cdot x(\tau_4)}\\
\end{aligned}\,.
\]
Accounting for interaction with the background gravity, we now have not only the linear vertex operator but also a new one that is quadratic in background field $h_{\mu\nu}$, which is derived from expanding Ricci scalar $R$ to second order and is called a pinch operator. 
The linear vertex operator is what we already saw in the 3-point case,
\[
\begin{aligned}
	&V_i(\tau_i)=-\frac{1}{2}\epsilon_{i\mu\nu}[\dot{x}^{\mu}(\tau_i)\dot{x}^{\nu}(\tau_i)+b^{\mu}(\tau_{i})c^{\nu}(\tau_{i})+a^{\mu}(\tau_{i})a^{\nu}(\tau_{i})]e^{i k_i \cdot x(\tau_i)}\ \ \ \ (i=2,3)\\ 
\end{aligned}\,,
\]
while the pinch operator is new and comes from the expansion of $R$,
\[
\begin{aligned}
	V_{23}(\tau)=&\frac{1}{8}R^{(2)}\\
	=&\frac{1}{8}\bigg[\frac{3}{4}(\partial_{\mu}h_{\alpha\beta})^2-\frac{1}{2}(\partial^{\alpha}h^{\beta\mu})(\partial_{\beta}h_{\mu\alpha})\bigg]\\
	=&\bigg[\frac{3t}{32}(\epsilon_2 \cdot \epsilon_3)^2+\frac{1}{8}(\epsilon_2 \cdot k_3)(\epsilon_3 \cdot k_2)(\epsilon_2 \cdot \epsilon_3)\bigg]e^{i(k_2+k_3)\cdot x(\tau)}\,.
\end{aligned}
\]
We represent the pinch operator diagrammatically as
\begin{figure}[H]
	\centering
	\includegraphics{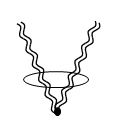}
	\caption{The pinch operator for scalar-graviton interaction}
	\label{fig:pinch operator}
\end{figure}
\noindent Diagrammatically, the scattering amplitude is given by
\begin{figure}[H]
	\centering
	\includegraphics[width=15cm]{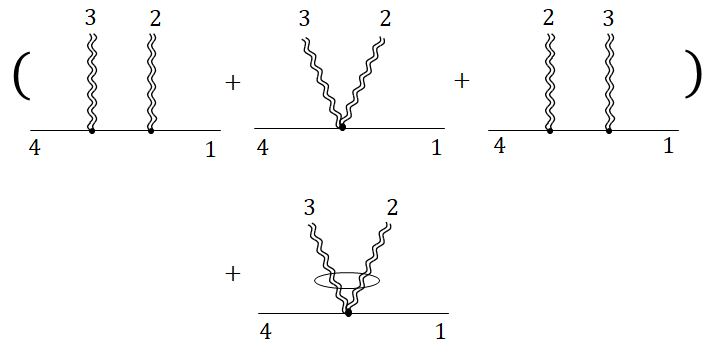}
	\caption{4-point amplitude for scalar-graviton interaction}
	\label{fig:scalarGravity4pt}
\end{figure}
and more explicitly by:
\ba
	A_{4,sg}&=&\lim_{\substack{\tau_4 \rightarrow +\infty\\ \tau_1 \rightarrow -\infty}}\int^{+\infty}_{-\infty} d\tau_{3}\langle\mathcal{T}\{ V_4(\tau_4)V_3(\tau_3)V_2(0)V(\tau_1)\}\rangle \nonumber\\
	&&+\lim_{\substack{\tau_4 \rightarrow +\infty\\ \tau_1 \rightarrow -\infty}} \langle V_4(\tau_4) V_{23}(\tau) V_1(\tau_1)\rangle \nonumber\\
	&=& \int^{+\infty}_{-\infty} d\tau \langle V_4(+\infty)\mathcal{T}\{V_3(\tau)V_2(0)\}V_1(-\infty)\rangle \nonumber\\
	&&+\langle V_4(+\infty) V_{32}(0) V_1(-\infty)\rangle \nonumber%\\
 \ea
 \ba
	&=& \int^{+\infty}_{-\infty} d\tau \langle e^{i k_4 \cdot x(+\infty)}\mathcal{T}\{[-\frac{1}{2}\epsilon_{3\mu\nu}\dot{x}^{\mu}(\tau) \dot{x}^{\nu}(\tau)]e^{i k_3 \cdot x(\tau)} [-\frac{1}{2}\epsilon_{2\bar{\mu}\bar{\nu}}\dot{x}^{\bar{\mu}}(0) \dot{x}^{\bar{\nu}}(0)]e^{i k_2 \cdot x(0)}\} e^{i k_1 \cdot x(-\infty)}\rangle \nonumber\\
	&&+\int^{+\infty}_{-\infty} d\tau \langle e^{i k_4 \cdot x(+\infty)}\mathcal{T}\{[-\frac{1}{2}\epsilon_{3\mu\nu}a^{\mu}(\tau) a^{\nu}(\tau)]e^{i k_3 \cdot x(\tau)} [-\frac{1}{2}\epsilon_{2\bar{\mu}\bar{\nu}}a^{\bar{\mu}}(0) a^{\bar{\nu}}(0)]e^{i k_2 \cdot x(0)}\} e^{i k_1 \cdot x(-\infty)}\rangle\nonumber \\
	&&+\int^{+\infty}_{-\infty} d\tau \langle e^{i k_4 \cdot x(+\infty)}\mathcal{T}\{[-\frac{1}{2}\epsilon_{3\mu\nu}b^{\mu}(\tau) c^{\nu}(\tau)]e^{i k_3 \cdot x(\tau)} [-\frac{1}{2}\epsilon_{2\bar{\mu}\bar{\nu}}b^{\bar{\mu}}(0) c^{\bar{\nu}}(0)]e^{i k_2 \cdot x(0)}\} e^{i k_1 \cdot x(-\infty)}\rangle \nonumber\\
	&&+\bigg[\frac{3t}{32}(\epsilon_2 \cdot \epsilon_3)^2+\frac{1}{8}(\epsilon_2 \cdot k_3)(\epsilon_3 \cdot k_2)(\epsilon_2 \cdot \epsilon_3)\bigg]\langle e^{i k_4 \cdot x(+\infty)} e^{i(k_2+k_3)\cdot x(0)} e^{i k_1 \cdot x(-\infty)}\rangle \label{ampl4sg}\,,
\ea
where in FIG.\ref{fig:scalarGravity4pt} we represented diagrammatically the  terms in \eqref{ampl4sg}.
To calculate the expression,  all we need are the correlation functions $\langle x(\tau)x(\tau')\rangle $, $\langle a(\tau)a(\tau')\rangle $, $\langle b(\tau)c(\tau')\rangle $ and a proper  regularization (we are using dimensional regularization in this paper) to handle those products of distributions which are ill-defined in $D=1$. For details on how  the calculation was preformed please see Appendix \ref{appendix: scalar-gravity}.
The final result is
\ba
	A_{4,sg}=-\frac{1}{2}\bigg[\frac{1}{s}\bigg((\epsilon_2 \cdot k_1)(\epsilon_3 \cdot k_4)-\frac{s}{2}(\epsilon_2 \cdot \epsilon_3)\bigg)^2+\frac{1}{u}\bigg((\epsilon_3 \cdot k_1)(\epsilon_2 \cdot k_4)-\frac{u}{2}(\epsilon_2 \cdot \epsilon_3)\bigg)^2\bigg]\,.
\ea
There is one more thing that is worth mentioning. Let's look at the expression of the amplitude again. If we use the spinor helicity formalism\footnote{For details about the spinor helicity formalism, please see \cite{Elvang:2015rqa}.}, we could get rid of the terms containing $(\epsilon_2 \cdot \epsilon_3)$ by appropriately choosing the reference twistors $|\pm\rangle$ and $|\pm]$. For example if particles 2 and 3 have the same helicity ($\epsilon^{(+)}_{2,3}\propto|-\rangle[k_{2,3}|$, $\epsilon^{(-)}_{2,3}\propto|k_{2,3}\rangle[+|$) then $\epsilon_2 \cdot \epsilon_3=0$. And if the helicities are opposite we can still arrange for $\epsilon_2 \cdot \epsilon_3=0$ by choosing e.g. $|-\rangle\propto |k_2\rangle$ if 3 has negative helicity. 
 Then  there are no more $\dot{x}\dot{x}$, $aa$, $bc$ contractions and we are left with
\[
\begin{aligned}
	A_{4,sg}=& \int^{+\infty}_{0} d\tau \langle e^{i k_4 \cdot x(+\infty)}[-\frac{1}{2}\epsilon_{3\mu}\epsilon_{3\nu}\dot{x}^{\mu}(\tau) \dot{x}^{\nu}(\tau)]e^{i k_3 \cdot x(\tau)} [-\frac{1}{2}\epsilon_{2\bar{\mu}}\epsilon_{2\bar{\nu}}\dot{x}^{\bar{\mu}}(0) \dot{x}^{\bar{\nu}}(0)]e^{i k_2 \cdot x(0)} e^{i k_1 \cdot x(-\infty)}\rangle \\
	&+\int^{0}_{-\infty} d\tau \langle e^{i k_4 \cdot x(+\infty)} [-\frac{1}{2}\epsilon_{2\bar{\mu}}\epsilon_{2\bar{\nu}}\dot{x}^{\bar{\mu}}(0) \dot{x}^{\bar{\nu}}(0)]e^{i k_2 \cdot x(0)} [-\frac{1}{2}\epsilon_{3\mu}\epsilon_{3\nu}\dot{x}^{\mu}(\tau) \dot{x}^{\nu}(\tau)]e^{i k_3 \cdot x(\tau)} e^{i k_1 \cdot x(-\infty)}\rangle \\
	=&\frac{1}{4}[\epsilon_3\cdot \frac{i}{2} (k_4-k_2-k_1)]^2 [\epsilon_2\cdot \frac{i}{2} (k_4+k_3-k_1)]^2 \int^{+\infty}_{0} d\tau \langle e^{i k_4 \cdot x(+\infty)} e^{i k_3 \cdot x(\tau)} e^{i k_2 \cdot x(0)} e^{i k_1 \cdot x(-\infty)}\rangle \\
	&\frac{1}{4}[\epsilon_2\cdot \frac{i}{2} (k_4-k_3-k_1)]^2 [\epsilon_3\cdot \frac{i}{2} (k_4+k_2-k_1)]^2 \int^{0}_{-\infty} d\tau \langle e^{i k_4 \cdot x(+\infty)} e^{i k_2 \cdot x(0)} e^{i k_3 \cdot x(\tau)} e^{i k_1 \cdot x(-\infty)}\rangle\,. \\
\end{aligned}
\]
In the above expression, the two factors $\int^{+\infty}_{0} d\tau \langle e^{i k_4 \cdot x(+\infty)} e^{i k_3 \cdot x(\tau)} e^{i k_2 \cdot x(0)} e^{i k_1 \cdot x(-\infty)}\rangle$ and $\int^{0}_{-\infty} d\tau \langle e^{i k_4 \cdot x(+\infty)} e^{i k_2 \cdot x(0)} e^{i k_3 \cdot x(\tau)} e^{i k_1 \cdot x(-\infty)}\rangle $  yield the poles  $-\frac{2}{s}$ and $-\frac{2}{u}$ respectively.\\
Let's compare that with the corresponding scalar QED case,
\[
\begin{aligned}
	A_{4,sb}=& \int^{+\infty}_{0} d\tau \langle e^{i k_4 \cdot x(+\infty)}[-\frac{i}{2}\epsilon_{3\mu}\dot{x}^{\mu}(\tau)]e^{i k_3 \cdot x(\tau)} [-\frac{i}{2}\epsilon_{2\bar{\mu}}\dot{x}^{\bar{\mu}}(0)]e^{i k_2 \cdot x(0)} e^{i k_1 \cdot x(-\infty)}\rangle \\
	&+\int^{0}_{-\infty} d\tau \langle e^{i k_4 \cdot x(+\infty)} [-\frac{i}{2}\epsilon_{2\bar{\mu}}\dot{x}^{\bar{\mu}}(0) ]e^{i k_2 \cdot x(0)} [-\frac{i}{2}\epsilon_{3\mu}\dot{x}^{\mu}(\tau)]e^{i k_3 \cdot x(\tau)} e^{i k_1 \cdot x(-\infty)}\rangle \\
	=&-\frac{1}{4}[\epsilon_3\cdot \frac{i}{2} (k_4-k_2-k_1)] [\epsilon_2\cdot \frac{i}{2} (k_4+k_3-k_1)] \int^{+\infty}_{0} d\tau \langle e^{i k_4 \cdot x(+\infty)} e^{i k_3 \cdot x(\tau)} e^{i k_2 \cdot x(0)} e^{i k_1 \cdot x(-\infty)}\rangle \\
	&-\frac{1}{4}[\epsilon_2\cdot \frac{i}{2} (k_4-k_3-k_1)] [\epsilon_3\cdot \frac{i}{2} (k_4+k_2-k_1)] \int^{0}_{-\infty} d\tau \langle e^{i k_4 \cdot x(+\infty)} e^{i k_2 \cdot x(0)} e^{i k_3 \cdot x(\tau)} e^{i k_1 \cdot x(-\infty)}\rangle \\
\end{aligned}\,.
\]
So, the double copy structure is shown in plain sight, that is,
\ba
	A_{4,sg} &=& -\frac{n_{s,sg}}{2s} - \frac{n_{u,sg}}{2u} \\
	A_{4,sb} &=& -\frac{n_{s,sb}}{2s} - \frac{n_{u,sb}}{2u} \\
	n_{s,sg} &=& n^2_{s, sb} \\
	n_{u,sg} &=& n^2_{u, sb} \,.
\ea
And this double copy relation is ready to be extended to $n$-point MHV-like (at most one background gauge boson/graviton has opposite helicity to others) amplitudes through exactly the same process.

%%%%%%%%%%%%%%%%%%%%%%%%%%%%%%%%%%5
\subsubsection{Photon worldline}
\noindent To create asymptotic photon states, we need the following vertex operators \cite{Dai:2008bh}
\ba
	V_1(\tau_1) &=& \epsilon_1 \cdot \bar{\Psi}(\tau_1) e^{i k_1 \cdot x(\tau_1)}\nonumber\\
	V_4(\tau_4) &=& \epsilon_4 \cdot \Psi(\tau_4) e^{i k_4 \cdot x(\tau_4)}\,.
\ea
We get the linear vertex operators and the pinch operator by expanding the interaction part of the Lagrangian to second order in the background field\footnote{The ghosts are actually part of the linear vertex operator, we separate them out as $V^{gh}$ only for the simplicity of the expressions.},
\ba
	V_i^{(1)}(\tau_i) &=& -\frac{1}{2}\epsilon_{i\mu\nu}(\tau_i) \dot{x}^{\mu}(\tau_i) [\dot{x}^{\nu}(\tau_i) + i k_{i \sigma} S^{\nu\sigma}(\tau)]e^{i k_i \cdot x(\tau_i)}\nonumber\\
	V^{(2)}(\tau)& =& -\frac{1}{2} \dot{x}^{\mu}(\tau) [\omega_{\mu a b}(\tau)]^{(2)} S^{ab}(\tau)-\frac{1}{2} [R_{ad}(\tau)]^{(2)} \bar{\Psi}^{a}(\tau) \Psi^d(\tau) + \frac{1}{8} [R(\tau)]^{(2)}\nonumber\\
	V^{gh}_i(\tau_i) &=& -\frac{1}{2} \epsilon_{i\mu\nu}(\tau_i) [b^{\mu}(\tau_{i})c^{\nu}(\tau_{i})+a^{\mu}(\tau_{i})a^{\nu}(\tau_{i})] e^{i k_i \cdot x(\tau_i)}\,.
\ea
The amplitude is given by
\begin{figure}[H]
	\centering
	\includegraphics[width=15cm]{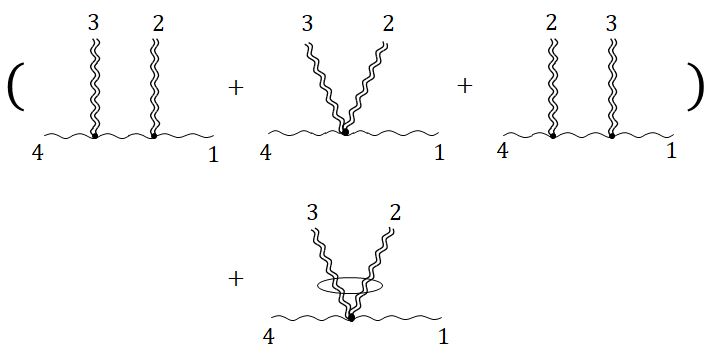}
	\caption{4-point amplitude for photon-graviton interaction}
	\label{fig:photon gravity 4pt}
\end{figure}
\ba
	A_{4,bg} &=& \int^{+\infty}_{-\infty} d\tau\langle V_4(+\infty) V^{(1)}(\tau) V^{(1)}(0) V_1(-\infty)\rangle \nonumber\\
	&& +\int^{+\infty}_{-\infty} d\tau\langle V_4(+\infty) V^{gh}_3(\tau) V^{gh}_2(0) V_1(-\infty)\rangle 
	\nonumber\\
 && + \langle V_4(+\infty) V^{(2)}(0) V_1(-\infty)\rangle \,.
\ea
Since the result is pretty long, we put it into the appendix. For details on the calculation and result, please see appendix \ref{appendix: photon-gravity}. Just like the scalar case, if we use spinor helicity formalism, we will see there is no contribution coming from the ghosts, the pinch operator, and $\dot{x}\dot{x}$ contraction. Thus the relation between different linear vertex operators is reflected in the final amplitude. Since
\ba
	V_{bg, i}(\tau_i) &=& -\frac{1}{2}\epsilon_{\mu\nu, i}(\tau_i) \dot{x}^{\mu}(\tau_i) [\dot{x}^{\nu}(\tau_i) + i k_{i \sigma} S^{\nu\sigma}(\tau)]e^{i k_i \cdot x(\tau_i)}\\
	V_{sb, i}(\tau_i)& =& -\frac{1}{2}\epsilon_{\mu, i} \dot{x}^{\mu} e^{i k_i \cdot x^(\tau_i)}\\
	V_{bb, i}(\tau_i) &=& -\frac{1}{2}\epsilon_{\nu, i} [\dot{x}^{\nu}(\tau_i) + i k_{i \sigma} S^{\nu\sigma}(\tau)] e^{i k_i \cdot x^(\tau_i)}\,,
\ea
ignoring the $e^{i k_i \cdot x^(\tau_i)}$ factor, we have the following relation
\be
V_{bg,i}(\tau_i) = V_{sb,i}(\tau_i) V_{bb,i}(\tau_i)\,.
\ee
The final amplitudes have the structure
\ba
	A_{4,bg} &=&\ \frac{n_{s,bg}}{2s} + \frac{n_{u,bg}}{2u} \\
	A_{4,sb} &=&\ \frac{n_{s,sb}}{2s} + \frac{n_{u,sb}}{2u} \\
	A_{4,bb}(1243)& =&\ \frac{n_{s,bb}}{2s} + \frac{n_{u,bb}}{2u} \\
	n_{s,bg} &=&\ n_{s, sb}\ n_{s, bb} \\
	n_{u,bg} &=&\ n_{u, sb}\ n_{u, bb}\,,
 \ea
where $A_{4,bb}(1243)$ is the color-ordered 4-point amplitude of gauge bosons. This relation is also ready to be extended to $n$ point MHV-like amplitudes.

%%%%%%%%%%%%%%%%%%%%%%%%%%%%%%%%%%%%%%%%%%%%%%%%%%%%%%%%%
\subsubsection{Graviton worldline}

\noindent For the graviton worldline, we need the following vertex operators to create the asymptotic  states \cite{Bonezzi:2018box}
\ba
	V_1(\tau_1) &=&\ \epsilon_{1\mu\nu} \bar{\Psi}^{\mu}(\tau_1) \bar{\tilde{\Psi}}^{\nu}(\tau_1) e^{i k_1 \cdot x(\tau_1)}\nonumber\\
	V_4(\tau_4) &=&\ \epsilon_{4\mu\nu} \tilde{\Psi}^{\mu}(\tau_4) \Psi^{\nu}(\tau_4) e^{i k_4 \cdot x(\tau_4)}\,.
\ea
 The linear vertex operator and the pinch operator are
\ba
	V_i^{(1)}(\tau_i) &=& -\frac{1}{2}\epsilon_{i\mu\nu}(\tau_i) [\dot{x}^{\mu}(\tau_i) + i k_{i \rho} S^{\mu\rho}(\tau_i)] [\dot{x}^{\nu}(\tau_i) + i k_{i \sigma} \tilde{S}^{\nu\sigma}(\tau)]e^{i k_i \cdot x(\tau_i)}\\
	V^{(2)}(\tau) &=& -\frac{1}{2} \dot{x}^{\mu}(\tau) [\omega_{\mu a b}(\tau)]^{(2)} [S^{ab}(\tau) + \tilde{S}^{ab}(\tau)]\nonumber\\
	&& +\frac{1}{2} [R_{ad}(\tau)]^{(2)} [\bar{\Psi}^{a}(\tau) \Psi^d(\tau) + \bar{\tilde{\Psi}}^{a}(\tau) \tilde{\Psi}^d(\tau)] \nonumber\\
	& &-\frac{3}{8} [R(\tau)]^{(2)}\\
	V^{gh}_i(\tau_i) &=& -\frac{1}{2} \epsilon_{i\mu\nu}(\tau_i) [b^{\mu}(\tau_{i})c^{\nu}(\tau_{i})+a^{\mu}(\tau_{i})a^{\nu}(\tau_{i})] e^{i k_i \cdot x(\tau_i)}\,.
\ea
Since now the background field is dynamical, similar to how the 4-point scattering of non-abelian gauge bosons is computed from the worldline, we also have to add a lower-order tree (the t-channel) to the worldline to get the correct four graviton scattering amplitude.

In principle, we need to solve the field equation of gravity for the higher-order perturbation coupled to the worldline. %However, as we have seen in 3-point vertex calculation, the worldline 3-point vertex is not so much different from the result from field theory except that it requires the endpoints of world line to be on-shell, which means actually it's 2 on-shell 1 off-shell vertex from the point of view of field theory. So we
This is equivalent to using two worldline three-point vertices connected with one field theory propagator to get contribution from the t-channel.

Thus, the 4-point amplitude of graviton is given by
\begin{figure}[H]
	\centering
	\includegraphics[width=15cm]{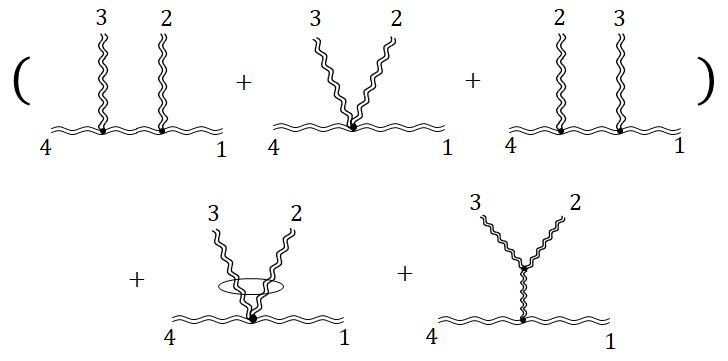}
	\caption{graviton 4-point amplitude}
	\label{fig:gravity 4pt}
\end{figure}
\[
\begin{aligned}
	A^{\text{full}}_{4,gg} =& \int^{+\infty}_{-\infty} d\tau\langle V_4(+\infty) V^{(1)}(\tau) V^{(1)}(0) V_1(-\infty)\rangle \\
	& +\int^{+\infty}_{-\infty} d\tau\langle V_4(+\infty) V^{gh}_3(\tau) V^{gh}_2(0) V_1(-\infty)\rangle \\
	& + \langle V_4(+\infty) V^{(2)}(0) V_1(-\infty)\rangle \\
	& + \langle V_2(+\infty) V^{(1)\mu\nu}(0) V_1(-\infty)\rangle  (- \frac{2}{t}) (\frac{1}{2}\eta_{\mu\rho}\eta_{\nu\sigma} + \frac{1}{2}\eta_{\mu\sigma}\eta_{\nu\rho} - \frac{1}{2}\eta_{\mu\nu}\eta_{\rho\sigma}) \langle V_4(+\infty) V^{(1)\rho\sigma}(0) V_3(-\infty)\rangle 
\end{aligned}
\]
For details on the calculation, please see appendix \ref{appendix: graviton-gravity}. The result is the same as the one from general relativity.

\subsection{Counterterms when background field is dynamical and source-free}\label{dynamicalbg}
\noindent Here we will prove that once we turn on the self-interaction among free background gravitons, several coefficients in the worldline action can be arbitrary.

Consider terms in the action that has $R_{\mu\nu}$ or $R$. Let's represent them simply as $R_{\#}$ and also write the metric as $g_{\mu\nu} = \eta_{\mu\nu} + h_{\mu\nu}$. The metric perturbation $h_{\mu\nu}$ satisfies the Einstein equation in the vacuum, $R_{\mu\nu}-\frac{1}{2}g_{\mu\nu}R = 0$. We can expand $h_{\mu\nu}$ as $h_{\mu\nu} = h^{(1)}_{\mu\nu} + h^{(2)}_{\mu\nu}+...$ and solve $h_{\mu\nu}$ order by order, where $h^{(1)}_{\mu\nu}\sim \epsilon_{\mu\nu}e^{i k\cdot x}$ is a plane wave with $k^2=0$ and the polarization tensor is tranverse and traceless. Diagrammatically, the higher order terms in $h_{\mu\nu}$ are constructed from the lower order ones through tree diagrams which account for the self-interaction of gravitons.

With the self-interaction coming from the background field, when we compute the contribution from $R_{\#}$, we are essentially expanding $R_{\#}$ in terms of $h_{\mu\nu}$ and further expanding $h_{\mu\nu}$, 
\begin{align}\label{Rexpandsion}
	R_{\#} =& R^{(1)}_{\#}(h_{\mu\nu}) + R^{(2)}_{\#}(h_{\mu\nu}, h_{\rho\sigma})+... \nonumber \\
	=& R^{(1)}_{\#}(h^{(1)}_{\mu\nu}) + \bigg( R^{(2)}_{\#}(h^{(2)}_{\mu\nu}) + R^{(2)}_{\#}(h^{(1)}_{\mu\nu}, h^{(1)}_{\rho\sigma}) \bigg) + ...
\end{align}
However, due to the Einstein equation, we know that $R_{\#} = 0$. Thus each order in the expansion vanishes, which means any term having $R_{\#}$ gives no contribution to the amplitude. 

Let's take the three terms in the expression(\ref{Rexpandsion}) and explain the idea a bit more in detail. The first term $R^{(1)}_{\#}(h^{(1)}_{\mu\nu})$ is needed when we compute the 3-point amplitude. However, since $h^{(1)}_{\mu\nu}$ is a plane wave, that term simply vanishes. The second term and third terms are needed when computing the 4-point amplitude. The second term represents graviton self-interaction. Diagrammatically, it is a 3-point graviton tree attached to the worldline. The third term is a pinch operator. Diagrammatically, it is represented with two gravitons attached to the worldline at the same point. If we compute the contribution from these two diagrams, we will notice that they simply cancel out each other.

Thus, with dynamical {source-free } (vacuum) background gravity, any term that has $R_{\#}$ gives no contribution and can have any coefficient. For the scalar worldline, the coefficient $\beta$ can be arbitrary. For the photon worldline, the coefficients $\alpha$ and $\beta$ can be arbitrary.\footnote{Due to the contraction from the fermions, $R_{abcd}$ is always turned into $R_{ad}$.} For the graviton worldline, the coefficients $\alpha_2$ and $\beta$ can be arbitrary.\footnote{Same reason as above.}

However, we need to point out that in general, the gravitons can be off-shell in which case the counterterms are fixed. For example, that would be the case of two scalar particles interacting with each other by exchanging gravitons, though in the classical limit the counterterm $\beta R$ does not contribute  \cite{Mogull:2020sak}.

\section{MHV amplitudes in the  worldline formalism}\label{mhv}
The worldline approach is most efficient for studying MHV amplitudes when combined with the spinor helicity formalism. With the spinor helicity formalism, once we make clever choices of the reference spinors, we could get rid of most $ (\epsilon_i \cdot \epsilon_j) $ terms. In the language of the worldline approach, it simply means there will be no contributions coming from $ \langle \dot{x} \dot{x} \rangle$ contractions, pinch operators, or any higher order ones. So the "double-copy" like relation of vertex operators we encountered in the previous section will directly result in a "double-copy" like relation in the scattering amplitudes.

In general, the application of the worldline formalism to tree-level scattering amplitudes has a number of restrictions: the two ends of the worldline have to be on-shell; the lower tree attached to the worldline has to be calculated separately (by solving the full non-linear equation of motion), which is equivalent to using only one worldline for each diagram (in the classical limit though it is possible to think of the scattering process of e.g. two scalar particles exchanging soft gravitons as two worldlines dressed by the off-shell gravitons which are subsequently integrated over); the choice of the worldline has to be the same in the different diagrams of a single scattering process. We will see that these restrictions are greatly eased for MHV amplitude.

For simplicity, let us make the convention that in an  $n$-point scattering MHV amplitude  particles 1 and $n$ have negative helicity while all the others have positive helicity. The clever choice that we take is that we choose the reference spinor\footnote{We recall that for a positive helicity gauge boson, the polarization vector is $\frac{1}{\langle \epsilon| p\rangle}|\epsilon\rangle[p|$ and for a negative helicity one its polarization vector is $\frac{1}{[\epsilon|p] }|p\rangle[\epsilon|$. Since the twisor product is antisymmetric, it is clear that the contraction of two polarization vectors for the same helicity type gauge bosons is zero. With the choice $|\epsilon\rangle=|1\rangle$ we are ensuring that all contractions of the type $\epsilon_1\cdot\epsilon_i$ vanish, leaving $\epsilon_n\cdot \epsilon_i$ the only non-vanishing contractions, with $i$ a positive helicity boson. }  $|+\rangle=|1\rangle$ where the momentum of the particle 1 is $k_1=|1\rangle[1|$.  %having the same momentum as particle 1.
Then any lower tree in one diagram can be seen as a separate diagram (with the polarization vector/tensor stripped for the attaching point) and it can be calculated by choosing another worldline connecting the attaching point and any other particle. This procedure can be applied recursively to cover the whole diagram.

We will explain the idea in detail for gauge boson and graviton cases respectively and we will see that the double-copy relation emerges naturally as a consequence of the double-copy relation between their linear vertex operators.

\subsection{MHV amplitudes for gauge boson scattering}
In an $n$-point tree-level scattering amplitude there will be $n$ polarization vectors $\epsilon_i^\mu$ and at most $n-2$ momenta $k_i^\mu$ from the cubic vertices. Their contractions will result in factors of the type $\epsilon_i\cdot \epsilon_j$, $k_i\cdot k_j$ and $\epsilon_i\cdot k_j$. In an MHV scattering amplitude, there can be at most two contractions of the type $\epsilon_i\cdot\epsilon_j$, and these must involve the two negative helicity polarization vectors.
{First, we will show that with our previous choice of reference spinors, the MHV diagram does not have any four point vertex.} {With our choice, the number of contractions of the type $\epsilon_i \cdot \epsilon_j$ in the amplitude is further reduced from two to one. %In terms of the number of momenta, each four-point vertex can be thought as two connected cubic vertices\footnote{not real cubic vertices} where the two momenta coming from cubic vertices contract with each other and cancel the propagator. Comparing with diagrams that contain only cubic vertices, the total number of polarization vectors $\epsilon_i$ and that of momenta $k_i$ in the numerator are the same\footnote{By artificially replacing each four-point vertex with two cubic vertices, we restore and keep the number of momenta in the denominator to be the same as that in diagrams that contain only cubic vertices}, thus one more pair of $k_i \cdot k_j$ contraction results in one more pair of $\epsilon_i \cdot \epsilon_j$ contraction. 
Since diagrams that contain only cubic vertices already have at least one $\epsilon_i \cdot \epsilon_j$ contraction, diagrams that have four-point vertices will have at least two $\epsilon_i \cdot \epsilon_j$ contractions which then vanish as a consequence of our choice of reference spinors. }%By simply counting the number of $\epsilon$ and $k$, each 4-point vertex will bring one extra $(\epsilon \cdot \epsilon)$, so in each term there will be at least two $(\epsilon \cdot \epsilon)$. However, in an MHV scattering amplitude, there can be at most one non-vanishing $(\epsilon \cdot \epsilon)$ in each term once we have chosen the appropriate reference spinor. 
Thus no four-point vertex can contribute to the MHV tree-level amplitude. In the worldline formalism, $\langle \dot{x} \dot{x} \rangle$ contractions and pinch operator represent the 4-point vertex, so they must disappear from the MHV amplitude. \par
Next, let's consider a lower tree attached to the worldline. A general structure of a lower tree can be written as 
\[
J^\mu(p) = \frac{2}{p^2}\Big( \mathcal{M}^\mu + p^\mu f(\{\epsilon\},\{k\}) \Big)
\]
Notice that when computing the on-shell amplitude, only the first terms contribute. The other term vanishes due to transversality. However, being as a lower tree, the second term in general doesn't vanish, which is the reason why we have to compute it (solving the background field equation of motion to higher order in the background fields, or equivalently, using Feynman diagrams). But in an MHV diagram, we can show that the second term gives zero contribution. The proof is given in Appendix \ref{appendix: lower tree}. Thus, we can calculate $J^\mu$ effectively by simply choosing a new worldline that connects the attaching point and an arbitrary particle in the lower tree. 
In conclusion, after choosing an arbitrary worldline, the amplitude for a specific pole structure (channel) can be represented as
\[
\begin{aligned}
\mathcal{M}_{\{s\}} = (\prod_{i=3}^{N-1} \int d\tau_i) &\langle [\epsilon_N \cdot \Psi(+\infty) e^{ik_N\cdot x(+\infty)}]\\
&[{J_{N-1}}_\mu(\dot{x}_{N-1}^\mu + i {k_{N-1}}_\rho S^{\mu\rho}(\tau_{N-1})) e^{ik_{N-1}\cdot x_{N-1}}]\\
&\dots [{J_2}_\nu(\dot{x}(0)^\nu + i {k_2}_\sigma S^{\nu\sigma}(0)) e^{ik_2\cdot x(0)}]\\
&[\epsilon_1 \cdot \bar{\Psi}(-\infty) e^{i k_1 \cdot x(-\infty)}]\rangle\,,
\end{aligned}
\]
where $N\leq n$ is the total number of vertices (including the two asymptotic states) on the worldline and $J_\mu$ represents a lower tree or simply a polarization vector $\epsilon_\mu$. We recall that on each worldline we have the freedom to fix three $\tau_i$ insertions, and we choose them in the usual fashion $-\infty, 0,+\infty$ leaving $N-3$ insertion points to be integrated over. {Since we want to stress the overall structure of the expression, we will write the expression schematically as the following,
\[
\mathcal{M}_{\{s\}} = (\prod_{i=3}^{N-1} \int d\tau_i) \langle [\epsilon \cdot \Psi e^{ik\cdot x}]_N [J_\mu(\dot{x}^\mu + i k_\rho S^{\mu\rho}) e^{ik\cdot x}]_{N-1}\dots [J_\nu(\dot{x}^\nu + i k_\sigma S^{\nu\sigma}) e^{ik\cdot x}]_2 [\epsilon \cdot \bar{\Psi} e^{i k \cdot x}]_1\rangle\,,
\]
where polarization vectors $\epsilon$, momenta $k$ and lower trees $J$ in each pair of square brackets should be understood as quantities belonging to the corresponding vertex $i$, which is marked after the brackets, while operators $x$, $\Psi$ and $\bar{\Psi}$ in the same pair of brackets depend on $\tau_i$. From now on, we will take this convention when we write expressions.}
\par
For each lower tree $J_\mu$, we choose a new worldline connecting the attaching point and an arbitrary particle, and thus we have
%\[
%\begin{aligned}
%J_\mu = \frac{2}{p^2} (\prod_{i=3}^{M-1} \int d\tau_i) &\langle [\epsilon'_M \cdot \Psi(+\infty) e^{ik'_M\cdot %x(+\infty)}]\\
%&[{J'_{M-1}}_\lambda(\dot{x}_{M-1}^\lambda + i {k'_{M-1}}_\rho S^{\lambda\rho}(\tau_{M-1})) e^{ik'_{M-1}\cdot x_{M-1}}]\\
%&\dots [{J'_2}_\nu(\dot{x}(0)^\nu + i {k'_2}_\sigma S^{\nu\sigma}(0)) e^{ik'_2\cdot x(0)}]\\
%&[\bar{\Psi}_{\mu}(-\infty) e^{i k'_1 \cdot x(-\infty)}]\rangle\,,
%\end{aligned}
%\]
{
\[
J_\mu = \frac{2}{p^2} (\prod_{i=3}^{M-1} \int d\tau_i) \langle [\epsilon \cdot \Psi e^{ik\cdot x}]_M [J'_\alpha(\dot{x}^\alpha + i k_\rho S^{\alpha\rho}) e^{ik\cdot x}]_{M-1}\dots [J'_\nu(\dot{x}^\nu + i k_\sigma S^{\nu\sigma}) e^{ik\cdot x}]_2 [\bar{\Psi}_\mu e^{i k \cdot x}]_1\rangle\,,
\]}
where $p$ is the momentum of the lower tree line which is attached to the worldline, $M$ is the total number of vertices (including the two ends) on the worldline, and $J'_\nu$ represents either $\epsilon$ or a possible further lower tree and will be calculated repeatedly with the same formula.

\subsection{MHV amplitudes for graviton scattering}
With almost the same argument as previously, by counting the $\epsilon$ and $k$, we can show that any higher-order vertex vanishes for MHV diagrams. Thus there will be no $ \langle \dot{x} \dot{x} \rangle$ contraction and no pinch/higher-order operator.

{A general structure of a lower tree of graviton would be
\[
J^{\mu\nu}(p) = \frac{2}{p^2}\Big( \mathcal{M}^{\mu\nu} + p^\mu f^\nu(\{\epsilon\},\{k\}) + p^\nu f^\mu(\{\epsilon\},\{k\}) + \eta^{\mu\nu} h(\{\epsilon\},\{k\})\Big)\,.
\]}
The last three terms will vanish for on-shell amplitudes due to the transversality and tracelessness of the polarization tensor. But as a lower tree, they generally do give contributions. However, we have the same conclusion as the gauge boson case that they also vanish for MHV diagrams. Please see Appendix \ref{appendix: lower tree} for the corresponding discussion.\par
Thus, we can represent the amplitude for a specific pole structure in a very similar way to that for the gauge boson case, 
\[
\begin{aligned}
	\mathcal{M}_{\{s\}} = (\prod_{i=3}^{N-1} \int d\tau_i) \langle &[\epsilon_\mu \tilde{\epsilon}_\nu \Psi^\mu \tilde{\Psi}^\nu e^{ik\cdot x}]_N\\ &[J_{\alpha\beta}(\dot{x}^\alpha + i k_\gamma S^{\alpha\gamma})(\dot{x}^\beta + i k_\theta \tilde{S}^{\beta\theta}) e^{ik\cdot x}]_{N-1}\\
	&\dots\\
	&[J_{\lambda\tau}(\dot{x}^\lambda + i k_\delta S^{\lambda\delta})(\dot{x}^\tau + i k_\xi \tilde{S}^{\tau\xi}) e^{ik\cdot x}]_2\\
	&[\epsilon_\rho \tilde{\epsilon}_\sigma \bar{\Psi}^\rho \bar{\tilde{\Psi}}^\sigma e^{ik\cdot x}]_1\rangle
\end{aligned}
\]
where $J_{\mu\nu}$ represents a lower tree or simply a polarization tensor $\epsilon_{\mu} \tilde{\epsilon}_{\nu}$. {Since in spinor helicity formalism, there is really no difference between $\epsilon_\mu$ and $\tilde{\epsilon}_\mu$, we will simply replace every $\tilde{\epsilon}_\mu$ with $\epsilon_\mu$ from now on.} For each $J_{\mu\nu}$, we choose a new worldline and have,
\[
\begin{aligned}
	J_{\mu\nu} = \frac{2}{p^2} (\prod_{i=3}^{M-1} \int d\tau_i) \langle &[\epsilon_\alpha \epsilon_\beta \Psi^\alpha \tilde{\Psi}^\beta e^{ik\cdot x}]_M\\
	&[J'_{\rho\sigma}(\dot{x}^\rho + i k_\gamma S^{\rho\gamma})(\dot{x}^\sigma + i k_\theta \tilde{S}^{\sigma\theta}) e^{ik\cdot x}]_{M-1}\\
	&\dots \\
	&[J'_{\lambda\tau}(\dot{x}^\lambda + i k_\delta S^{\lambda\delta})(\dot{x}^\tau + i k_\xi \tilde{S}^{\sigma\xi}) e^{ik\cdot x}]_2 \\
	&[\bar{\Psi}_\mu \bar{\tilde{\Psi}}_\nu e^{i k \cdot x}]_1\rangle
\end{aligned}
\]
where $J'_{\alpha\beta}$ represents a polarization tensor $\epsilon_{\mu} \epsilon_{\nu}$ or a further lower tree and will be calculated repeatedly with the same formula.\par 
We should now see that the calculation for graviton MHV is much easier and more well-organized than the field theory calculation. In field theory, even when using the spinor helicity formalism, we could only simplify the calculation after going through the tedious contractions of vertices and propagators. But with the worldline approach, those vanishing terms are already dropped at the beginning, and only necessary terms are kept. However, %as you may have already noticed, 
what is more important is that the worldline approach expressions manifestly display  the close relationship between gravity and gauge boson scattering amplitudes, namely the double copy relation.

\subsection{Double copy and BCJ relations}
The worldline expressions for gauge boson and graviton scattering amplitudes display the double copy relation for MHV diagrams in plain sight. \\
The gauge boson tree-level scattering amplitude can be written as 
{
\[
\mathcal{M} = \sum_{\{s\}} \mathcal{M}_{\{s\}} = \sum_{\{s\}} \frac{c_{\{s\}}n_{gb,\{s\}}}{\prod_\alpha p_{\{s\}_\alpha}^2}
\]
where $c_{\{s\}}$ is color factor,
\ba
n_{gb,\{s\}} &=& \frac{(\prod_{i=3}^{N-1} \int d\tau_i) \langle [\epsilon \cdot \Psi e^{ik\cdot x}]_N [n_\mu(\dot{x}^\mu + i k_\rho S^{\mu\rho}) e^{ik\cdot x}]_{N-1}\dots [n_\nu(\dot{x}^\nu + i k_\sigma S^{\nu\sigma}) e^{ik\cdot x}]_2 [\epsilon \cdot \bar{\Psi} e^{i k \cdot x}]_1\rangle}{(\prod_{j=3}^{N-1} \int d\tau_j)\langle  e^{ik_N\cdot x_N} e^{ik_{N-1}\cdot x_{N-1}}\dots e^{ik_2\cdot x_2} e^{i k_1 \cdot x_1}\rangle}
\nonumber\\
&=&\frac{\langle [\epsilon \cdot \Psi e^{ik\cdot x}]_N [n_\mu(\dot{x}^\mu + i k_\rho S^{\mu\rho}) e^{ik\cdot x}]_{N-1}\dots [n_\nu(\dot{x}^\nu + i k_\sigma S^{\nu\sigma}) e^{ik\cdot x}]_2 [\epsilon \cdot \bar{\Psi} e^{i k \cdot x}]_1\rangle}{\langle  e^{ik_N\cdot x_N} e^{ik_{N-1}\cdot x_{N-1}}\dots e^{ik_2\cdot x_2} e^{i k_1 \cdot x_1}\rangle}
\nonumber\\
n_\mu &= &
\frac{\langle [\epsilon \cdot \Psi e^{ik\cdot x}]_M [n'_\nu(\dot{x}^\nu + i k_\rho S^{\nu\rho}) e^{ik\cdot x}]_{M-1}\dots [n'_\lambda(\dot{x}^\lambda + i k_\sigma S^{\lambda\sigma}) e^{ik\cdot x}]_2 [\bar{\Psi}_\mu e^{i k \cdot x}]_1\rangle}{\langle  e^{ik_M\cdot x_M} e^{ik_{M-1}\cdot x_{M-1}}\dots e^{ik_2\cdot x_2} e^{i k_1 \cdot x_1}\rangle}
 \label{nsforbcj}
\ea
and where we highlighted the sum over the different pole structures. Since the ordering of all the vertices is fixed for a specific pole structure, $\tau_N > \tau_{N-1} > \dots > \tau_3 > 0$ and $\tau_M > \tau_{M-1} > \dots > \tau_3 > 0$, the dependence on worldline time $\tau_i$ will only remain in contractions among exponentials $e^{i k_i \cdot x_i}$, which just give the corresponding pole structure, similar to what we encountered in going from \eqref{a4interm}} to \eqref{a4sgfinal}. Thus we can drop the integral over the worldline times $\tau_i$ when calculating the numerators $n_{gb,\{s\}}$ and $n_\mu$.
Similar expressions can be written for gravitons 
\[
\mathcal{M} = \sum_{\{s\}} \mathcal{M}_{\{s\}} = \sum \frac{n_{gr,\{s\}}}{\prod_\alpha p_{\{s\}_\alpha}^2}
\]
where
\[
\begin{aligned}
	n_{gr,\{s\}} = \langle &[\epsilon_\mu \epsilon_\nu \Psi^\mu \tilde{\Psi}^\nu e^{ik\cdot x}]_N\\ &[n_{\alpha\beta}(\dot{x}^\alpha + i k_\gamma S^{\alpha\gamma})(\dot{x}^\beta + i k_\theta \tilde{S}^{\beta\theta}) e^{ik\cdot x}]_{N-1}\\
	&\dots\\
	&[n_{\lambda\tau}(\dot{x}^\lambda + i k_\delta S^{\lambda\tau})(\dot{x}^\tau + i k_\xi \tilde{S}^{\tau\xi}) e^{ik\cdot x}]_2\\
	&[\epsilon_\rho \epsilon_\sigma \bar{\Psi}^\rho \bar{\tilde{\Psi}}^\sigma e^{ik\cdot x}]_1\rangle\\
	&/\langle  e^{ik_N\cdot x_N} e^{ik_{N-1}\cdot x_{N-1}}\dots e^{ik_2\cdot x_2} e^{i k_1 \cdot x_1}\rangle\\
	n_{\mu\nu} = \langle &[\epsilon_\rho \epsilon_\sigma \Psi^\rho \tilde{\Psi}^\sigma e^{ik\cdot x}]_M\\
	&[n'_{\alpha\beta}(\dot{x}^\alpha + i k_\gamma S^{\alpha\gamma})(\dot{x}^\beta + i k_\theta \tilde{S}^{\beta\theta}) e^{ik\cdot x}]_{M-1}\\
	&\dots \\
	&[n'_{\lambda\tau}(\dot{x}^\lambda + i k_\delta S^{\lambda\delta})(\dot{x}^\tau + i k_\xi \tilde{S}^{\tau\xi}) e^{ik\cdot x}]_2 \\
	&[\bar{\Psi}_\mu \bar{\tilde{\Psi}}_\nu e^{i k \cdot x}]_1\rangle\\
	&/\langle  e^{ik_M\cdot x_M} e^{ik_{M-1}\cdot x_{M-1}}\dots e^{ik_2\cdot x_2} e^{i k_1 \cdot x_1}\rangle\\
\end{aligned}
\]
Comparing the two cases, we can easily see that
\[
\begin{aligned}
	n_{\mu\nu} &= n_\mu n_\nu\\
	n_{gr, \{s\}} &= n_{gb, \{s\}}^2
\end{aligned}
\]
which is just the double-copy relation! It is interesting because usually we get the double copy from combing the KLT relation and BCJ relation. But with the worldline approach, we jumped over the two relations and directly found the double copy relation.

Thus a natural question that arises is whether those numerators also satisfy the BCJ relations, since we know that once the numerators satisfy the BCJ relation, they produce the double copy relation. The answer is affirmative, as shown in Appendix \ref{appendix: BCJ}. %We could simply compute and show that the BCJ relation holds for those numerators. But it is less obvious and needs careful analysis, so we put the proof in Appendix \ref{appendix: BCJ}.

\section{Conclusions}\label{conclusions3}
\noindent In this paper we analyzed the interaction of different spin particle ($S=0,1,2)$ with background gravity using the worldline formalism. We focused on tree amplitudes and explicitly computed 3-point and 4-point amplitudes with background gravitons. We noticed that the spin-coupling coefficients and the counterterms determined by matching with QFT diagrams (3-point vertices are enough) are different from those calculated by matching with a spinning particle Hamiltonian, as it was done in the past literature. In the special case when the background gravitons are dynamical and source-free (i.e. corresponding to vacuum solutions), we proved that many coefficients can be arbitrarily chosen.

One goal of our study was to investigate if the worldline formalism, which offers a first quantized perspective much like in string theory, can illuminate the double-copy relations among the scattering amplitudes involving particles with different spins, relations that are not obvious from a  field theory perspective. One basic building block in worldline formalism is the linear vertex operator, which describes how a background particle interacts with the  worldline particle. We can see from the following table that there are relations among different linear vertex operators,
\begin{table}[H]
	\centering
	\begin{tabular}{|c|c|c|c|} \hline
		spin & 0 & 1 & 2 \\ \hline
		0 & $e^{i k \cdot x}$   & $\frac{1}{2}\epsilon_{\mu}\dot{x}^{\mu}e^{i k \cdot x}$ & $\frac{1}{2}\epsilon_{\mu\nu} \dot{x}^{\mu}\dot{x}^{\nu}e^{i k \cdot x}$ \\ \hline
		1 & NA  & $\frac{1}{2}\epsilon_{\mu}(\dot{x}^{\mu} + i k_{\nu}S^{\mu\nu})e^{i k \cdot x}$ & $\frac{1}{2}\epsilon_{\mu\nu}\dot{x}^{\mu}(\dot{x}^{\nu} + i k_{\rho}S^{\nu\rho})e^{i k \cdot x}$ \\ \hline
		2 & NA & NA & $\frac{1}{2}\epsilon_{\mu\nu}(\dot{x}^{\mu} + i k_{\rho}S^{\mu\rho})(\dot{x}^{\nu} + i k_{\sigma}\tilde{S}^{\nu\sigma})e^{i k \cdot x}$\\ \hline
	\end{tabular}
	\caption{The linearized vertex operators of different interactions}
\end{table}
\noindent where the first row represents background fields with different spins while the first column represents different worldlines. Although similar to string theory, in many aspects, e.g. in that the higher point vertex can emerge from contracting linear vertex operators, the worldline vertices could still have pinch operators and possibly even higher order vertex operators (such as the lower trees that need to attach to the worldline; in string theory we simply have strings splitting and joining instead of the lower tree structures).

We also studied one special class of amplitudes, the MHV amplitudes. This special class of amplitudes allows us to get rid of all the contractions of linear vertex operators and all the higher-order vertex operators by making clever choices of the reference spinor. Thus, the relations among linear vertex operators are also directly shown in the final expressions of the tree amplitudes. For the interaction among gauge bosons and those among gravitons, this relation is just the double copy relation. We also proved that the numerators for these MHV amplitudes constructed through the worldline formalism satisfy the BCJ relations.

Beyond MHV, one generally needs not only the contractions and higher-order vertex operators from the worldline formalism but also quantum field theory to calculate any lower tree that is attached to the worldline, which renders the formalism less powerful and more convoluted. Thus it is natural to seek a way to realize all the calculations with only linear vertex operators, just as in string theory. One such attempt is the idea of replacing the notion of a "worldline" with that of a "worldgraph" \cite{Dai:2008bh}, and have worldline junctions. However, even for the non-abelian gauge boson tree-level amplitudes considered in the paper, the method seems to break down when computing 6-point amplitudes where there could be "snow-flake" diagrams, and the diffeomorphism  and susy ghosts on the internal lines cannot be properly accounted for. Another problem arises for gravity when  replacing the usual partial derivative in the vertex operator with the worldgraph derivative of \cite{Dai:2008bh} and following the worldgraph rules does not seem to produce correct 4-point amplitude. But the "worldgraph" construction could be of further interest.

\section{Acknowledgements}

The work of D.V. and Y.D. was supported in part by the U.S. Department of Energy under Grant No. DE-SC0007984.

\appendix
\section{Scalar QED}\label{appendix: scalar QED}
\noindent First, we want to promote $-\frac{i}{2}\epsilon_2 \cdot \dot{x}_2$ into the exponential,
\begin{align*}
	A_{3,sb}=&\langle e^{i k_3 \cdot x_3} (-\frac{i}{2}\epsilon_2 \cdot \dot{x}_2) e^{i k_2\cdot x_2} e^{i k_1 \cdot x_1}\rangle \\
	=&-\frac{1}{2} \langle e^{i k_3 \cdot x_3} e^{i k_2\cdot x_2 + i \epsilon_2 \cdot \dot{x}_2} e^{i k_1 \cdot x_1}\rangle |_{m.l.}
\end{align*}
where the $m.l.$ means take multi-linear terms of different $\epsilon$. Here we only have one $\epsilon$, so that means we will take the term which is linear in $\epsilon_2$ at the end. The advantage of this form is that we could easily apply the formula
\[
\langle \prod_{i}:e^{A_i}:\rangle  = e^{\sum_{i < j} \langle A_i A_j\rangle }
\]
Thus, we have
\begin{align*}
	A_{3,sb}=&-\frac{1}{2} exp\{\langle (i k_3 \cdot x_3)(i\epsilon_2 \cdot \dot{x}_2)\rangle  + \langle (i\epsilon_2 \cdot \dot{x}_2)(i k_1 \cdot x_1)\rangle \}|_{m.l.}\\
	&\exp\{\langle (i k_3 \cdot x_3)(i k_2\cdot x_2)\rangle +\langle (i k_3 \cdot x_3)(i k_1 \cdot x_1)\rangle +\langle (i k_2\cdot x_2)(i k_1 \cdot x_1)\rangle \}\\
	=& -\frac{1}{2} \exp\{- k_{3\mu} \epsilon_{2\nu} \langle x^{\mu}_3 \dot{x}^{\nu}_2\rangle  - \epsilon_{2\mu} k_{1\nu} \langle \dot{x}^{\mu}_2 x^{\nu}_1\rangle \}|_{m.l.}\\
	&\exp\{-k_{3\mu} k_{2\nu} \langle x^{\mu}_3 x^{\nu}_2\rangle  - k_{3\mu} k_{1\nu} \langle x^{\mu}_3 x^{\nu}_1\rangle  - k_{2\mu} k_{1\nu} \langle x^{\mu}_2 x^{\nu}_1\rangle \}\\
	=& -\frac{1}{2} \exp\{-\frac{1}{2} k_{3} \cdot \epsilon_{2} + \frac{1}{2} \epsilon_{2} \cdot k_{1}\}|_{m.l.}\\
	&\exp\{\frac{1}{2} k_{3} \cdot k_{2} (\tau_3-\tau_2) + \frac{1}{2} k_{3} \cdot k_{1} (\tau_3-\tau_1) + \frac{1}{2} k_{2} \cdot k_{1} (\tau_2-\tau_1)\}\\
	=& -\frac{1}{2} \times (-\frac{1}{2}) \epsilon_2 \cdot (k_3 - k_1) \times 1\\
	=& \frac{1}{4} \epsilon_2 \cdot (k_3 - k_1)
\end{align*}
where in the last steps we used the mass shell condition for the scalar particle $k_3^2=k_1^2=0$. The dependence on $\tau_2$ drops out for the same reason: $\exp(\frac 12 \tau_2 k_2\cdot(k_1-k_3))=\exp(-\frac 12 \tau_2 (k_1+k_3)\cdot(k_1-k_3))=1$.

\section{Scalar interacting with background gravity}
\label{appendix: scalar-gravity}
\noindent Recall the expression for the amplitude
\begin{align}
	A_{4,sg}
	=& \int^{+\infty}_{-\infty} d\tau \langle e^{i k_4 \cdot x(+\infty)}\mathcal{T}\{[-\frac{1}{2}\epsilon_{3\mu\nu}\dot{x}^{\mu}(\tau) \dot{x}^{\nu}(\tau)]e^{i k_3 \cdot x(\tau)} [-\frac{1}{2}\epsilon_{2\bar{\mu}\bar{\nu}}\dot{x}^{\bar{\mu}}(0) \dot{x}^{\bar{\nu}}(0)]e^{i k_2 \cdot x(0)}\} e^{i k_1 \cdot x(-\infty)}\rangle  \label{eq:1} \\
	&+\int^{+\infty}_{-\infty} d\tau \langle e^{i k_4 \cdot x(+\infty)}\mathcal{T}\{[-\frac{1}{2}\epsilon_{3\mu\nu}a^{\mu}(\tau) a^{\nu}(\tau)]e^{i k_3 \cdot x(\tau)} [-\frac{1}{2}\epsilon_{2\bar{\mu}\bar{\nu}}a^{\bar{\mu}}(0) a^{\bar{\nu}}(0)]e^{i k_2 \cdot x(0)}\} e^{i k_1 \cdot x(-\infty)}\rangle  \label{eq:2} \\
	&+\int^{+\infty}_{-\infty} d\tau \langle e^{i k_4 \cdot x(+\infty)}\mathcal{T}\{[-\frac{1}{2}\epsilon_{3\mu\nu}b^{\mu}(\tau) c^{\nu}(\tau)]e^{i k_3 \cdot x(\tau)} [-\frac{1}{2}\epsilon_{2\bar{\mu}\bar{\nu}}b^{\bar{\mu}}(0) c^{\bar{\nu}}(0)]e^{i k_2 \cdot x(0)}\} e^{i k_1 \cdot x(-\infty)}\rangle  \label{eq:3} \\
	&+[\frac{3t}{32}(\epsilon_2 \cdot \epsilon_3)^2+\frac{1}{8}(\epsilon_2 \cdot k_3)(\epsilon_3 \cdot k_2)(\epsilon_2 \cdot \epsilon_3)]\langle e^{i k_4 \cdot x(+\infty)} e^{i(k_2+k_3)\cdot x(0)} e^{i k_1 \cdot x(-\infty)}\rangle  \label{eq:4}
\end{align}
with basic unregularized correlation functions
\[
\begin{aligned}
	\langle x^{\mu}(\tau)x^{\nu}(\tau')\rangle  =& -\frac{1}{2}\eta^{\mu\nu}|\tau-\tau'|\\
	\langle a^{\mu}(\tau)a^{\nu}(\tau')\rangle  =& \eta^{\mu\nu}\delta(\tau-\tau')\\
	\langle b^{\mu}(\tau)c^{\nu}(\tau')\rangle  =& -2\eta^{\mu\nu}\delta(\tau-\tau')
\end{aligned}
\]
Other correlation functions can be deduced from the basic ones
\[
\begin{aligned}
	\langle \dot{x}^{\mu}(\tau)x^{\nu}(\tau')\rangle  =& -\frac{1}{2}\eta^{\mu\nu}sign(\tau-\tau')\\
	\langle \dot{x}^{\mu}(\tau)\dot{x}^{\nu}(\tau')\rangle  =& \eta^{\mu\nu} \delta(\tau-\tau')\\
	\langle c^{\mu}(\tau)b^{\nu}(\tau')\rangle  =& +2\eta^{\mu\nu}\delta(\tau-\tau')
\end{aligned}
\]
Then all we need is just do the Wick contractions for $x$, $a$, $b$ and $c$. Just as we did in the scalar QED case we can exponentiate $A_{4,sg}$ to facilitate performing the Wick contractions
\[
\begin{aligned}
	A_{4,sg}
	=& \frac{1}{4}\int^{+\infty}_{-\infty} d\tau 
	\langle e^{i k_4 \cdot x_4}
	\mathcal{T}\{e^{i k_3 \cdot x_3 + \epsilon_3 \cdot \dot{x}_3+\tilde{\epsilon}_3 \cdot \dot{x}_3} 
	e^{i k_2 \cdot x_2 + \epsilon_2 \cdot \dot{x}_2+\tilde{\epsilon}_2 \cdot \dot{x}_2}\} 
	e^{i k_1 \cdot x_1}\rangle |_{m.l.}\\
	&+\frac{1}{4}\int^{+\infty}_{-\infty} d\tau 
	\langle e^{i k_4 \cdot x_4}
	\mathcal{T}\{e^{i k_3 \cdot x_3 + \epsilon_3 \cdot a_3 + \tilde{\epsilon}_3 \cdot a_3}
	e^{i k_2 \cdot x_2 + \epsilon_2 \cdot a_2 + \tilde{\epsilon}_2 \cdot a_2}\} 
	e^{i k_1 \cdot x_1}\rangle |_{m.l.}\\
	&-\frac{1}{4}\int^{+\infty}_{-\infty} d\tau \int d\xi_3\ d\tilde{\xi_3}\ d\xi_2 d\tilde{\xi_2}\\
	&\langle e^{i k_4 \cdot x_4}
	\mathcal{T}\{e^{i k_3 \cdot x_3 + \xi_3 \epsilon_3 \cdot b_3 + \tilde{\xi}_3 \tilde{\epsilon}_3 \cdot c_3} 
	e^{i k_2 \cdot x_2 + \xi_2 \epsilon_2 \cdot b_2 + \tilde{\xi}_2 \tilde{\epsilon}_2 \cdot c_2}\} 
	e^{i k_1 \cdot x_1}\rangle |_{m.l.}\\
	&+[\frac{3t}{32}(\epsilon_2 \cdot \epsilon_3)^2+\frac{1}{8}(\epsilon_2 \cdot k_3)(\epsilon_3 \cdot k_2)(\epsilon_2 \cdot \epsilon_3)]\langle e^{i k_4 \cdot x_4} e^{i(k_2+k_3)\cdot x_2} e^{i k_1 \cdot x_1}\rangle \\
\end{aligned}\,.
\]
Here, we have written $\epsilon_{\mu\nu}$ as $\epsilon_{\mu} \tilde{\epsilon}_{\nu}$ while $\epsilon$ and $\tilde{\epsilon}$ are actually the same thing (recall that the on-shell graviton polarizations are ++ and --). We write them differently just for bookkeeping, which will help us correctly take the multi-linear terms at the end. We also abbreviate $x(+\infty)$, $x(\tau)$, $x(0)$ and $x(-\infty)$ as $x_4$, $x_3$, $x_2$ and $x_1$ respectively. We introduced $\xi$ and $\tilde{\xi}$ as Grassman-odd parameters to keep the order of $b$ and $c$ ghosts. The $m.l.$ means take multi-linear terms of different $\epsilon$ and $\tilde{\epsilon}$ at the end.
In practice, there is no need to actually do the promotion at the beginning and the selection at the end. All we need is to sum up all possible contractions.\par
In principle at this stage, all we have left to do is the Wick contractions.  However, this is naive. By doing the Wick contractions for the amplitude, one encounters ill-defined integrals such as $\int d\tau\  \delta^2(\tau) e^{\Sigma}$ and $\int d\tau\ \delta(\tau) sign^2(\tau) e^{\Sigma}$. It is known that properly to handle such ill-defined integrals, we need to regularize them. There are different regularization schemes but we will only consider dimensional regularization (dim-reg hereafter) right now.\par
In dim-reg one promotes the worldline action to $D$ dimension,
\[
S \rightarrow \int_{\Omega} d^{D}\tau\ \bigg[\frac{1}{2}g_{\mu\nu}(\partial_{I} x^{\mu} \partial_{I} x^{\nu} + a^{\mu}a^{\nu} + b^{\mu}c^{\nu}) - \frac{1}{8}R\bigg]
\]
which will result in the following correlation functions
\begin{align*}
	\langle x^{\mu}(\tau)x^{\nu}(\tau')\rangle  =& \eta^{\mu\nu} \Delta(\tau-\tau')\\
	\langle a^{\mu}(\tau)a^{\nu}(\tau')\rangle  =& \eta^{\mu\nu} \Delta_{gh}(\tau-\tau')\\
	\langle b^{\mu}(\tau)c^{\nu}(\tau')\rangle  =& -2\eta^{\mu\nu}\Delta_{gh}(\tau-\tau') = -2\eta^{\mu\nu}\delta^{D}(\tau-\tau')\\
	\langle \partial_I x^{\mu}(\tau)x^{\nu}(\tau')\rangle  =& \eta^{\mu\nu} \Delta_I(\tau-\tau')\\
	\langle \partial_I x^{\mu}(\tau) \partial_J x^{\nu}(\tau')\rangle  =& -\eta^{\mu\nu} \Delta_{I J}(\tau-\tau')\\
	\langle c^{\mu}(\tau)b^{\nu}(\tau')\rangle  =& +2\eta^{\mu\nu}\Delta_{gh}(\tau-\tau')= +2\eta^{\mu\nu}\delta^{D}(\tau-\tau')
\end{align*}
where $\Delta(\tau-\tau')$ is a Green's function, the solution to the equation\\
${\Box G(\tau) \equiv \partial_I \partial_I G(\tau) = -\delta^D(\tau)}$. ${\Delta_I(\tau) \equiv \partial_I \Delta(\tau)}$ and ${\Delta_{I J}(\tau) \equiv \partial_I \partial_{J} \Delta(\tau)}$ are two abbreviations. And ${\Delta_{gh}(\tau) = \delta^D(\tau)}$. %From now on, we will simply write $d\tau$ even when it's actually $d^D \tau$.

Notice that with dim-reg, $\Delta_{I J}(\tau) \ne \Delta_{I I}(\tau)$ in general $D$ dimensions, but they reduce to be the same when $D=1$, which shows us why there is ambiguity in $D=1$ case but not in general $D$ dimension cases. Our goal is that by manipulating the expression(integration by parts) in general $D$ dimension case, we get a form that is no longer ambiguous when reduced to $D=1$.\par
In practice, we could start from the $D=1$ case, calculate the well-defined terms, promote those problematic terms into general $D$-dimensions, manipulate them until no more ambiguity exists, go back to $D=1$, and finish the calculation.\par
Now, we are finally ready to tackle the calculation amplitude, we will proceed term by term. 

First consider the term \eqref{eq:1}
\begin{align*}
	& \int^{+\infty}_{-\infty} d\tau \langle e^{i k_4 \cdot x(+\infty)}\mathcal{T}\{[-\frac{1}{2}\epsilon_{3\mu\nu}\dot{x}^{\mu}(\tau) \dot{x}^{\nu}(\tau)]e^{i k_3 \cdot x(\tau)} [-\frac{1}{2}\epsilon_{2\bar{\mu}\bar{\nu}}\dot{x}^{\bar{\mu}}(0) \dot{x}^{\bar{\nu}}(0)]e^{i k_2 \cdot x(0)}\} e^{i k_1 \cdot x(-\infty)}\rangle  \\
	\Rightarrow& \frac{1}{4} [\epsilon_3 \cdot \frac{i}{2}(k_4 - k_2 - k_1)]^2 [\epsilon_2 \cdot \frac{i}{2}(k_4 + k_3 - k_1)]^2 \int_{0}^{\infty} d\tau\ \langle e^{i k_4 \cdot x_4}e^{i k_3 \cdot x_3}e^{i k_2 \cdot x_2}e^{i k_1 \cdot x_1}\rangle \\
	&+\frac{1}{4} [\epsilon_2 \cdot \frac{i}{2}(k_4 - k_3 - k_1)]^2 [\epsilon_3 \cdot \frac{i}{2}(k_4 + k_2 - k_1)]^2 \int_{-\infty}^{0} d\tau\ \langle e^{i k_4 \cdot x_4}e^{i k_2 \cdot x_2}e^{i k_3 \cdot x_3}e^{i k_1 \cdot x_1}\rangle \\
	&+\frac{1}{4}\int_{\Omega} d^D\tau\ 4 (\epsilon_2 \cdot \epsilon_3)[-\Delta_{I J}(\tau)] [\epsilon_3 \cdot (\frac{i}{2}k_4 + i k_2 \Delta_I(\tau) - \frac{i}{2}k_1)][\epsilon_2 \cdot (\frac{i}{2}k_4 + i k_3 \Delta_J(-\tau) - \frac{i}{2}k_1)]\\
	&\ \ \ \ \ \ \ \ \langle e^{i k_4 \cdot x_4}\mathcal{T}\{e^{i k_3 \cdot x_3}e^{i k_2 \cdot x_2}\}e^{i k_1 \cdot x_1}\rangle \\
	&+\frac{1}{4} \int_{\Omega} d^D\tau\ 2 (\epsilon_2 \cdot \epsilon_3)^2 \Delta_{I J}(\tau) \Delta_{I J}(\tau) \langle e^{i k_4 \cdot x_4}\mathcal{T}\{e^{i k_3 \cdot x_3}e^{i k_2 \cdot x_2}\}e^{i k_1 \cdot x_1}\rangle \\
	=& \frac{1}{4} (\epsilon_3 \cdot k_4)^2 (\epsilon_2 \cdot k_1)^2 (-\frac{2}{s})
	+\frac{1}{4} (\epsilon_2 \cdot k_4)^2 (\epsilon_3 \cdot k_1)^2 (-\frac{2}{u})\\
	&+\Big\{\int^{+\infty}_{-\infty} d\tau\ (\epsilon_2 \cdot \epsilon_3)\delta(\tau) [\epsilon_3 \cdot (\frac{i}{2}k_4 - \frac{i}{2}k_1)][\epsilon_2 \cdot (\frac{i}{2}k_4 - \frac{i}{2}k_1)]e^{\Sigma}\\
	&+\int_{\Omega} d^D\tau\ (\epsilon_2 \cdot \epsilon_3)[-\Delta_{I J}(\tau)] [\epsilon_3 \cdot i k_2 \Delta_I(\tau)][\epsilon_2 \cdot (\frac{i}{2}k_4 - \frac{i}{2}k_1)]e^{\Sigma}\\
	&+\int_{\Omega} d^D\tau\ (\epsilon_2 \cdot \epsilon_3)[-\Delta_{I J}(\tau)] [\epsilon_3 \cdot (\frac{i}{2}k_4 - \frac{i}{2}k_1)][\epsilon_2 \cdot i k_3 \Delta_J(-\tau)]e^{\Sigma}\\
	&+\int_{\Omega} d^D\tau\ (\epsilon_2 \cdot \epsilon_3)[-\Delta_{I J}(\tau)] [\epsilon_3 \cdot i k_2 \Delta_I(\tau) ][\epsilon_2 \cdot i k_3 \Delta_J(-\tau)] e^{\Sigma}\Big\} \\
	&+\frac{1}{2} \int_{\Omega} d^D\tau\ (\epsilon_2 \cdot \epsilon_3)^2 \Delta_{I J}(\tau) \Delta_{I J}(\tau) e^{\Sigma}\\
	=& -\frac{1}{2s} (\epsilon_3 \cdot k_4)^2 (\epsilon_2 \cdot k_1)^2
	-\frac{1}{2u} (\epsilon_2 \cdot k_4)^2 (\epsilon_3 \cdot k_1)^2 \\
	&-\frac{1}{4} (\epsilon_2 \cdot \epsilon_3) [\epsilon_3 \cdot (k_4 - k_1)][\epsilon_2 \cdot (k_4 - k_1)]\\
	&+\frac{1}{2}(\epsilon_2 \cdot \epsilon_3) (\epsilon_3 \cdot k_2) [\epsilon_2 \cdot (k_4 - k_1)]\int_{\Omega} d^D\tau\ \Delta_{I J}(\tau) \Delta_I(\tau) e^{\Sigma}\\
	&-\frac{1}{2}(\epsilon_2 \cdot \epsilon_3) [\epsilon_2 \cdot (k_4 - k_1)] (\epsilon_2 \cdot k_3)\int_{\Omega} d^D\tau\ \Delta_{I J}(\tau) \Delta_J(\tau)e^{\Sigma}\\
	&-(\epsilon_2 \cdot \epsilon_3) (\epsilon_3 \cdot k_2) (\epsilon_3 \cdot k_2) \int_{\Omega} d^D\tau\ \Delta_{I J}(\tau) \Delta_I(\tau) \Delta_J(\tau) e^{\Sigma}\\
	&+\frac{1}{2} (\epsilon_2 \cdot \epsilon_3)^2 \int_{\Omega} d^D\tau\  \Delta_{I J}(\tau) \Delta_{I J}(\tau) e^{\Sigma}\\
\end{align*}
where
\begin{align*}
	&\Sigma \equiv \sum_{4 \ge i>j \ge 1} (-k_i \cdot k_j) \Delta^{(i,j)}\\
	&\Delta^{(i,j)} \equiv \Delta(\tau_i-\tau_j)\,.
\end{align*}
Here are three potentially problematic integrals over products of distributions
\begin{enumerate}
	\item $\int_{\Omega} d^D\tau\ \Delta_{I J}(\tau) \Delta_I(\tau) e^{\Sigma} \equiv \int_{\Omega} d^D\tau\ \Delta^{(3,2)}_{I J} \Delta^{(3,2)}_I e^{\Sigma}$
	\item $\int_{\Omega} d^D\tau\ \Delta_{I J}(\tau) \Delta_I(\tau) \Delta_J(\tau) e^{\Sigma} \equiv \int_{\Omega} d\tau\ \Delta^{(3,2)}_{I J} \Delta^{(3,2)}_I \Delta^{(3,2)}_J e^{\Sigma}$
	\item $\int_{\Omega} d\tau\ \Delta_{I J}(\tau) \Delta_{I J}(\tau) e^{\Sigma} \equiv \int_{\Omega} d\tau\  \Delta^{(3,2)}_{I J} \Delta^{(3,2)}_{I J} e^{\Sigma}\,.$
\end{enumerate}
We will calculate them one by one to show how dim-reg solves the singularity and ambiguity
\begin{enumerate}
	\item $\int_{\Omega} d^D\tau\ \Delta^{(3,2)}_{I J} \Delta^{(3,2)}_I e^{\Sigma}$
	\begin{align*}
		&\int_{\Omega} d^D\tau\ \Delta^{(3,2)}_{I J} \Delta^{(3,2)}_I e^{\Sigma}\\
		=&\int_{\Omega} d^D\tau\ \frac{1}{2} \partial_J (\Delta^{(3,2)}_I)^2 e^{\Sigma}\\
		=&-\frac{1}{2}\int_{\Omega} d^D\tau\ (\Delta^{(3,2)}_{I})^2 e^{\Sigma} \partial_J \Sigma\,.
	\end{align*}
	The last expression no longer has ambiguities, and can be reduced to $D=1$.
	\begin{align*}
		\int_{\Omega} d^D\tau\ \Delta^{(3,2)}_{I J} \Delta^{(3,2)}_I e^{\Sigma}
		=&-\frac{1}{2}\int^{+\infty}_{0} d\tau\ (-\frac{1}{2}sign(\tau))^2 e^{\frac{s}{2} \tau} \partial_\tau (\frac{s}{2} \tau)\\
		&-\frac{1}{2}\int^{0}_{-\infty} d\tau\ (-\frac{1}{2}sign(\tau))^2 e^{-\frac{u}{2} \tau} \partial_\tau (-\frac{u}{2} \tau)\\
		=&-\frac{1}{2}\times\frac{1}{4}\times(-\frac{2}{s})\times \frac{s}{2} -\frac{1}{2}\times\frac{1}{4}\times(-\frac{2}{u})\times (-\frac{u}{2})\\
		=&\ 0\,.
	\end{align*}
	\item $\int_{\Omega} d^D\tau\ \Delta^{(3,2)}_{I J} \Delta^{(3,2)}_I \Delta^{(3,2)}_J e^{\Sigma}$
	\begin{align*}
		&\int_{\Omega} d^D\tau\ \Delta^{(3,2)}_{I J} \Delta^{(3,2)}_I \Delta^{(3,2)}_J e^{\Sigma}\\
		=&\int_{\Omega} d^D\tau\ \frac{1}{2} \partial_J (\Delta^{(3,2)}_I)^2 \Delta^{(3,2)}_J e^{\Sigma}\\
		=&-\frac{1}{2}\int_{\Omega} d^D\tau\ (\Delta^{(3,2)}_{I})^2 \Delta^{(3,2)}_{J J} e^{\Sigma}  -\frac{1}{2}\int_{\Omega} d^D\tau\ (\Delta^{(3,2)}_{I})^2 \Delta^{(3,2)}_J e^{\Sigma} \partial_J \Sigma\,.
	\end{align*}
	Recall that $\Delta_{J J} = -\delta^{D}(\tau)$ is well defined in general $D$ dimensions and $\Delta_I(0) = 0$. The latter can be understood by simply looking at the expansion in momentum space
	\[
	\Delta_I(\tau) 
	= \partial_I \int \frac{d^D p}{(2 \pi)^D} \frac{e^{i p \cdot \tau}}{p^2 + i\epsilon}
	= \int \frac{d^D p}{(2 \pi)^D} \frac{i p_I}{p^2 + i\epsilon} e^{i p \cdot \tau}\,.
	\]
	As we can see, when $\tau = 0$, the integrand becomes odd and the integral vanishes. So only the second term is left
	\[
	\int_{\Omega} d^D\tau\ \Delta^{(3,2)}_{I J} \Delta^{(3,2)}_I \Delta^{(3,2)}_J e^{\Sigma} = 0 -\frac{1}{2}\int_{\Omega} d^D\tau\ (\Delta^{(3,2)}_{I})^2 \Delta^{(3,2)}_J e^{\Sigma} \partial_J \Sigma\,.
	\]
	Since this is no longer problematic in $D = 1$, we finish the calculation in $D = 1$
	\begin{align*}
		\int_{\Omega} d^D\tau\ \Delta^{(3,2)}_{I J} \Delta^{(3,2)}_I \Delta^{(3,2)}_J e^{\Sigma}
		=&-\frac{1}{2}\int^{+\infty}_{0} d\tau\ (-\frac{1}{2}sign(\tau))^3 e^{\frac{s}{2} \tau} \partial_\tau (\frac{s}{2} \tau)\\
		&-\frac{1}{2}\int^{0}_{-\infty} d\tau\ (-\frac{1}{2}sign(\tau))^3 e^{-\frac{u}{2} \tau} \partial_\tau (-\frac{u}{2} \tau)\\
		=&-\frac{1}{2}\times(-\frac{1}{8})\times(-\frac{2}{s})\times \frac{s}{2} -\frac{1}{2}\times(\frac{1}{8})\times(-\frac{2}{u})\times (-\frac{u}{2})\\
		=&-\frac{1}{8}\,.
	\end{align*}
	\item $\int_{\Omega} d^D\tau\  \Delta^{(3,2)}_{I J} \Delta^{(3,2)}_{I J} e^{\Sigma}$
	\begin{align*}
		&\int_{\Omega} d^D\tau\ \Delta^{(3,2)}_{I J} \Delta^{(3,2)}_{I J} e^{\Sigma}\\
		=&-\int_{\Omega} d^D\tau\ \Delta^{(3,2)}_{I I J} \Delta^{(3,2)}_J e^{\Sigma}
		- \int_{\Omega} d^D\tau\ \Delta^{(3,2)}_{I J} \Delta^{(3,2)}_{J} e^{\Sigma} \partial_I \Sigma\\
		=&\Big\{ \int_{\Omega} d^D\tau\ \Delta^{(3,2)}_{I I} \Delta^{(3,2)}_{J J} e^{\Sigma}
		+\int_{\Omega} d^D\tau\ \Delta^{(3,2)}_{I I} \Delta^{(3,2)}_{J} e^{\Sigma} \partial_{J} \Sigma \Big\}\\
		&-\frac{1}{2} \int_{\Omega} d^D\tau\ \partial_{I} (\Delta^{(3,2)}_{J})^2 e^{\Sigma} \partial_I \Sigma\\
		=&\int_{\Omega} d^D\tau\ \Delta^{(3,2)}_{I I} \Delta^{(3,2)}_{J J} e^{\Sigma}
		+0\\
		&+\frac{1}{2} \int_{\Omega} d^D\tau\ (\Delta^{(3,2)}_{J})^2 e^{\Sigma} (\partial_I \Sigma)^2
		+\frac{1}{2} \int_{\Omega} d^D\tau\ (\Delta^{(3,2)}_{J})^2 e^{\Sigma} (\partial^2_I \Sigma)\\
		=&\int_{\Omega} d^D\tau\ \Delta^{(3,2)}_{I I} \Delta^{(3,2)}_{J J} e^{\Sigma}\\
		&+\frac{1}{2} \int_{\Omega} d^D\tau\ (\Delta^{(3,2)}_{J})^2 e^{\Sigma} (\partial_I \Sigma)^2
		+\Big\{ \frac{1}{2} \int_{\Omega} d^D\tau\ (\Delta^{(3,2)}_{J})^2 e^{\Sigma} (-k_3 \cdot k_2 \Delta^{(3,2)}_{I I})\\
		&+\frac{1}{2} \int_{\Omega} d^D\tau\ (\Delta^{(3,2)}_{J})^2 e^{\Sigma} (-k_4 \cdot k_3 \Delta^{(4,3)}_{I I})
		+\frac{1}{2} \int_{\Omega} d^D\tau\ (\Delta^{(3,2)}_{J})^2 e^{\Sigma} (-k_3 \cdot k_1 \Delta^{(3,1)}_{I I})\Big\} \\
		=&\int_{\Omega} d^D\tau\ \Delta^{(3,2)}_{I I} \Delta^{(3,2)}_{J J} e^{\Sigma}
		+\frac{1}{2} \int_{\Omega} d^D\tau\ (\Delta^{(3,2)}_{J})^2 e^{\Sigma} (\partial_I \Sigma)^2
		+0+0+0\\
		=&\int_{\Omega} d^D\tau\ \Delta^{(3,2)}_{I I} \Delta^{(3,2)}_{J J} e^{\Sigma}
		+\frac{1}{2} \int_{\Omega} d^D\tau\ (\Delta^{(3,2)}_{J})^2 e^{\Sigma} (\partial_I \Sigma)^2\,.
	\end{align*}
	There are terms vanishing in the intermediate steps because either $\Delta_I(0) = 0$ or $e^\Sigma$ converges to 0 at infinity in $D = 1$. In the final expression above, the first term  is a divergent part and we will soon see the ghosts' contribution coming to rescue us. The second term is no longer problematic:
	\begin{align*}
		\int_{\Omega} d^D\tau\  \Delta^{(3,2)}_{I J} \Delta^{(3,2)}_{I J} e^{\Sigma}
		=&\int_{\Omega} d^D\tau\ \Delta^{(3,2)}_{I I} \Delta^{(3,2)}_{J J} e^{\Sigma}\\
		&+\Big\{\frac{1}{2} \int^{+\infty}_{0} d^D\tau\ (-\frac{1}{2}sign(\tau))^2 e^{\frac{s}{2}\tau} [\partial_{\tau} (\frac{s}{2}\tau)]^2\\
		&+\frac{1}{2} \int^{0}_{-\infty} d^D\tau\ (-\frac{1}{2}sign(\tau))^2 e^{-\frac{u}{2}\tau} [\partial_{\tau} (-\frac{u}{2}\tau)]^2\Big\}\\
		=&\int_{\Omega} d^D\tau\ \Delta^{(3,2)}_{I I} \Delta^{(3,2)}_{J J} e^{\Sigma} + \frac{t}{16}\,.
	\end{align*}
\end{enumerate}
We have finished the calculation for term \eqref{eq:1}:
\begin{align*}
	& \int^{+\infty}_{-\infty} d\tau \langle e^{i k_4 \cdot x(+\infty)}\mathcal{T}\{[-\frac{1}{2}\epsilon_{3\mu\nu}\dot{x}^{\mu}(\tau) \dot{x}^{\nu}(\tau)]e^{i k_3 \cdot x(\tau)} [-\frac{1}{2}\epsilon_{2\bar{\mu}\bar{\nu}}\dot{x}^{\bar{\mu}}(0) \dot{x}^{\bar{\nu}}(0)]e^{i k_2 \cdot x(0)}\} e^{i k_1 \cdot x(-\infty)}\rangle  \\
	\Rightarrow& -\frac{1}{2s} (\epsilon_3 \cdot k_4)^2 (\epsilon_2 \cdot k_1)^2
	-\frac{1}{2u} (\epsilon_2 \cdot k_4)^2 (\epsilon_3 \cdot k_1)^2 \\
	&-\frac{1}{4} (\epsilon_2 \cdot \epsilon_3) [\epsilon_3 \cdot (k_4 - k_1)][\epsilon_2 \cdot (k_4 - k_1)]\\
	&+\frac{1}{8}(\epsilon_2 \cdot \epsilon_3) (\epsilon_3 \cdot k_2) (\epsilon_3 \cdot k_2)\\
	&+\Big\{ \frac{t}{32} (\epsilon_2 \cdot \epsilon_3)^2
	+\frac{1}{2} (\epsilon_2 \cdot \epsilon_3)^2 \int_{\Omega} d^D\tau\  \delta^D(\tau) \delta^D(\tau) e^{\Sigma}\Big\}\,.\\
\end{align*}
The terms \eqref{eq:2}, \eqref{eq:3} and \eqref{eq:4} are much easier to calculate.
Consider first \eqref{eq:2}
\begin{align*}
	&\int^{+\infty}_{-\infty} d\tau \langle e^{i k_4 \cdot x(+\infty)}\mathcal{T}\{[-\frac{1}{2}\epsilon_{3\mu\nu}a^{\mu}(\tau) a^{\nu}(\tau)]e^{i k_3 \cdot x(\tau)} [-\frac{1}{2}\epsilon_{2\bar{\mu}\bar{\nu}}a^{\bar{\mu}}(0) a^{\bar{\nu}}(0)]e^{i k_2 \cdot x(0)}\} e^{i k_1 \cdot x(-\infty)}\rangle \\
	\Rightarrow&\frac{1}{4}\int_{\Omega} d^D\tau\ 2 (\epsilon_2 \cdot \epsilon_3)^2 \Delta^2_{gh} e^{\Sigma}\\
	=&\frac{1}{2} (\epsilon_2 \cdot \epsilon_3)^2 \int_{\Omega} d^D\tau\ \Delta^2_{gh} e^{\Sigma}\\
	=&\frac{1}{2} (\epsilon_2 \cdot \epsilon_3)^2 \int_{\Omega} d^D\tau\  \delta^D(\tau) \delta^D(\tau) e^{\Sigma}\,.
\end{align*}
Next, consider  \eqref{eq:3}:
\begin{align*}
	&\int^{+\infty}_{-\infty} d\tau \langle e^{i k_4 \cdot x(+\infty)}\mathcal{T}\{[-\frac{1}{2}\epsilon_{3\mu\nu}b^{\mu}(\tau) c^{\nu}(\tau)]e^{i k_3 \cdot x(\tau)} [-\frac{1}{2}\epsilon_{2\bar{\mu}\bar{\nu}}b^{\bar{\mu}}(0) c^{\bar{\nu}}(0)]e^{i k_2 \cdot x(0)}\} e^{i k_1 \cdot x(-\infty)}\rangle \\
	\Rightarrow&\frac{1}{4}\int_{\Omega} d^D\tau\ (-4) (\epsilon_2 \cdot \epsilon_3)^2 \Delta^2_{gh} e^{\Sigma}\\
	=&-(\epsilon_2 \cdot \epsilon_3)^2 \int_{\Omega} d^D\tau\ \Delta^2_{gh} e^{\Sigma}\\
	=&-(\epsilon_2 \cdot \epsilon_3)^2 \int_{\Omega} d^D\tau\  \delta^D(\tau) \delta^D(\tau) e^{\Sigma}\,.
\end{align*}
Lastly, we can calculate the term \eqref{eq:4}:
\begin{align*}
	&[\frac{3t}{32}(\epsilon_2 \cdot \epsilon_3)^2+\frac{1}{8}(\epsilon_2 \cdot k_3)(\epsilon_3 \cdot k_2)(\epsilon_2 \cdot \epsilon_3)]\langle e^{i k_4 \cdot x(+\infty)} e^{i(k_2+k_3)\cdot x(0)} e^{i k_1 \cdot x(-\infty)}\rangle \\
	=&\frac{3t}{32}(\epsilon_2 \cdot \epsilon_3)^2+\frac{1}{8}(\epsilon_2 \cdot k_3)(\epsilon_3 \cdot k_2)(\epsilon_2 \cdot \epsilon_3)\,.
\end{align*}
Summing up the terms \eqref{eq:1}, \eqref{eq:2}, \eqref{eq:3} and \eqref{eq:4}, we can see that the divergent part cancels out completely and we arrive at the final expression for the 2-scalar - 2-background-graviton scattering amplitude:
\begin{align*}
	A_{4,sg}=
	&-\frac{1}{2s} (\epsilon_3 \cdot k_4)^2 (\epsilon_2 \cdot k_1)^2
	-\frac{1}{2u} (\epsilon_2 \cdot k_4)^2 (\epsilon_3 \cdot k_1)^2 \\
	&-\frac{1}{4} (\epsilon_2 \cdot \epsilon_3) [\epsilon_3 \cdot (k_4 - k_1)][\epsilon_2 \cdot (k_4 - k_1)]\\
	&+\frac{1}{8}(\epsilon_2 \cdot \epsilon_3) (\epsilon_3 \cdot k_2) (\epsilon_3 \cdot k_2)\\
	&+\frac{t}{32} (\epsilon_2 \cdot \epsilon_3)^2\\
	&+\frac{3t}{32}(\epsilon_2 \cdot \epsilon_3)^2+\frac{1}{8}(\epsilon_2 \cdot k_3)(\epsilon_3 \cdot k_2)(\epsilon_2 \cdot \epsilon_3)\\
	=&-\frac{1}{2}\{\frac{1}{s}[(\epsilon_2 \cdot k_1)(\epsilon_3 \cdot k_4)-\frac{s}{2}(\epsilon_2 \cdot \epsilon_3)]^2+\frac{1}{u}[(\epsilon_3 \cdot k_1)(\epsilon_2 \cdot k_4)-\frac{u}{2}(\epsilon_2 \cdot \epsilon_3)]^2\}\,.
\end{align*}
\pagebreak

\section{Photon worldline interacting with background gravity}
\label{appendix: photon-gravity}
\noindent The scattering amplitude for a process where an on-shell photon, with incoming momentum $k_1$ and polarization $\epsilon_1$, interacts with two background gravitons and becomes an on-shell outgoing photon with momentum $k_4$ and polarization $\epsilon_4$ is given by
\begin{align*}
	A_{4,bg}
	=& \int^{+\infty}_{-\infty} d\tau 
	\langle \epsilon_4 \cdot \Psi(+\infty) e^{i k_4 \cdot x(+\infty)} \nonumber \\
	&\hspace{2cm} \mathcal{T}\Big\{\big[-\frac{1}{2}\epsilon_{3\mu\nu} \dot{x}^{\mu}(\tau) [\dot{x}^{\nu}(\tau) + i k_{3 \sigma} S^{\nu\sigma}(\tau)]\big]e^{i k_3 \cdot x(\tau)} \nonumber \\
	& \hspace{2.5cm} \big[-\frac{1}{2}\epsilon_{2\bar{\mu}\bar{\nu}} \dot{x}^{\bar{\mu}}(0) [\dot{x}^{\bar{\nu}}(0) + i k_{2 \bar{\sigma}} S^{\bar{\nu}\bar{\sigma}}(0)]\big]e^{i k_2 \cdot x(0)}\Big\} \nonumber \\
	& \hspace{1.5cm} \epsilon_1 \cdot \bar{\Psi}(-\infty) e^{i k_1 \cdot x(-\infty)}\rangle  \\
	&+\int^{+\infty}_{-\infty} d\tau 
	\langle\epsilon_4 \cdot \Psi(+\infty) e^{i k_4 \cdot x(+\infty)}\\
	&\hspace{2cm} \mathcal{T}\{[-\frac{1}{2}\epsilon_{3\mu\nu}a^{\mu}(\tau) a^{\nu}(\tau)]e^{i k_3 \cdot x(\tau)} [-\frac{1}{2}\epsilon_{2\bar{\mu}\bar{\nu}}a^{\bar{\mu}}(0) a^{\bar{\nu}}(0)]e^{i k_2 \cdot x(0)}\} \\
	&\hspace{2cm} \epsilon_1 \cdot \bar{\Psi}(-\infty) e^{i k_1 \cdot x(-\infty)}\rangle  \\
	&+\int^{+\infty}_{-\infty} d\tau 
	\langle\epsilon_4 \cdot \Psi(+\infty) e^{i k_4 \cdot x(+\infty)}\\
	&\hspace{2cm} \mathcal{T}\{[-\frac{1}{2}\epsilon_{3\mu\nu}b^{\mu}(\tau) c^{\nu}(\tau)]e^{i k_3 \cdot x(\tau)} [-\frac{1}{2}\epsilon_{2\bar{\mu}\bar{\nu}}b^{\bar{\mu}}(0) c^{\bar{\nu}}(0)]e^{i k_2 \cdot x(0)}\} \\
	&\hspace{2cm} \epsilon_1 \cdot \bar{\Psi}(-\infty) e^{i k_1 \cdot x(-\infty)}\rangle  \\
	&+\langle\epsilon_4 \cdot \Psi(+\infty) e^{i k_4 \cdot x(+\infty)}\\
	&\hspace{0.5cm} \Big\{-\frac{1}{2} \dot{x}^{\mu}(0) \omega_{\mu a b}^{(2\epsilon)} S^{ab}(0)-\frac{1}{2} R_{ad}^{(2\epsilon)} \bar{\Psi}^{a}(0) \Psi^d(0) + \frac{1}{8} R^{(2\epsilon)} \Big\} e^{i(k_2+k_3)\cdot x(0)}\\
	&\hspace{0.5cm} \epsilon_1 \cdot \bar{\Psi}(-\infty) e^{i k_1 \cdot x(-\infty)}\rangle\,,
\end{align*}
where we used a subscript $(2\epsilon)$ to denote that we are expanding in plane waves, taking multilinear terms and striping the common factor $e^{i (k_2 + k_3) \cdot x(0)}$. e.g.
\begin{align*}
	R^{(2)}(\tau) =& \frac{3}{4} (\partial_{\mu} h_{\alpha \beta}(\tau))^2 - \frac{1}{2} \partial_{\alpha} h_{\beta \mu}(\tau) \partial^{\beta} h^{\alpha \mu}(\tau)\\
	\Rightarrow& [\frac{3t}{4}(\epsilon_2 \cdot \epsilon_3)^2+(\epsilon_2 \cdot k_3)(\epsilon_3 \cdot k_2)(\epsilon_2 \cdot \epsilon_3)]e^{i(k_2+k_3)\cdot x(\tau)}\\
	R^{(2\epsilon)} =& \frac{3t}{4}(\epsilon_2 \cdot \epsilon_3)^2+(\epsilon_2 \cdot k_3)(\epsilon_3 \cdot k_2)(\epsilon_2 \cdot \epsilon_3)\,.
\end{align*}
By simplifying the expression and rearranging the terms we get
\begin{align*}
	A_{4,bg}
	=& \Big\{ (\epsilon_4 \cdot \epsilon_1) \int^{+\infty}_{-\infty} d\tau 
	\langle e^{i k_4 \cdot x(+\infty)} \nonumber \\
	&\hspace{2cm} \mathcal{T}\Big\{\big[-\frac{1}{2}\epsilon_{3\mu\nu} \dot{x}^{\mu}(\tau) \dot{x}^{\nu}(\tau)e^{i k_3 \cdot x(\tau)} %\nonumber \\
	%& \hspace{2.5cm} 
 \big[-\frac{1}{2}\epsilon_{2\bar{\mu}\bar{\nu}} \dot{x}^{\bar{\mu}}(0) \dot{x}^{\bar{\nu}}(0) \big]e^{i k_2 \cdot x(0)}\Big\} %\nonumber \\
	%& \hspace{2.5cm} 
 e^{i k_1 \cdot x(-\infty)}\rangle  \\
	&+ (\epsilon_4 \cdot \epsilon_1) \int^{+\infty}_{-\infty} d\tau 
	\langle e^{i k_4 \cdot x(+\infty)}\\
	&\hspace{2.5cm} \mathcal{T}\{[-\frac{1}{2}\epsilon_{3\mu\nu}a^{\mu}(\tau) a^{\nu}(\tau)]e^{i k_3 \cdot x(\tau)} [-\frac{1}{2}\epsilon_{2\bar{\mu}\bar{\nu}}a^{\bar{\mu}}(0) a^{\bar{\nu}}(0)]e^{i k_2 \cdot x(0)}\} \\
	&\hspace{2.5cm} e^{i k_1 \cdot x(-\infty)}\rangle  \\
	&+ (\epsilon_4 \cdot \epsilon_1) \int^{+\infty}_{-\infty} d\tau 
	\langle e^{i k_4 \cdot x(+\infty)}\\
	&\hspace{2.5cm} \mathcal{T}\{[-\frac{1}{2}\epsilon_{3\mu\nu}b^{\mu}(\tau) c^{\nu}(\tau)]e^{i k_3 \cdot x(\tau)} [-\frac{1}{2}\epsilon_{2\bar{\mu}\bar{\nu}}b^{\bar{\mu}}(0) c^{\bar{\nu}}(0)]e^{i k_2 \cdot x(0)}\} \\
	&\hspace{2.5cm} e^{i k_1 \cdot x(-\infty)}\rangle  \\
	&+ (\epsilon_4 \cdot \epsilon_1) \frac{1}{8} R^{(2\epsilon)} \Big\} \\
	&+ \int^{+\infty}_{-\infty} d\tau 
	\langle \epsilon_4 \cdot \Psi(+\infty) e^{i k_4 \cdot x(+\infty)} \nonumber \\
	&\hspace{2cm} \mathcal{T}\Big\{\big[-\frac{1}{2}\epsilon_{3\mu\nu} \dot{x}^{\mu}(\tau) \dot{x}^{\nu}(\tau)\big]e^{i k_3 \cdot x(\tau)} \nonumber \\
	& \hspace{2.5cm} \big[-\frac{1}{2}\epsilon_{2\bar{\mu}\bar{\nu}} \dot{x}^{\mu}(0) [i k_{2 \bar{\sigma}} S^{\bar{\nu}\bar{\sigma}}(0)]\big]e^{i k_2 \cdot x(0)}\Big\} \nonumber \\
	& \hspace{1.5cm} \epsilon_1 \cdot \bar{\Psi}(-\infty) e^{i k_1 \cdot x(-\infty)}\rangle  \\
	&+ \int^{+\infty}_{-\infty} d\tau 
	\langle \epsilon_4 \cdot \Psi(+\infty) e^{i k_4 \cdot x(+\infty)} \nonumber \\
	&\hspace{2cm} \mathcal{T}\Big\{\big[-\frac{1}{2}\epsilon_{3\mu\nu} \dot{x}^{\mu}(\tau) [i k_{i \sigma} S^{\nu\sigma}(\tau)]\big]e^{i k_3 \cdot x(\tau)} \nonumber \\
	& \hspace{2.5cm} \big[-\frac{1}{2}\epsilon_{2\bar{\mu}\bar{\nu}} \dot{x}^{\bar{\mu}}(0) \dot{x}^{\bar{\nu}}(0) \big]e^{i k_3 \cdot x(\tau)}\Big\} \nonumber \\
	& \hspace{1.5cm} \epsilon_1 \cdot \bar{\Psi}(-\infty) e^{i k_1 \cdot x(-\infty)}\rangle  \\
	&+ \int^{+\infty}_{-\infty} d\tau 
	\langle \epsilon_4 \cdot \Psi(+\infty) e^{i k_4 \cdot x(+\infty)} \nonumber \\
	&\hspace{2cm} \mathcal{T}\Big\{\big[-\frac{1}{2}\epsilon_{3\mu\nu} \dot{x}^{\mu}(\tau_i) [i k_{i \sigma} S^{\nu\sigma}(\tau)]\big]e^{i k_3 \cdot x(\tau)} \nonumber \\
	& \hspace{2.5cm} \big[-\frac{1}{2}\epsilon_{2\bar{\mu}\bar{\nu}} \dot{x}^{\bar{\mu}}(0) [i k_{i \bar{\sigma}} S^{\bar{\nu}\bar{\sigma}}(0)]\big]e^{i k_2 \cdot x(0)}\Big\} \nonumber \\
	& \hspace{1.5cm} \epsilon_1 \cdot \bar{\Psi}(-\infty) e^{i k_1 \cdot x(-\infty)}\rangle  \\
	&+\langle\epsilon_4 \cdot \Psi(+\infty) e^{i k_4 \cdot x(+\infty)}\\
	&\hspace{0.5cm} \Big\{-\frac{1}{2} \dot{x}^{\mu}(\tau) \omega_{\mu a b}^{(2\epsilon)} S^{ab}(0)-\frac{1}{2} R_{ad}^{(2\epsilon)} \bar{\Psi}^{a}(0) \Psi^d(0) \Big\} e^{i(k_2+k_3)\cdot x(0)}\\
	&\hspace{0.5cm} \epsilon_1 \cdot \bar{\Psi}(-\infty) e^{i k_1 \cdot x(-\infty)}\rangle\,.
\end{align*}
We can easily recognize the part which is the amplitude of scalar interacting with background gravity
\begin{align}
	A_{4,bg}
	=& (\epsilon_4 \cdot \epsilon_1) A_{4,sg} \nonumber \\
	&+ \int^{+\infty}_{-\infty} d\tau 
	\langle \epsilon_4 \cdot \Psi(+\infty) e^{i k_4 \cdot x(+\infty)} \nonumber \\
	&\hspace{2cm} \mathcal{T}\Big\{\big[-\frac{1}{2}\epsilon_{3\mu\nu} \dot{x}^{\mu}(\tau) \dot{x}^{\nu}(\tau)\big]e^{i k_3 \cdot x(\tau)} \nonumber \\
	& \hspace{2.5cm} \big[-\frac{1}{2}\epsilon_{2\bar{\mu}\bar{\nu}} \dot{x}^{\mu}(0) [i k_{2 \bar{\sigma}} S^{\bar{\nu}\bar{\sigma}}(0)]\big]e^{i k_2 \cdot x(0)}\Big\} \nonumber \\
	& \hspace{1.5cm} \epsilon_1 \cdot \bar{\Psi}(\tau_1) e^{i k_1 \cdot x(-\infty)}\rangle \label{eq:b1} \\
	&+ \int^{+\infty}_{-\infty} d\tau 
	\langle \epsilon_4 \cdot \Psi(+\infty) e^{i k_4 \cdot x(+\infty)} \nonumber \\
	&\hspace{2cm} \mathcal{T}\Big\{\big[-\frac{1}{2}\epsilon_{3\mu\nu} \dot{x}^{\mu}(\tau) [3 k_{3 \sigma} S^{\nu\sigma}(\tau)]\big]e^{i k_3 \cdot x(\tau)} \nonumber \\
	& \hspace{2.5cm} \big[-\frac{1}{2}\epsilon_{2\bar{\mu}\bar{\nu}} \dot{x}^{\bar{\mu}}(0) \dot{x}^{\bar{\nu}}(0) \big]e^{i k_2 \cdot x(\tau)}\Big\} \nonumber \\
	& \hspace{1.5cm} \epsilon_1 \cdot \bar{\Psi}(-\infty) e^{i k_1 \cdot x(-\infty)}\rangle \label{eq:b2} \\
	&+ \int^{+\infty}_{-\infty} d\tau 
	\langle \epsilon_4 \cdot \Psi(+\infty) e^{i k_4 \cdot x(+\infty)} \nonumber \\
	&\hspace{2cm} \mathcal{T}\Big\{\big[-\frac{1}{2}\epsilon_{3\mu\nu} \dot{x}^{\mu}(\tau) [i k_{3 \sigma} S^{\nu\sigma}(\tau)]\big]e^{i k_3 \cdot x(\tau)} \nonumber \\
	& \hspace{2.5cm} \big[-\frac{1}{2}\epsilon_{2\bar{\mu}\bar{\nu}} \dot{x}^{\bar{\mu}}(0) [i k_{2 \bar{\sigma}} S^{\bar{\nu}\bar{\sigma}}(0)]\big]e^{i k_2 \cdot x(0)}\Big\} \nonumber \\
	& \hspace{1.5cm} \epsilon_1 \cdot \bar{\Psi}(-\infty) e^{i k_1 \cdot x(-\infty)}\rangle  \label{eq:b3}\\
	&+\langle\epsilon_4 \cdot \Psi(+\infty) e^{i k_4 \cdot x(+\infty)} \nonumber \\
	&\hspace{0.5cm} \Big\{-\frac{1}{2} \dot{x}^{\mu}(\tau) \omega_{\mu a b}^{(2\epsilon)} S^{ab}(0)-\frac{1}{2} R_{ad}^{(2\epsilon)} \bar{\Psi}^{a}(0) \Psi^d(0) \Big\} e^{i(k_2+k_3)\cdot x(0)} \nonumber\\
	&\hspace{0.5cm} \epsilon_1 \cdot \bar{\Psi}(-\infty) e^{i k_1 \cdot x(-\infty)}\rangle \label{eq:b4}\,.
\end{align}
The additional "naive" correlation function that we need is the one for fermions, %which is also easily got from the Lagrangian
\[
\langle \mathcal{T}\{ \Psi^a(\tau) \bar{\Psi}^b(\tau') \} \rangle = \eta^{a b} \Theta(\tau-\tau')\,,
\]
where the vacuum is defined to be annihilated by $\Psi^a$.
We will also introduce a new convention to mark an anti-symmetric structure
\[
... \overbrace{ A B }^{a.s.} ... \equiv (... A B ...) - (... B A ...)
\]
which will help us keep terms tractable when there are more fermion contractions.\par
The expressions \eqref{eq:b1}, \eqref{eq:b2} and \eqref{eq:b4} are straightforward to calculate (remember that we had already defined  $\int^{+\infty}_{-\infty} d\tau\ sign(\tau)\delta(\tau)e^\Sigma = 0$ with dim-reg in Appendix \ref{appendix: scalar-gravity}). 
We have the following for the term \eqref{eq:b1}
\begin{align*}
	& \int^{+\infty}_{-\infty} d\tau 
	\langle \epsilon_4 \cdot \Psi(+\infty) e^{i k_4 \cdot x(+\infty)} \nonumber \\
	&\hspace{2cm} \mathcal{T}\Big\{\big[-\frac{1}{2}\epsilon_{3\mu\nu} \dot{x}^{\mu}(\tau) \dot{x}^{\nu}(\tau)\big]e^{i k_3 \cdot x(\tau)} \nonumber \\
	& \hspace{2.5cm} \big[-\frac{1}{2}\epsilon_{2\bar{\mu}\bar{\nu}} \dot{x}^{\bar{\mu}}(0) [i k_{2 \bar{\sigma}} S^{\bar{\nu}\bar{\sigma}}(0)]\big]e^{i k_2 \cdot x(0)}\Big\} \nonumber \\
	& \hspace{2cm} \epsilon_1 \cdot \bar{\Psi}(-\infty) e^{i k_1 \cdot x(-\infty)}\rangle\\
	=& \frac{1}{4} (\epsilon_4 \cdot \overbrace{ \epsilon_{2})( ik_2 }^{a.s.} \cdot \epsilon_1)
	\big[(\epsilon_3 \cdot \frac{i}{2} k_4)^2 [\epsilon_2 \cdot \frac{i}{2} (-k_1)] (-\frac{s}{2})\\
	&\hspace{2.75cm} +(\epsilon_2 \cdot \frac{i}{2} k_4) [\epsilon_3 \cdot \frac{i}{2} (-k_1)]^2 (-\frac{u}{2})\\
	&\hspace{2.75cm} +2 [\epsilon_3 \cdot \frac{i}{2} (k_4 - k_1)] (\epsilon_3 \cdot \epsilon_2)\big]\,, \\
\end{align*}
the term \eqref{eq:b2}
\begin{align*}
	& \int^{+\infty}_{-\infty} d\tau 
	\langle \epsilon_4 \cdot \Psi(+\infty) e^{i k_4 \cdot x(+\infty)} \nonumber \\
	&\hspace{2cm} \mathcal{T}\Big\{\big[-\frac{1}{2}\epsilon_{3\mu\nu} \dot{x}^{\mu}(\tau) [i k_{3 \sigma} S^{\nu\sigma}(\tau)]\big]e^{i k_3 \cdot x(\tau)} \nonumber \\
	& \hspace{2.5cm} \big[-\frac{1}{2}\epsilon_{2\bar{\mu}\bar{\nu}} \dot{x}^{\bar{\mu}}(0) \dot{x}^{\bar{\nu}}(0) \big]e^{i k_2 \cdot x(0)}\Big\} \nonumber \\
	& \hspace{1.5cm} \epsilon_1 \cdot \bar{\Psi}(-\infty) e^{i k_1 \cdot x(-\infty)}\rangle\\
	=& \frac{1}{4} (\epsilon_4 \cdot \overbrace{ \epsilon_{3})( ik_3 }^{a.s.} \cdot \epsilon_1)
	\big[(\epsilon_3 \cdot \frac{i}{2} k_4) [\epsilon_2 \cdot \frac{i}{2} (-k_1)]^2 (-\frac{s}{2})\\
	&\hspace{2.75cm} +(\epsilon_2 \cdot \frac{i}{2} k_4)^2 [\epsilon_3 \cdot \frac{i}{2} (-k_1)] (-\frac{u}{2})\\
	&\hspace{2.75cm} +2 [\epsilon_2 \cdot \frac{i}{2} (k_4 - k_1)] (\epsilon_3 \cdot \epsilon_2)\big]\,, \\
\end{align*}
and the term \eqref{eq:b4}
\begin{align*}
	&\langle\epsilon_4 \cdot \Psi(+\infty) e^{i k_4 \cdot x(+\infty)} \nonumber \\
	&\hspace{0.5cm} \Big\{-\frac{1}{2} \dot{x}^{\mu}(0) \omega_{\mu a b}^{(2\epsilon)} S^{ab}(0)-\frac{1}{2} R_{ad}^{(2\epsilon)} \bar{\Psi}^{a}(0) \Psi^d(0) \Big\} e^{i(k_2+k_3)\cdot x(0)} \nonumber\\
	&\hspace{0.5cm} \epsilon_1 \cdot \bar{\Psi}(-\infty) e^{i k_1 \cdot x(-\infty)}\rangle\\
	=& -\frac{1}{2} \times \frac{i}{2} (k_4-k_1)^{\mu} \omega_{\mu a b}^{(2\epsilon)} \overbrace{ \epsilon_{4}^{a} \epsilon_{1}^{b} }^{a.s.} - \frac{1}{2} R_{ad}^{(2\epsilon)} \epsilon_{4}^{a} \epsilon_{1}^{b}\,.
\end{align*}
The expansion of $\omega_{\mu a b}$ and $R_{a d}$ can be found in Appendix \ref{appendix: expansion}.\\
The term \eqref{eq:b3} has a new kind of integral over product of distributions, which comes from contracting a pair of $\dot{x}$ and all the fermions. To render it well-defined we use dim-reg once more. We give here a quick review of how dim-reg works for fermions. Start with the action in $D$ dimensions:
\[
\begin{aligned}
	S|_{DR}=\int_{\Omega} d^D\tau\ \bigg[&\frac{1}{2}g_{\mu\nu}(\partial_I x^{\mu} \partial_I x^{\nu}+b^{\mu}c^{\nu}+a^{\mu}a^{\nu}) + \bar{\Psi}_a \gamma^I \partial_I \Psi^a\\
	&+\frac{1}{2}\partial_I x^\mu \omega_{\mu a b} (\bar{\Psi}^a \gamma^I \Psi^b - \bar{\Psi}^b \gamma^I \Psi^a) -\frac{1}{8} R_{abcd}S^{ab}S^{cd}-\frac{1}{8} R\bigg]\,.
\end{aligned}
\]
The fermion correlation function is $\langle \Psi^a(\tau) \bar{\Psi}^b(\tau')\rangle \equiv \eta^{a b}\Big(\Delta_{F}(\tau-\tau')+\frac{1}{2}\Big)$ (see Appendix \ref{appendix: fermion} where
\[
\gamma^{I} \partial_I \Delta_{F}(\tau) = \delta^D(\tau)\,.
\]
$\Delta_{F}(\tau-\tau')$ will reduce to $-\frac{1}{2}sign(\tau-\tau')$ in $D=1$. As pointed out in \cite{Bastianelli:2006rx}, we won't need to use the $\gamma^I$ matrices algebra. The $\gamma^I$ matrices are used as a bookkeeping device to keep track of the $"I"$ indices. The one new integral we encounter is
\begin{align*}
	&\int_{\Omega} \Delta_{I J} \gamma^I \Delta_{F} \gamma^J e^\Sigma\\
	=&-\int_{\Omega} \Delta_{J} \gamma^I \partial_I \Delta_{F} \gamma^J e^\Sigma
	-\int_{\Omega} \Delta_{J} \gamma^I \Delta_{F} \gamma^J e^\Sigma \partial_I \Sigma\\
	\Rightarrow& 0 - (-\frac{1}{4})\times(-\frac{2}{s})\times \frac{s}{2} - (-\frac{1}{4})\times(-\frac{2}{u})\times(-\frac{u}{2})\\
	=&0\,.
\end{align*}
Now we are ready to calculate the term \eqref{eq:b3}:
\begin{align*}
	&\int^{+\infty}_{-\infty} d\tau 
	\langle \epsilon_4 \cdot \Psi(+\infty) e^{i k_4 \cdot x(+\infty)} \nonumber \\
	&\hspace{2cm} \mathcal{T}\Big\{\big[-\frac{1}{2}\epsilon_{3\mu\nu} \dot{x}^{\mu}(\tau) [i k_{3 \sigma} S^{\nu\sigma}(\tau)]\big]e^{i k_3 \cdot x(\tau)} \nonumber \\
	& \hspace{2.5cm} \big[-\frac{1}{2}\epsilon_{2\bar{\mu}\bar{\nu}} \dot{x}^{\bar{\mu}}(0) [i k_{2 \bar{\sigma}} S^{\bar{\nu}\bar{\sigma}}(0)]\big]e^{i k_2 \cdot x(0)}\Big\} \nonumber \\
	& \hspace{1.5cm} \epsilon_1 \cdot \bar{\Psi}(-\infty) e^{i k_1 \cdot x(-\infty)}\rangle\\
	=& \frac{1}{4} (\epsilon_3 \cdot \frac{i}{2} k_4) [\epsilon_2 \cdot \frac{i}{2}(-k_1)] (\epsilon_4 \cdot \overbrace{ \epsilon_3)(ik_3 }^{a.s.} \cdot \overbrace{ \epsilon_2)(ik_2 }^{a.s.} \cdot \epsilon_1)(-\frac{2}{s})\\
	&+\frac{1}{4} (\epsilon_2 \cdot \frac{i}{2} k_4) [\epsilon_3 \cdot \frac{i}{2}(-k_1)] (\epsilon_4 \cdot \overbrace{ \epsilon_2)(ik_2 }^{a.s.} \cdot \overbrace{ \epsilon_3)(ik_3 }^{a.s.} \cdot \epsilon_1)(-\frac{2}{u})\\
	&+\frac{1}{4} (\epsilon_3 \cdot \epsilon_2) \frac{1}{2} \big[(\epsilon_4 \cdot \overbrace{ \epsilon_3)(ik_3 }^{a.s.} \cdot \overbrace{ \epsilon_2)(ik_2 }^{a.s.} \cdot \epsilon_1) + (\epsilon_4 \cdot \overbrace{ \epsilon_2)(ik_2 }^{a.s.} \cdot \overbrace{ \epsilon_3)(ik_3 }^{a.s.} \cdot \epsilon_1) \big]\,.
\end{align*}
We have finished evaluating all the terms. We are ready to sum up $(\epsilon_4 \cdot \epsilon_1) A_{4,sg}$ and the terms \eqref{eq:b1}, \eqref{eq:b2}, \eqref{eq:b3} and \eqref{eq:b4} to evaluate the 4-point tree level scattering amplitude for two photons and two gravitons. This matches the field theory result.

\section{Graviton interacting with background gravity}
\label{appendix: graviton-gravity}
\noindent In previous calculations, in Appendices \ref{appendix: scalar-gravity} and \ref{appendix: photon-gravity}, we %actually discussed and 
worked out all possible integrals of products of distributions. Using previous results, we compute \par
\begin{align*}
	A^{\text{full}}_{4,gg}
	=& \int^{+\infty}_{-\infty} d\tau 
	\langle \epsilon_{4\alpha\beta} \tilde{\Psi}^\alpha(+\infty) \Psi^\beta(+\infty) e^{i k_4 \cdot x(+\infty)} \nonumber \\
	&\hspace{2cm} \mathcal{T}\Big\{\big[-\frac{1}{2}\epsilon_{3\mu\nu} [\dot{x}^{\mu}(\tau) + i k_{3 \rho} S^{\mu\rho}(\tau)] [\dot{x}^{\nu}(\tau) + i k_{3 \sigma} \tilde{S}^{\nu\sigma}(\tau)]\big]e^{i k_3 \cdot x(\tau)} \nonumber \\
	& \hspace{2.5cm} \big[-\frac{1}{2}\epsilon_{2\bar{\mu}\bar{\nu}} [\dot{x}^{\bar{\mu}}(0)+ i k_{2 \bar{\rho}} S^{\bar{\mu}\bar{\rho}}(0)] [\dot{x}^{\bar{\nu}}(0) + i k_{2 \bar{\sigma}} \tilde{S}^{\bar{\nu}\bar{\sigma}}(0)]\big]e^{i k_2 \cdot x(0)}\Big\} \nonumber \\
	& \hspace{1.5cm} \epsilon_{1\bar{\alpha} \bar{\beta}} \Psi^{\bar{\alpha}}(-\infty) \tilde{\Psi}^{\bar{\beta}}(-\infty)e^{i k_1 \cdot x(-\infty)} \rangle  \\
	&+\int^{+\infty}_{-\infty} d\tau 
	\langle \epsilon_{4\alpha\beta} \tilde{\Psi}^\alpha(+\infty) \Psi^\beta(+\infty) e^{i k_4 \cdot x(+\infty)}\\
	&\hspace{2cm} \mathcal{T}\{[-\frac{1}{2}\epsilon_{3\mu\nu}a^{\mu}(\tau) a^{\nu}(\tau)]e^{i k_3 \cdot x(\tau)} [-\frac{1}{2}\epsilon_{2\bar{\mu}\bar{\nu}}a^{\bar{\mu}}(0) a^{\bar{\nu}}(0)]e^{i k_2 \cdot x(0)}\} \\
	&\hspace{2cm} \epsilon_{1\bar{\alpha} \bar{\beta}} \Psi^{\bar{\alpha}}(-\infty) \tilde{\Psi}^{\bar{\beta}}(-\infty)e^{i k_1 \cdot x(-\infty)} \rangle  \\
	&+\int^{+\infty}_{-\infty} d\tau 
	\langle \epsilon_{4\alpha\beta} \tilde{\Psi}^\alpha(+\infty) \Psi^\beta(+\infty) e^{i k_4 \cdot x(+\infty)}\\
	&\hspace{2cm} \mathcal{T}\{[-\frac{1}{2}\epsilon_{3\mu\nu}b^{\mu}(\tau) c^{\nu}(\tau)]e^{i k_3 \cdot x(\tau)} [-\frac{1}{2}\epsilon_{2\bar{\mu}\bar{\nu}}b^{\bar{\mu}}(0) c^{\bar{\nu}}(0)]e^{i k_2 \cdot x(0)}\} \\
	&\hspace{2cm} \epsilon_{1\bar{\alpha} \bar{\beta}} \Psi^{\bar{\alpha}}(-\infty) \tilde{\Psi}^{\bar{\beta}}(-\infty)e^{i k_1 \cdot x(-\infty)} \rangle  \\
	&+\langle \epsilon_{4\alpha\beta} \tilde{\Psi}^\alpha(+\infty) \Psi^\beta(+\infty) e^{i k_4 \cdot x(+\infty)}\\
	&\hspace{0.5cm} \Big\{-\frac{1}{2} \dot{x}^{\mu}(0) \omega_{\mu a b}^{(2\epsilon)} S^{ab}(0)+\frac{1}{2} R_{ad}^{(2\epsilon)} \big[ \bar{\Psi}^{a}(0) \Psi^d(0) + \bar{\tilde{\Psi}}^{a}(0) \tilde{\Psi}^d(0) \big]\\
	&\hspace{0.5cm} +\frac{1}{4} R^{(2\epsilon)}_{a b c d} S^{a b}(0) \tilde{S}^{c d}(0) - \frac{3}{8} R^{(2\epsilon)} \Big\} e^{i(k_2+k_3)\cdot x(0)}\\
	&\hspace{0.5cm} \epsilon_{1\bar{\alpha} \bar{\beta}} \Psi^{\bar{\alpha}}(-\infty) \tilde{\Psi}^{\bar{\beta}}(-\infty)e^{i k_1 \cdot x(-\infty)} \rangle\\
	& + \langle V_2(+\infty) V^{(1)\mu\nu}(0) V_1(-\infty) \rangle\\ 
	&\hspace{0.5cm}(- \frac{2}{t}) (\frac{1}{2}\eta_{\mu\rho}\eta_{\nu\sigma} + \frac{1}{2}\eta_{\mu\sigma}\eta_{\nu\rho} - \frac{1}{2}\eta_{\mu\nu}\eta_{\rho\sigma}) \\
	&\hspace{0.5cm}\langle V_4(+\infty) V^{(1)\rho\sigma}(0) V_3(-\infty)\rangle\,.\\
\end{align*}
After simplifying and rearranging the terms we find
\begin{align}
	A^{\text{full}}_{4,gg}
	=& (\epsilon_4 \cdot \epsilon_1) A_{4, bg} \nonumber \\
	&+\frac{1}{4}\int^{+\infty}_{-\infty} d\tau 
	\langle \epsilon_{4\alpha\beta} \tilde{\Psi}^\alpha(+\infty) \Psi^\beta(+\infty) e^{i k_4 \cdot x(+\infty)} \nonumber \\
	&\hspace{2cm} \mathcal{T}\Big\{\big[\epsilon_{3\mu\nu} \dot{x}^{\mu}(\tau) [\dot{x}^{\nu}(\tau) + i k_{3 \sigma} \tilde{S}^{\nu\sigma}(\tau)]\big]e^{i k_3 \cdot x(\tau)} \nonumber \\
	& \hspace{2.5cm} \big[\epsilon_{2\bar{\mu}\bar{\nu}} i k_{2 \bar{\rho}} S^{\bar{\mu}\bar{\rho}}(0) [\dot{x}^{\bar{\nu}}(0) + i k_{2 \bar{\sigma}} \tilde{S}^{\bar{\nu}\bar{\sigma}}(0)]\big]e^{i k_2 \cdot x(0)}\Big\} \nonumber \\
	& \hspace{2cm} \epsilon_{1\bar{\alpha} \bar{\beta}} \Psi^{\bar{\alpha}}(-\infty) \tilde{\Psi}^{\bar{\beta}}(-\infty)e^{i k_1 \cdot x(-\infty)} \rangle  \label{eq:c1} \\
	&+\frac{1}{4}\int^{+\infty}_{-\infty} d\tau 
	\langle \epsilon_{4\alpha\beta} \tilde{\Psi}^\alpha(+\infty) \Psi^\beta(+\infty) e^{i k_4 \cdot x(+\infty)} \nonumber \\
	&\hspace{2cm} \mathcal{T}\Big\{\big[\epsilon_{3\mu\nu} i k_{3 \rho} S^{\mu\rho}(\tau) [\dot{x}^{\nu}(\tau) + i k_{3 \sigma} \tilde{S}^{\nu\sigma}(\tau)]\big]e^{i k_3 \cdot x(\tau)} \nonumber \\
	& \hspace{2.5cm} \big[\epsilon_{2\bar{\mu}\bar{\nu}} \dot{x}^{\bar{\nu}}(0) [\dot{x}^{\bar{\nu}}(0) + i k_{2 \bar{\sigma}} \tilde{S}^{\bar{\nu}\bar{\sigma}}(0)]\big]e^{i k_2 \cdot x(0)}\Big\} \nonumber \\
	& \hspace{2cm} \epsilon_{1\bar{\alpha} \bar{\beta}} \Psi^{\bar{\alpha}}(-\infty) \tilde{\Psi}^{\bar{\beta}}(-\infty)e^{i k_1 \cdot x(-\infty)} \rangle \label{eq:c2}\\
	&+\frac{1}{4}\int^{+\infty}_{-\infty} d\tau 
	\langle \epsilon_{4\alpha\beta} \tilde{\Psi}^\alpha(+\infty) \Psi^\beta(+\infty) e^{i k_4 \cdot x(+\infty)} \nonumber \\
	&\hspace{2cm} \mathcal{T}\Big\{\big[\epsilon_{3\mu\nu} i k_{3 \rho} S^{\mu\rho}(\tau) [\dot{x}^{\nu}(\tau) + i k_{3 \sigma} \tilde{S}^{\nu\sigma}(\tau)]\big]e^{i k_3 \cdot x(\tau)} \nonumber \\
	& \hspace{2.5cm} \big[\epsilon_{2\bar{\mu}\bar{\nu}}  i k_{2 \bar{\rho}} S^{\bar{\mu}\bar{\rho}}(0) [\dot{x}^{\bar{\nu}}(0) + i k_{2 \bar{\sigma}} \tilde{S}^{\bar{\nu}\bar{\sigma}}(0)]\big]e^{i k_2 \cdot x(0)}\Big\} \nonumber \\
	& \hspace{2cm} \epsilon_{1\bar{\alpha} \bar{\beta}} \Psi^{\bar{\alpha}}(-\infty) \tilde{\Psi}^{\bar{\beta}}(-\infty)e^{i k_1 \cdot x(-\infty)} \rangle \label{eq:c3}\\
	&+\langle \epsilon_{4\alpha\beta} \tilde{\Psi}^\alpha(+\infty) \Psi^\beta(+\infty) e^{i k_4 \cdot x(+\infty)} \nonumber \\
	&\hspace{0.5cm} \Big\{-\frac{1}{2} \dot{x}^{\mu}(0) \omega_{\mu a b}^{(2\epsilon)} S^{ab}(0)+\frac{1}{2} R_{ad}^{(2\epsilon)} \big[ \bar{\Psi}^{a}(0) \Psi^d(0) + \bar{\tilde{\Psi}}^{a}(0) \tilde{\Psi}^d(0) \big] \nonumber \\
	&\hspace{0.5cm} +\frac{1}{4} R^{(2\epsilon)}_{a b c d} S^{a b}(0) \tilde{S}^{c d}(0) - \frac{1}{2} R^{(2\epsilon)} \Big\} e^{i(k_2+k_3)\cdot x(0)} \nonumber \\
	&\hspace{0.5cm} \epsilon_{1\bar{\alpha} \bar{\beta}} \Psi^{\bar{\alpha}}(-\infty) \tilde{\Psi}^{\bar{\beta}}(-\infty)e^{i k_1 \cdot x(-\infty)} \rangle \label{eq:c4} \\
	& + \langle V_2(+\infty) V^{(1)\mu\nu}(0) V_1(-\infty)\rangle \nonumber \\ 
	&\hspace{0.5cm}(- \frac{2}{t}) (\frac{1}{2}\eta_{\mu\rho}\eta_{\nu\sigma} + \frac{1}{2}\eta_{\mu\sigma}\eta_{\nu\rho} - \frac{1}{2}\eta_{\mu\nu}\eta_{\rho\sigma}) \nonumber\\
	&\hspace{0.5cm}\langle V_4(+\infty) V^{(1)\rho\sigma}(0) V_3(-\infty)\rangle \,.\label{eq:c5}
\end{align}
Again, let us compute each term. First, the term \eqref{eq:c1}
\begin{align*}
	&\frac{1}{4}\int^{+\infty}_{-\infty} d\tau 
	\langle \epsilon_{4\alpha\beta} \tilde{\Psi}^\alpha(+\infty) \Psi^\beta(+\infty) e^{i k_4 \cdot x(+\infty)} \nonumber \\
	&\hspace{2cm} \mathcal{T}\Big\{\big[\epsilon_{3\mu\nu} \dot{x}^{\mu}(\tau) [\dot{x}^{\nu}(\tau) + i k_{3 \sigma} \tilde{S}^{\nu\sigma}(\tau)]\big]e^{i k_3 \cdot x(\tau)} \nonumber \\
	& \hspace{2.5cm} \big[\epsilon_{2\bar{\mu}\bar{\nu}} i k_{2 \bar{\rho}} S^{\bar{\mu}\bar{\rho}}(0) [\dot{x}^{\bar{\nu}}(0) + i k_{2 \bar{\sigma}} \tilde{S}^{\bar{\nu}\bar{\sigma}}(0)]\big]e^{i k_2 \cdot x(0)}\Big\} \nonumber \\
	& \hspace{2cm} \epsilon_{1\bar{\alpha} \bar{\beta}} \Psi^{\bar{\alpha}}(-\infty) \tilde{\Psi}^{\bar{\beta}}(-\infty)e^{i k_1 \cdot x(-\infty)} \rangle\\
	& +\frac{1}{4} (\epsilon_4 \cdot \overbrace{ \epsilon_2)(ik_2 }^{a.s.} \cdot \epsilon_1) \\
	&\hspace{0.5cm} \Big\{ 2 (\epsilon_3 \cdot \epsilon_2) [\epsilon_3 \cdot \frac{i}{2}(k_4 - k_1)]\\
	&\hspace{0.75cm} + (\epsilon_3 \cdot ik_4)^2 (\epsilon_4 \cdot \overbrace{ \epsilon_2)(ik_2 }^{a.s.} \cdot  \epsilon_1) (-\frac{2}{s})\\
	&\hspace{0.75cm} + [\epsilon_3 \cdot (-ik_1)]^2 (\epsilon_4 \cdot \overbrace{ \epsilon_2)(ik_2 }^{a.s.} \cdot  \epsilon_1) (-\frac{2}{u})\\
	&\hspace{0.75cm} + (\epsilon_3 \cdot \epsilon_2) (\epsilon_4 \cdot \overbrace{ \epsilon_3)(ik_3 }^{a.s.} \cdot  \epsilon_1)\\
	&\hspace{0.75cm} + (\epsilon_3 \cdot ik_4) [\epsilon_2 \cdot (-ik_1)] (\epsilon_4 \cdot \overbrace{ \epsilon_3)(ik_3 }^{a.s.} \cdot  \epsilon_1) (-\frac{2}{s})\\
	&\hspace{0.75cm} + (\epsilon_2 \cdot ik_4) [\epsilon_3 \cdot (-ik_1)] (\epsilon_4 \cdot \overbrace{ \epsilon_3)(ik_3 }^{a.s.} \cdot  \epsilon_1) (-\frac{2}{u})\\
	&\hspace{0.75cm} + (\epsilon_3 \cdot ik_4) (\epsilon_4 \cdot \overbrace{ \epsilon_3)(ik_3 }^{a.s.} \cdot  \overbrace{ \epsilon_2)(ik_2 }^{a.s.} \cdot \epsilon_1) (-\frac{2}{s})\\
	&\hspace{0.75cm} + [\epsilon_3 \cdot (-ik_1)] (\epsilon_4 \cdot \overbrace{ \epsilon_2)(ik_2 }^{a.s.} \cdot  \overbrace{ \epsilon_3)(ik_3 }^{a.s.} \cdot \epsilon_1) (-\frac{2}{u}) \Big\}\,.
\end{align*}
The term \eqref{eq:c2} can be quickly obtained by interchanging the particle index 2 and 3 in term \eqref{eq:c1}.\\
The term \eqref{eq:c3} will require the integral over products of distributions which we calculated when considering the term \eqref{eq:b3} in Appendix \ref{appendix: photon-gravity}
\begin{align*}
	&\frac{1}{4}\int^{+\infty}_{-\infty} d\tau 
	\langle \epsilon_{4\alpha\beta} \tilde{\Psi}^\alpha(+\infty) \Psi^\beta(+\infty) e^{i k_4 \cdot x(+\infty)} \nonumber \\
	&\hspace{2cm} \mathcal{T}\Big\{\big[\epsilon_{3\mu\nu} i k_{3 \rho} S^{\mu\rho}(\tau) [\dot{x}^{\nu}(\tau) + i k_{3 \sigma} \tilde{S}^{\nu\sigma}(\tau)]\big]e^{i k_3 \cdot x(\tau)} \nonumber \\
	& \hspace{2.5cm} \big[\epsilon_{2\bar{\mu}\bar{\nu}}  i k_{2 \bar{\rho}} S^{\bar{\mu}\bar{\rho}}(0) [\dot{x}^{\bar{\nu}}(0) + i k_{2 \bar{\sigma}} \tilde{S}^{\bar{\nu}\bar{\sigma}}(0)]\big]e^{i k_2 \cdot x(0)}\Big\} \nonumber \\
	& \hspace{2cm} \epsilon_{1\bar{\alpha} \bar{\beta}} \Psi^{\bar{\alpha}}(-\infty) \tilde{\Psi}^{\bar{\beta}}(-\infty)e^{i k_1 \cdot x(-\infty)} \rangle \\
	=& \frac{1}{4} \times \frac{1}{2} (\epsilon_3 \cdot \epsilon_2) (\epsilon_4 \cdot \epsilon_1) \big[(\epsilon_4 \cdot \overbrace{ \epsilon_3)(ik_3 }^{a.s.} \cdot  \overbrace{ \epsilon_2)(ik_2 }^{a.s.} \cdot \epsilon_1) + (\epsilon_4 \cdot \overbrace{ \epsilon_2)(ik_2 }^{a.s.} \cdot  \overbrace{ \epsilon_3)(ik_3 }^{a.s.} \cdot \epsilon_1) \big]\\
	&+\frac{1}{4} (\epsilon_3 \cdot ik_4) (\epsilon_4 \cdot \overbrace{ \epsilon_3)(ik_3 }^{a.s.} \cdot  \overbrace{ \epsilon_2)(ik_2 }^{a.s.} \cdot \epsilon_1) (\epsilon_4 \cdot \overbrace{ \epsilon_2)(ik_2 }^{a.s.} \cdot \epsilon_1) (-\frac{2}{s})\\
	&+\frac{1}{4} [\epsilon_3 \cdot (-ik_1)] (\epsilon_4 \cdot \overbrace{ \epsilon_2)(ik_2 }^{a.s.} \cdot  \overbrace{ \epsilon_3)(ik_3 }^{a.s.} \cdot \epsilon_1) (\epsilon_4 \cdot \overbrace{ \epsilon_2)(ik_2 }^{a.s.} \cdot \epsilon_1) (-\frac{2}{u})\\	
	&+\frac{1}{4} (\epsilon_2 \cdot (-ik_1)) (\epsilon_4 \cdot \overbrace{ \epsilon_3)(ik_3 }^{a.s.} \cdot  \overbrace{ \epsilon_2)(ik_2 }^{a.s.} \cdot \epsilon_1) (\epsilon_4 \cdot \overbrace{ \epsilon_3)(ik_3 }^{a.s.} \cdot \epsilon_1) (-\frac{2}{s})\\
	&+\frac{1}{4} (\epsilon_2 \cdot ik_4) (\epsilon_4 \cdot \overbrace{ \epsilon_2)(ik_2 }^{a.s.} \cdot  \overbrace{ \epsilon_3)(ik_3 }^{a.s.} \cdot \epsilon_1) (\epsilon_4 \cdot \overbrace{ \epsilon_3)(ik_3 }^{a.s.} \cdot \epsilon_1) (-\frac{2}{u})\\
	&+\frac{1}{4} [(\epsilon_4 \cdot \overbrace{ \epsilon_3)(ik_3 }^{a.s.} \cdot  \overbrace{ \epsilon_2)(ik_2 }^{a.s.} \cdot \epsilon_1)]^2 (-\frac{2}{s})
	+\frac{1}{4} [(\epsilon_4 \cdot \overbrace{ \epsilon_2)(ik_2 }^{a.s.} \cdot  \overbrace{ \epsilon_3)(ik_3 }^{a.s.} \cdot \epsilon_1)]^2 (-\frac{2}{u})\,.
\end{align*}
Term \eqref{eq:c4}
\begin{align*}
	&\langle \epsilon_{4\alpha\beta} \tilde{\Psi}^\alpha(+\infty) \Psi^\beta(+\infty) e^{i k_4 \cdot x(+\infty)} \nonumber \\
	&\hspace{0.5cm} \Big\{-\frac{1}{2} \dot{x}^{\mu}(0) \omega_{\mu a b}^{(2\epsilon)} S^{ab}(0)+\frac{1}{2} R_{ad}^{(2\epsilon)} \big[ \bar{\Psi}^{a}(0) \Psi^d(0) + \bar{\tilde{\Psi}}^{a}(0) \tilde{\Psi}^d(0) \big] \nonumber \\
	&\hspace{0.5cm} +\frac{1}{4} R^{(2\epsilon)}_{a b c d} S^{a b}(0) \tilde{S}^{c d}(0) - \frac{1}{2} R^{(2\epsilon)} \Big\} e^{i(k_2+k_3)\cdot x(0)} \nonumber \\
	&\hspace{0.5cm} \epsilon_{1\bar{\alpha} \bar{\beta}} \Psi^{\bar{\alpha}}(-\infty) \tilde{\Psi}^{\bar{\beta}}(-\infty)e^{i k_1 \cdot x(-\infty)} \rangle\\
	=& -\frac{1}{2} (\epsilon_4 \cdot \epsilon_1) \frac{i}{2}(k_4 - k_1)^\mu \omega_{\mu a b}^{(2\epsilon)} \overbrace{ \epsilon_{4}^{a} \epsilon_{1}^{b} }^{a.s.}
	+ (\epsilon_4 \cdot \epsilon_1) R^{(2\epsilon)}_{a d} \overbrace{ \epsilon_{4}^{a} \epsilon_{1}^{d} }^{a.s.}\\
	&+\frac{1}{4} R^{(2\epsilon)}_{a b c d} \overbrace{ \epsilon_{4}^{a} \epsilon_{1}^{b} }^{a.s.} \overbrace{ \epsilon_{4}^{c} \epsilon_{1}^{d} }^{a.s.} - \frac{1}{2} (\epsilon_4 \cdot \epsilon_1)^2 R^{(2\epsilon)}\,.
\end{align*}
The second order expansions $(2\epsilon)$ of $\omega_{\mu a b}$, $R$, $R_{a d}$ and $R_{a b c d}$ can be found in Appendix \ref{appendix: expansion}.\\
The term \eqref{eq:c5} can be obtained by taking the result of the graviton three point vertex (which is already calculated) and relabeling the particle index.\par
Summing up $(\epsilon_4 \cdot \epsilon_1) A_{4, bg}$ and the terms \eqref{eq:c1}, \eqref{eq:c2}, \eqref{eq:c3}, \eqref{eq:c4} and \eqref{eq:c5}, we find that it matches the field theory result of the four graviton scattering amplitude. 
\pagebreak

\section{Dimension regularization for the Dirac fermion correlation function}
\label{appendix: fermion}

Consider the action of one pair of free Dirac fermions in $D$-dimensions 
\[
S = \int d^D\tau\ \bar{\Psi} \slash\!\!\!{\partial} \Psi
\]
The Feynman propagator is
\[
\langle \Omega | \mathcal{T}\{\Psi(\tau) \bar{\Psi}(\tau') \} | \Omega \rangle =  \int \frac{d^D p}{(2\pi)^D} \frac{i \slash\!\!\!p}{p^2 + i\epsilon} e^{i p \cdot (\tau - \tau')}
\]
When reduced to $D = 1$, it becomes
\begin{align*}
	\langle \Omega | \mathcal{T}\{\Psi(\tau) \bar{\Psi}(\tau') \} | \Omega \rangle &=  \int \frac{d p}{(2\pi)} \frac{i p}{p^2 + i\epsilon} e^{i p (\tau - \tau')}\\
	& \rightarrow -\frac{1}{2}  sign(\tau - \tau')
\end{align*}
which can be easily recognized to be the derivative of the bosonic correlation function.
However,  the Feynman propagator does not reduce to the fermion correlation function that we used in our earlier calculations
\[
\langle \mathcal{T}\{\Psi^a(\tau) \bar{\Psi}^b(\tau') \} \rangle = \eta^{a b} \Theta(\tau-\tau')
\,,\]
which  differs from the Feynman propagator by a constant shift of $\frac{1}{2}$. \par
The reason for the difference is that we have chosen a different vacuum. The vacuum that we chose for the worldline formalism is the one annihilated by $\Psi^a$: $\Psi^a | \tilde{\Omega} \rangle = 0 $. This vacuum $| \tilde{\Omega} \rangle$ is not the one we used for Feynman propagator $| \Omega \rangle$.\par
We will explicitly work out the $D = 1$ case to show the difference. %For simplicity, we will only consider one set of Dirac fermion. 
The most general solution to the equation of motion will give us
\begin{align*}
	\Psi(\tau) &= \Psi = \frac{1}{\sqrt{2}}(\psi_1 + i \psi_2)\\
	\bar{\Psi}(\tau) &= \bar{\Psi} = \frac{1}{\sqrt{2}}(\psi_1 - i \psi_2)
\end{align*}
in which $\psi_1$ and $\psi_2$ are real. In quantizing, we promote $\Psi$ and $\bar{\Psi}$ to operators and impose canonical anti-commutation relations
\begin{align*}
	&\{\Psi, \bar{\Psi}\} = 1\\
	&\{\Psi, \Psi\} = \{\bar{\Psi}, \bar{\Psi}\} = 0\,.
\end{align*}
By solving $\psi_1$ and $\psi_2$ in terms of $\Psi$ and $\bar{\Psi}$ we can determine the commutation relation for $\psi_1$ and $\psi_2$
\[
\{ \psi_i, \psi_j \} = \delta_{i j}\,.
\]
One possible set of solutions is
\begin{align*}
	&\psi_1 = \frac{\sigma_1}{\sqrt{2}} = \frac{1}{\sqrt{2}}
	\begin{pmatrix}
		0 & 1 \\
		1 & 0
	\end{pmatrix},
	\psi_2 = \frac{\sigma_2}{\sqrt{2}} = \frac{1}{\sqrt{2}}
	\begin{pmatrix}
		0 & -i \\
		i & 0
	\end{pmatrix}\,,\\
	&\Psi =
	\begin{pmatrix}
		0 & 1 \\
		0 & 0
	\end{pmatrix},
	\bar{\Psi} =
	\begin{pmatrix}
		0 & 0 \\
		1 & 0
	\end{pmatrix}\,.
\end{align*}
A general state is of the form
\[
| f(c) \rangle = \frac{1}{\sqrt{1 + |c|^2}} \begin{pmatrix} 1 \\ 0 \end{pmatrix} + \frac{c}{\sqrt{1 + |c|^2}} \begin{pmatrix} 0 \\ 1 \end{pmatrix}\,.
\]
It is easy to see that $| \tilde{\Omega} \rangle = | f(0) \rangle$ since $\Psi | f(c) \rangle = 0$ holds only when c = 0.
In general, we have
\begin{align*}
	\langle f(c) | \mathcal{T}\{ \Psi(\tau) \bar{\Psi}(\tau') \} | f(c) \rangle &= \frac{1}{1+|c|^2} \Theta(\tau-\tau') - \frac{|c|^2}{1+|c|^2} \Theta(\tau'-\tau)\\
	&= \Theta(\tau-\tau') - \frac{|c|^2}{1 + |c|^2}\\
	&= \langle \tilde{\Omega} | \mathcal{T}\{ \Psi(\tau) \bar{\Psi}(\tau') \} | \tilde{\Omega} \rangle - \frac{|c|^2}{1 + |c|^2}\,,
\end{align*}
which tells us that the correlation function with different boundary conditions will relate to each other by a constant shift. In particular, $| \Omega \rangle = | f(1) \rangle$ and
\[
\langle \Omega | \mathcal{T}\{ \Psi(\tau) \bar{\Psi}(\tau') \} | \Omega \rangle = \langle \tilde{\Omega} | \mathcal{T}\{ \Psi(\tau) \bar{\Psi}(\tau') \} | \tilde{\Omega} \rangle - \frac{1}{2}\,.
\]\par
Let's call vacuum $| \tilde{\Omega} \rangle$ the "worldline vacuum" and $| \Omega \rangle $ the "ordinary vacuum". This relation of constant shift helps us quickly find the way to apply dim-reg to fermions in worldline approach: split worldline correlation function into ordinary vacuum part and constant part and all we need to do is to apply dim-reg to ordinary vacuum part, which is easy because it will be just ordinary Feynman propagator in $D$ dimension. The constant part will keep untouched since it has nothing to do with ambiguity.\par
In practice, we promote the correlation function as $\langle \Psi^a(\tau) \bar{\Psi}^b(\tau')\rangle \rightarrow \eta^{a b}\Big(\Delta_{F}(\tau-\tau')+\frac{1}{2}\Big)$, where $\Delta_F(\tau-\tau')$ is Feynman propagator in $D$ dimensions and satisfies
\[
\gamma^{I} \partial_I \Delta_{F}(\tau) = \delta^D(\tau)\,.
\]\par
%It may look a bit strange at first because $\Delta_{F}(\tau-\tau')+\frac{1}{2}$ may not have a definite meaning in $D$ %dimension, but it will give the correct result when reduced to $D = 1$, \textcolor{red}{because ordinary vacuum part %has absorbed all ambiguity that worldline correlation function has.}
%\pagebreak

\section{Expansion in background fields of the Riemann curvature}
\label{appendix: expansion}
We are using the following definitions (where the curved indices are raised/lowered by $g^{\mu \nu}$/$g_{\mu \nu}$ while the flat indices are raised/lowered by $\eta^{\mu \nu}$/$\eta_{\mu \nu}$ ):
\begin{align*}
	\omega_{\mu a b} \equiv& \frac{1}{2}e_a^\nu (\partial_\mu e_{\nu b} - \partial_\nu e_{\mu b}) 
	- \frac{1}{2}e_b^\nu (\partial_\mu e_{\nu a} - \partial_\nu e_{\mu a}) \\
	& -\frac{1}{2} e_a^\nu e_b^\sigma (\partial_\nu e_{\sigma c} - \partial_\sigma e_{\nu c}) e_\mu^c\\
	\Gamma^\mu_{\alpha \beta} \equiv& \frac{1}{2} g^{\mu \lambda} (\partial_\alpha g_{\beta \lambda} + \partial_\beta g_{\lambda \alpha} - \partial_\lambda g_{\alpha \beta})\\
	R^\mu_{\nu \rho \sigma} \equiv& \partial_\rho \Gamma^\mu_{\nu \sigma} - \partial_\sigma \Gamma^\mu_{\nu \rho} + \Gamma^\mu_{\lambda \rho} \Gamma^\lambda_{\nu \sigma} - \Gamma^\mu_{\lambda \sigma} \Gamma^\lambda_{\nu \rho}\\
	R_{\mu \nu} \equiv& R^\rho_{\mu \rho \nu}\\
	R \equiv& g^{\mu \nu} R_{\mu \nu}\,.
\end{align*}
Up to the second order in $h$, we have the following expansions (where all the indices on the right hand side of the equal sign are now raised or lowered by the flat Minkowski metric):
\begin{align*}
	g_{\mu \nu} =& \eta_{\mu\nu} + h_{\mu\nu}\\
	g^{\mu \nu} =& \eta^{\mu\nu} - h^{\mu\nu} + h^{\mu}_{\lambda} h^{\lambda \nu} + o(h^2)\\
	e_\mu^a =& \delta_\mu^a + \frac{1}{2} h_\mu^a - \frac{1}{8} h_{\mu \lambda} h^{\lambda a} + o(h^2)\\
	e^\mu_a =& \delta^\mu_a - \frac{1}{2} h^\mu_a + \frac{3}{8} h_{a \lambda} h^{\lambda \mu} + o(h^2)\\
	\omega_{\mu a b} =& \frac{1}{2}(\partial_b h_{a \mu} - \partial_a h_{b \mu})\\
	& +\frac{1}{8} (h^\nu_b \partial_\mu h_{a \nu} - h^\nu_a \partial_\mu h_{b \nu})\\
	& -\frac{1}{4} (h^\nu_b \partial_\nu h_{a \mu} - h^\nu_a \partial_\nu h_{b \mu})\\
	& +\frac{1}{4} (h^\nu_b \partial_a h_{\mu \nu} - h^\nu_a \partial_b h_{\mu \nu}) + o(h^2)\,.
\end{align*}
The following expansion, in which $h_{\mu\nu}$ could be off-shell, is  used in the 3-point vertex
calculation
\begin{align*}
	R^{(1)}_{a b c d} =& -\frac{1}{2}(\partial_a\partial_c h_{b d} + \partial_b\partial_d h_{a c} -\partial_a\partial_d h_{b c} -\partial_b\partial_c h_{a d})\\
	R^{(1)}_{a d} =& \frac{1}{2}(\partial_a \partial^b h_{d b} + \partial_d \partial^b h_{a b} - \partial_a \partial_d h - \eta^{b c} \partial_b \partial_c h_{a d})\\
	R^{(1)} =& (\partial^a \partial^b h_{a b} - \eta^{a b} \partial_a \partial_b h)\,.
\end{align*}
The following expansion, in which $h_{\mu \nu}$ are all on-shell, is  used in the 4-point amplitude calculation
\begin{align*}
	R^{(2)}_{a b c d} =& -\frac{1}{2} \Big\{ \partial_c[h_a^\lambda (\partial_b h_{d \lambda} + \partial_d h_{\lambda b} - \partial_\lambda h_{b d})] - (c \leftrightarrow d)\Big\}\\
	& +\frac{1}{4} \Big\{ (\partial_c h_{\lambda a} + \partial_\lambda h_{a c} - \partial_a h_{c \lambda})(\partial_b h_d^\lambda + \partial_d h^\lambda_b - \partial^\lambda h_{b d}) - (c \leftrightarrow d)\Big\}\\
	& +\frac{1}{4} h_{\mu a} \Big\{ \partial_c (\partial_b h_d^\mu + \partial_d h^\mu_b - \partial^\mu h_{b d}) - (c \leftrightarrow d)\Big\}\\
	&- \frac{1}{4} h^\nu_b \Big\{ \partial_c (\partial_\nu h_{d a} + \partial_d h_{a \nu} - \partial_a h_{\nu d}) - (c \leftrightarrow d)\Big\}\\
	&- \frac{1}{4} h^\rho_c \Big\{ \partial_\rho (\partial_b h_{d a} + \partial_d h_{a b} - \partial_a h_{b d}) - (\rho \leftrightarrow d)\Big\}\\
	&- \frac{1}{4} h^\sigma_d \Big\{ \partial_c (\partial_b h_{\sigma a} + \partial_\sigma h_{a b} - \partial_a h_{b \sigma}) - (c \leftrightarrow \sigma)\Big\}\\
	R^{(2)}_{a d} =& -\frac{1}{2} h^{\alpha \lambda} \partial_\alpha (\partial_a h_{d \lambda} + \partial_d h_{\lambda a} - \partial_\lambda h_{a d})\\
	& +\frac{1}{2} h^{\alpha \lambda} \partial_d (\partial_a h_{\alpha \lambda} + \partial_\alpha h_{\lambda a} - \partial_\lambda h_{a \alpha})\\
	& -\frac{1}{4} (\partial_\lambda h_d^\alpha - \partial_d h^\alpha_\lambda - \partial^\alpha h_{\lambda d}) (\partial_a h_\alpha^\lambda + \partial_\alpha h^\lambda_a - \partial^\lambda h_{a \alpha})\\
	R^{(2)} =& \frac{3}{4} (\partial_\mu h_{\alpha \beta})^2 - \frac{1}{2} \partial_\alpha h_{\beta \mu} \partial^\beta h^{\alpha \mu}\,.
\end{align*}

\section{Lower tree}
\label{appendix: lower tree}
\noindent First, consider the gauge boson case, 
\[
J^\mu(p) = \frac{2}{p^2}\Big( \mathcal{M}^\mu + p^\mu f(\{\epsilon\},\{k\}) \Big),
\]
{we want to prove that the second term does not give any contribution with our choice of reference spinors.
Recall that due to our choice of reference spinors, there can be only one $\epsilon_i \cdot \epsilon_j$ where either $i$ or $j$ must be $n$. Thus let's discuss based on the position of particle $n$ in the diagram.} {For any chosen worldline, if particle $n$ is not in the lower tree}, then as we know, $f(\{\epsilon\}, \{k \})$ has at least one $(\epsilon_i \cdot \epsilon_j)$, which then vanishes due to our choice of reference spinors.
However, if particle $n$ is in the lower tree, consider the corresponding term in the amplitude,
\[
(\prod_i^{N-3} \int d\tau_i) \langle \dots [\frac{2}{p^2} f(\{\epsilon\}, \{k \}) p_\mu (\dot{x}^\mu + i p_\rho S^{\mu\rho}) e^{i p \cdot x}]\dots\rangle
\]
There are two ways to see that the $\dot{x}_\mu$ term would not give any contribution. The first one is to notice that the $\dot{x}_\mu$ term would lead to a $(p \cdot k)$ term, of which the consequence would be two $(\epsilon_i \cdot \epsilon_j)$ in final expression vanishing. The other way is to count the $\epsilon$ and $k$ in the part without the lower tree. Equivalently, we will find that there must be a $(\epsilon_i \cdot \epsilon_j)$ if we pick the $\dot{x}_\mu$ term, which would vanish since particle $n$ is already in the lower tree. \\
Thus, all we need to consider is just
\[
\langle \dots  [\frac{2}{p^2} f(\{\epsilon\}, \{k \}) p_\mu( i p_\rho S^{\mu\rho}) e^{i p \cdot x}]\dots\rangle
\]
Since $p_\mu p_\rho$ is symmetric while $S^{\mu\rho}$ is antisymmetric in $\mu, \rho$, this expression vanishes.\\
Thus, we have proved that only the on-shell part give contributions to MHV amplitude,
\[
J^\mu_{\text{eff}}(p) = \frac{2}{p^2} \mathcal{M}^\mu
\]
The graviton case is very similar,
\[
J^{\mu\nu}(p) = \frac{2}{p^2}\Big( \mathcal{M}^{\mu\nu} + p^\mu f^\nu(\{\epsilon\},\{k\}) + p^\nu f^\mu(\{\epsilon\},\{k\}) + \eta^{\mu\nu} h(\{\epsilon\},\{k\})\Big)
\]
The quickest way to see the $p^\mu f^\nu(\{\epsilon\},\{k\})$(or $p^\nu f^\mu(\{\epsilon\},\{k\})$) term vanishing is to separately consider the part with $\mu$ index and $\nu$ index in the corresponding term,
\[
	(\prod_{i=3}^{M-1} \int d\tau_i) \langle\dots
    [\frac{2}{p^2} f_\nu(\{\epsilon\}, \{k \}) p_\mu (\dot{x}^\mu + i k_\rho S^{\mu\rho})(\dot{x}^\nu + i k_\sigma \tilde{S}^{\nu\sigma}) e^{i p \cdot x}]
    \dots\rangle
\]
Then we could apply exactly the same argument for the gauge boson case here for the $\mu$ part.
The last term also vanishes when we analyze the structure of $(\epsilon_i \cdot \epsilon_j)$. We know that there is at least one $(\epsilon_i \cdot \epsilon_j)$ term in $h(\{\epsilon\}, \{k \})$. After counting the $\epsilon$ and $k$ in the part without the lower tree, we will know that there is also at least one $(\epsilon_k \cdot \epsilon_l)$. Since the four $\epsilon_i$ coming from four different gravitons while non-vanishing $(\epsilon_i \cdot \epsilon_j)$ must always have the $\epsilon_n$ of particle $n$, this term vanishes.
Thus, we have also proved that for graviton,
\[
J^{\mu\nu}_{\text{eff}}(p) = \frac{2}{p^2} \mathcal{M}^{\mu\nu}
\]
Since the on-shell part is produced with the worldline formalism, that means we could choose a new worldline that connects the attaching point and an arbitrary particle to calculate the lower tree.

\section{BCJ relation for MHV worldline numerator}
\label{appendix: BCJ}
\noindent To quickly prove that the BCJ relation holds for MHV amplitude, we will use three properties for MHV amplitude in the worldline formalism.\par
The first one is the freedom to choose the worldline, which we already knew in our previous discussion. As we shall see, it simplifies the proof a lot since once we have a proof for a specific choice of worldline configuration, then it holds for all different configurations.\par
The second one is about the cut of a worldline. Since for MHV amplitude, there is no 4-point vertex, which means no $\langle \dot{x} \dot{x} \rangle$ contraction. Then for any diagram, we could freely do a cut at an arbitrary propagator on a single worldline so that it can be seen as two different parts joining together, thus we have the following relation
\[
\begin{aligned}
	& (\prod_{a=3}^{N-1} \int d\tau_a) \langle \dots [J_{\mu}(\dot{x}^{\mu} + i k_{\rho} S^{\mu\rho}) e^{i k_ \cdot x}]_i [J'_{\nu}(\dot{x}^{\nu} + i k_{\sigma} S^{\nu\sigma}) e^{i k \cdot x}]_{i-1} \dots\rangle \\
	=\frac{2}{p^2}&  (\prod_{a=3}^{N_L-1} \int d\tau_{a})\langle \dots [J_{\mu}(\dot{x}^{\mu} + i k_{\rho} S^{\mu\rho}) e^{i k \cdot x}]_2 [\bar{\Psi}_\lambda(-\infty) e^{i p \cdot x(-\infty)}]\rangle \\
    &(\prod_{b=1}^{N_R-1} \int d\tau_{b})\langle [\Psi^\lambda(+\infty) e^{-i p \cdot x(+\infty)}] [J'_{\nu}(\dot{x}^{\nu} + i k_{\sigma} S^{\mu\sigma}) e^{i k \cdot x}]_{N_R-1} \dots\rangle 
\end{aligned}
\]
for the gauge boson case. This can be straightforwardly extended to the graviton case.\par
{The third one is about the lower tree. The claim we are making here is that if particle $n$ is not involved in the lower tree, then the lower tree $J_\mu(p)$ satisfies $p \cdot J(p) = 0$, which means the $J_\mu$ can be effectively seen as a polarization vector that satisfies transversality.
Any specific lower tree $J_\mu$ can be written as
\[
\begin{aligned}
J_\mu &=\frac{2}{p^2} \langle [\mathcal{E} \cdot \Psi e^{i k \cdot x}]_3 [J'_\nu (\dot{x}^{\nu} + i q_{\rho} S^{\nu\rho})e^{i q \cdot x}]_2 [\bar{\Psi}_\mu e^{i p \cdot x}]_1\rangle \\
\mathcal{E}_\mu &= \frac{2}{k^2} \langle [\mathcal{E}' \cdot \Psi e^{i k' \cdot x}]_3 [J''_\nu (\dot{x}^{\nu} + i q'_{\rho} S^{\nu\rho})e^{i q' \cdot x}]_2 [\bar{\Psi}_\mu e^{i k \cdot x}]_1\rangle,
\end{aligned}
\]
where we have used the second property and all the lower trees $J'_\mu$, $J''_\mu$, $\mathcal{E}_\mu$ and $\mathcal{E}'_\mu$ will be calculated repetitively with the same formulas until they represent polarization vectors of external particles. Since particle $n$ is not in the lower tree, the open index $\mu$ in $\mathcal{E}_\mu$(or $J'_\mu$) must be carried by a polarization vector $\epsilon_i$, otherwise there will be a $\epsilon_i \cdot \epsilon_j$ in $\mathcal{E}_\mu$(or $J'_\mu$), which vanishes. Thus, $J_\mu$ can be calculated as
\[
J_\mu = \frac{2}{p^2} [J' \cdot \frac{i}{2}(k - p) \mathcal{E}_{\mu} - i (\mathcal{E} \cdot q) J'_\mu].
\]
When contracted with momentum $p^\mu$,
\[
\begin{aligned}
	p^\mu J_\mu =& \frac{2}{p^2} [J' \cdot \frac{i}{2}(k - p) (\mathcal{E} \cdot p) - i (\mathcal{E} \cdot q) (J' \cdot p)]\\
	=& \frac{2}{p^2} [- i J' \cdot (p + \frac{q}{2}) (\mathcal{E} \cdot p) + i (\mathcal{E} \cdot p) (J' \cdot p) + i (\mathcal{E} \cdot k) (J' \cdot p) ]\\
	=& -\frac{i}{p^2} (q \cdot J') (\mathcal{E} \cdot p) + \frac{2 i}{p^2} (\mathcal{E} \cdot k) (J' \cdot p).
\end{aligned}
\]
We can clearly see that if the two further lower tree $J'_\mu$ and $\mathcal{E}_\mu$ satisfy $J'(q) \cdot q = 0$ and $\mathcal{E}(k) \cdot k = 0$, then $J(p) \cdot p = 0$. This relation holds recursively until we reach the edge of the lower tree, where $J'_\mu$ or $\mathcal{E}_\mu$ or both of them simply represent polarization vectors of external particles. Since external particles satisfy transverse condition, we get to the conclusion that any lower tree $J_\mu(p)$ satisfies $p\cdot J(p) = 0$ if particle $n$ is not in it.}
\par 
With these three properties stated, we can finally prove the BCJ relation for the worldline numerator. For simplicity, we will choose particle $n$ to be in the lower-tree $J_4$ in the following expression,
\[
\mathcal{M}(12;34) = \int^{+\infty}_{0} d\tau \langle [J_4 \cdot \Psi_4 e^{i k_4 \cdot x_4}] [J_{3\mu} (\dot{x}_3^{\mu} + i k_{3\rho} S_3^{\mu\rho})e^{i k_3 \cdot x_3}] [J_{2\nu} (\dot{x}_2^{\nu} + i k_{2\sigma} S_2^{\nu\sigma})e^{i k_2 \cdot x_2}] [J_1 \cdot \bar{\Psi}_1 e^{i k_1 \cdot x_1}] \rangle 
\]
where $x_4 \equiv x(+\infty)$, $x_3 \equiv x(\tau)$,$x_2 \equiv x(0)$, $x_1 \equiv x(-\infty)$ and $J_i(i = 1,2,3)$ satisfies $k_i \cdot J_i = 0$. The corresponding numerator is
\[
\begin{aligned}
	n(12;34) =& \frac{\int^{+\infty}_{0} d\tau \langle [n_4 \cdot \Psi_4 e^{i k_4 \cdot x_4}] [n_{3\mu} (\dot{x}^{\mu}_3 + i k_{3\rho} S_3^{\mu\rho})e^{i k_3 \cdot x_3}] [n_{2\nu} (\dot{x}^{\nu}_2 + i k_{2\sigma} S^{\nu\sigma})e^{i k_2 \cdot x_2}] [n_1 \cdot \bar{\Psi}_1 e^{i k_1 \cdot x_1}] \rangle}{\int^{+\infty}_{0} d\tau \langle e^{i k_4 \cdot x_4}e^{i k_3 \cdot x_3}e^{i k_2 \cdot x_2}e^{i k_1 \cdot x_1}\rangle}\\
	=& \frac{\langle [n_4 \cdot \Psi e^{i k_4 \cdot x_4}] [n_{3\mu} (\dot{x}^{\mu}_3 + i k_{3\rho} S_3^{\mu\rho})e^{i k_3 \cdot x_3}] [n_{2\nu} (\dot{x}^{\nu}_2 + i k_{2\sigma} S^{\nu\sigma})e^{i k_2 \cdot x_2}] [n_1 \cdot \bar{\Psi}_1 e^{i k_1 \cdot x_1}] \rangle}{\langle e^{i k_4 \cdot x_4}e^{i k_3 \cdot x_3}e^{i k_2 \cdot x_2}e^{i k_1 \cdot x_1}\rangle}
\end{aligned}
\]
and $n_i(i = 1,2,3)$ satisfies $k_i \cdot n_i = 0$. By directly computing the expression (and remembering that those non-vanishing $(n_i \cdot n_j)$ must involve $n_4$), we get
\[
\begin{aligned}
	n(12;34)
	=& (n_4 \cdot n_1) (n_3 \cdot i k_4) (n_2 \cdot -i k_1)\\
	& + (n_4 \cdot n_3) (i k_3 \cdot n_1) (n_2 \cdot -i k_1)\\
	& + (n_4 \cdot n_2) (i k_2 \cdot n_1) (n_3 \cdot i k_4)\\
	& + (n_4 \cdot n_3) (i k_3 \cdot n_2) (i k_2 \cdot n_1)\\
	=& (n_4 \cdot n_1) (n_3 \cdot k_4) (n_2 \cdot k_1)\\
	& + (n_4 \cdot n_3) ( n_1 \cdot k_3) (n_2 \cdot k_1)\\
	& - (n_4 \cdot n_2) (n_1 \cdot k_2) (n_3 \cdot k_4)\\
	& - (n_4 \cdot n_3) (n_2 \cdot k_3) (n_1 \cdot k_2)\,.\\
\end{aligned}
\]
The expressions denoted by $n_{1,2,3,4}$ were previously defined in
\eqref{nsforbcj}.
By permuting $(1,2,3)$, we can get $n(23,14)$ and $n(31,24)$. Summing these numerators, we have
\[
\begin{aligned}
	&n(12,34) + n(23,14) + n(31,24)\\
	=& -(n_4\cdot n_1)(n_2\cdot k_3)(n_3\cdot k_1) - (n_4\cdot n_1) (n_2\cdot
	k_4) (n_3\cdot k_1)\\
	& + (n_4\cdot n_1)(n_2\cdot k_1)(n_3\cdot k_2) + (n_4\cdot n_1)(n_2\cdot k_1)(
	n_3\cdot k_4)\\
	& -(n_4\cdot n_2)(n_1\cdot k_2)(n_3\cdot k_1) + (n_4\cdot n_2)(n_1\cdot k_3)(n_3\cdot
	k_2) \\
	& +(n_4\cdot n_2)(n_1\cdot k_4)(n_3\cdot k_2) - (n_4\cdot n_2)(n_1\cdot k_2)(n_3\cdot
	k_4)\\
	& +(n_4\cdot n_3)(n_1\cdot k_3)(n_2\cdot k_1) - (n_4\cdot n_3)(n_1\cdot k_2)(n_2\cdot
	k_3) \\
	& -(n_4\cdot n_3)(n_1\cdot k_4)(n_2\cdot k_3) + (n_4\cdot n_3)(n_1\cdot k_3)(n_2\cdot k_4)\\
	=& +(n_4 \cdot n_1)(n_2 \cdot k_1)[n_3 \cdot (k_1 + k_2 + k_4)]\\
	& +(n_4 \cdot n_2)(n_3 \cdot k_2)[n_1 \cdot (k_2 + k_3 + k_4)]\\
	& +(n_4 \cdot n_3)(n_1 \cdot k_3)[n_2 \cdot (k_1 + k_3 + k_4)]\\
	=&\ 0\,.
\end{aligned}
\]
We would like to stress that our proof is in full generality and we do not take the lines labelled $1,2,3$ or $4$ on-shell. 
This concludes our proof that the BCJ relation holds for the MHV worldline numerators.

\end{document}